\documentclass[bibyear]{aa} 
\usepackage{natbib}

\usepackage{CJK}
\usepackage{enumitem}
\usepackage{verbatim}
\usepackage{amsmath}
\usepackage{xcolor}
\usepackage{graphicx}
\usepackage{txfonts}
\usepackage[colorlinks=true,
            linkcolor=blue,
            urlcolor=blue,
            citecolor=blue]{hyperref}
\newcommand{\todo}{\ifmmode \text{\color{purple}\Huge{\(\bullet\)}} \else {\color{purple}{\Huge$\bullet$}}\fi}

\newcommand{\mstar}{M_{\star}}
\newcommand{\msun}{M_{\odot}}
\newcommand{\mbh}{M_\mathrm{BH}}
\newcommand{\mbhfp}{M_\mathrm{BH,FP}}
\newcommand{\mbhmstar}{M_\mathrm{BH,\star}}

\newcommand{\mdotbh}{\dot{M}_\mathrm{BH}}

\newcommand{\lbol}{L_\mathrm{AGN,bol}}
\newcommand{\ledd}{L_{\rm Edd}}

\newcommand{\liragn}{L_\mathrm{AGN,IR}}

\newcommand{\ltwelve}{L_{12 \mu {\rm m}}}
\newcommand{\lsix}{L_{6 \mu {\rm m}}}
\newcommand{\lxsoftabs}{L_{0.5-2\mathrm{keV}}^\mathrm{(abs,corr)}}
\newcommand{\lxhardabs}{L_{2-10\mathrm{keV}}^\mathrm{(abs,corr)}}
\newcommand{\etar}{\eta_\mathrm{rad}}

\newcommand{\lfirst}{L_\mathrm{1.4GHz}}
\newcommand{\lfirstunit}{L_\mathrm{1.4GHz}/\mathrm{W}~\mathrm{Hz}^{-1} }
\newcommand{\lvfirst}{\nu L_\nu (\mathrm{1.4GHz})}

\newcommand{\ljet}{P_\mathrm{jet}}
\newcommand{\ljetunit}{P_\mathrm{jet}/\mathrm{erg}~\mathrm{s}^{-1}}
\newcommand{\etaj}{\eta_\mathrm{jet}}

\newcommand{\lambdaedd}{\lambda_\mathrm{Edd}}

\newcommand{\Fint}{F_\mathrm{int}}

\newcommand{\Robs}{\mathcal{R_{\rm obs}}}

\newcommand{\whz}{\mathrm{W}~\mathrm{Hz}^{-1}}

\newcommand{\fsoft}{f_\mathrm{0.5-2keV}}
\newcommand{\fsoftunit}{\mathrm{erg}~\mathrm{s}^{-1}~\mathrm{cm}^{-2}}
\newcommand{\NH}{N_\mathrm{H}}
\newcommand{\NHunit}{N_\mathrm{H}/\mathrm{cm}^{-2}}

\begin{document} 
\title{eROSITA Final Equatorial-Depth Survey (eFEDS):}
\subtitle{
eFEDS X-ray view of WERGS Radio Galaxies selected by the Subaru/HSC and VLA/FIRST survey}

\author{
Kohei Ichikawa\inst{1,2,3},
Takuji Yamashita\inst{4},
Andrea Merloni\inst{1},
Junyao Li\inst{5},
Teng Liu\inst{1},
Mara Salvato\inst{1},
Masayuki Akiyama\inst{3},
Riccardo Arcodia\inst{1},
Tom Dwelly\inst{1},
Xiaoyang Chen\inst{6},
Masatoshi Imanishi\inst{4},
Kohei Inayoshi\inst{7},
Toshihiro Kawaguchi\inst{8},
Taiki Kawamuro\inst{9,4},
Mitsuru Kokubo\inst{10},
Yoshiki Matsuoka\inst{11},
Tohru Nagao\inst{11},
Malte Schramm\inst{12},
Hyewon Suh\inst{13},
Masayuki Tanaka\inst{4},
Yoshiki Toba\inst{14,15,11},
Yoshihiro Ueda\inst{14}
}

\titlerunning{eROSITA X-ray properties of WERGS radio galaxies in eFEDS}
\authorrunning{Kohei Ichikawa et al.}

\institute{
Max Planck Institut f\"ur Extraterrestrische Physik, Giessenbachstrasse 1, 85748 Garching bei M\"unchen, Germany 
\and
Frontier Research Institute for Interdisciplinary Sciences, Tohoku University, Sendai 980-8578, Japan
\and
Astronomical Institute, Tohoku University, Aramaki, Aoba-ku, Sendai, Miyagi 980-8578, Japan
\and 
National Astronomical Observatory of Japan, Mitaka, Tokyo 181-8588, Japan
\and
Department of Astronomy, University of Illinois at Urbana-Champaign, Urbana, IL 61801, USA
\and
ALMA Project, National Astronomical Observatory of Japan, 2-21-1, Osawa, Mitaka, Tokyo 181-8588, Japan
\and
Kavli Institute for Astronomy and Astrophysics, Peking University, Beijing 100871, China
\and
Department of Economics, Management and Information Science, Onomichi City University, Hisayamada 1600-2, Onomichi, Hiroshima 722-8506, Japan
\and
Nu\'{c}leo de Astronom\'{i}a de la Facultad de Ingenier\'{i}a, Universidad Diego Portales, Av. Ej\'{e}ercito Libertador 441, Santiago, Chile
\and
Department of Astrophysical Sciences, Princeton University, 4 Ivy Lane, Princeton, NJ 08544, USA
\and
Research Center for Space and Cosmic Evolution, Ehime University, 2-5 
Bunkyo-cho, Matsuyama, Ehime 790-8577, Japan
\and
Graduate school of Science and Engineering, Saitama Univ. 255 Shimo-Okubo, Sakura-ku, Saitama City, Saitama 338-8570, Japan
\and
Gemini Observatory/NSF's NOIRLab, 670 N. A'ohoku Place, Hilo, Hawaii, 96720, USA
\and
Department of Astronomy, Kyoto University, Kitashirakawa-Oiwake-cho, Sakyo-ku, Kyoto 606-8502, Japan
\and
Academia Sinica Institute of Astronomy and Astrophysics, 11F of Astronomy-Mathematics Building, AS/NTU, No.1, Section 4, Roosevelt Road, Taipei 10617, Taiwan
}

\date{Received on June 15, 2022; accepted on February 14, 2023}

\abstract{
We construct the eROSITA X-ray catalog of radio galaxies discovered by the WERGS survey that is made by the cross-matching of the wide area Subaru/Hyper Suprime-Cam (HSC) optical survey and VLA/FIRST 1.4~GHz radio survey. 
We find 393 eROSITA detected radio galaxies in the 0.5--2~keV band in the eFEDS field covering 140~deg$^2$.
Thanks to the wide and medium depth eFEDS X-ray survey down to $f_\mathrm{0.5-2keV} = 6.5 \times 10^{-15}$~erg~s$^{-1}$~cm$^{-2}$,
the sample contains the rare and most X-ray luminous radio galaxies above the knee of the X-ray luminosity function, spanning $44 < \log(\lxsoftabs/\mathrm{erg}~\mathrm{s}^{-1}) < 46.5$ at $1<z<4$.
The sample also contains the sources around and below the knee for the sources
$41 < \log(\lxsoftabs/\mathrm{erg}~\mathrm{s}^{-1}) < 45$ at $z<1$.
Based on the X-ray properties obtained by the spectral fitting, 
37 sources show obscured AGN signature with $\log (\NHunit)>22$.
Those obscured and radio AGN reside in $0.4<z<3.2$,
indicating that they are obscured counterparts of the radio-loud quasar, which are missed in the previous optical
quasar surveys. 
By combining radio and X-ray luminosities, we also investigate the jet production efficiency $\etaj = \etar \ljet/\lbol$ by utilizing the jet power of $\ljet$. 
We find that there are 14 sources with extremely high jet production efficiency at $\etaj\approx1$. 
This high $\etaj$ value might be a result of
1) the decreased radiation efficiency of $\etar<0.1$ due to the low accretion rate for those sources and/or
2) the boosting due to the decline of $\lbol$
by a factor of 10--100 by keeping $\ljet$ constant in the previous Myr, indicating the experience of the AGN feedback.
Finally, inferring the BH masses from the stellar-mass, we find that X-ray luminous sources show the excess of the radio emission with respect to the value estimated from the fundamental plane.
Such radio emission excess cannot be explained by the Doppler booming alone, and therefore disk-jet connection of X-ray luminous eFEDS-WERGS is fundamentally different from the conventional fundamental plane which mainly covers low accretion regime.
}


\maketitle
%

\section{Introduction}\label{intro}

Relativistic jets launched by supermassive black holes (SMBHs) are the most energetic particle accelerator in the universe.
The galaxies hosting such radio jets are called radio galaxies\footnote{In general, the term
``radio galaxies'' represents the edge-on view (type-2) radio AGN whose host galaxies are detectable in the optical band \citep{ant93,urr95}.
In this study, we use the term radio galaxies as AGN with significant radio detection, and therefore
radio galaxies in this study are not necessarily type-2 AGN. For more details, see Section~\ref{sec:sample:crossmatch}.} or radio active galactic nuclei (AGN). Roughly 10\% of the accreting SMBHs or AGN in the local universe are known as radio galaxies \citep{kel89,ive02,sik07,ho08}, and the jets are considered to disturb the surrounding gas and thus affect the host galaxy evolution 
by injecting energy and momentum, which results in suppression of star-formation \citep[e.g.,][]{fab12}.

Local radio galaxies at $z<0.5$ have been studied over the last 20~yrs as a showcase in the act of ``AGN feedback'' \citep{sad02,bes05,mau07,bes12}.
Thanks to the combination of multi-wavelength data including optical, infrared (IR), and X-ray bands, the obtained ubiquitous features are
that those radio galaxies are massive galaxies with $\mstar > 10^{11} \msun$ and they show very low star-formation rate with the presence of dispersing interstellar medium \citep{mor05,hol08,nes17} and/or X-ray cavities \citep[e.g.,][]{raf06,mcn07,bla19}. They are also associated with low accretion rate onto the SMBHs, i.e., Eddington ratio of $\lambda_\mathrm{Edd} < 10^{-2}$, 
suggesting that the energy release is dominated by the kinetic power by jets, not by the radiation from AGN accretion disks \citep[e.g.,][]{bes12}.

On the other hand, high-$z$ radio galaxies show a different picture.
Using over $10^3$ radio AGN selected from the
Very Large Array (VLA)-COSMOS 3~GHz large project \citep{smo17a,smo17b},
\cite{del18} demonstrated that
SMBH accretion in radio AGN are most radiatively
efficient ($\lambda_\mathrm{Edd}>10^{-2}$) at $z>1$
and they reside in star-forming galaxies, which indicates the presence of plentiful cold gas in the host galaxies.
They might be tracing a rapidly growing phase of the SMBHs before the AGN feedback, shedding light on understanding the BH growth.
This picture of radio AGN is completely different from those seen in the local universe
in the same radio luminosity range \citep[e.g.,][]{hic09}. 
Still, the survey volume of VLA-COSMOS surveys is small with the survey area of $2$~deg$^2$ so they may be missing a rare, but radio-bright population in that redshift range of $z>1$. 
On the other hand, wide area radio surveys such as VLA/FIRST, even though its sensitivity is shallow ($>1$~mJy), still miss most of the optical counterparts
in the wide-area surveys. 
Roughly 70\% of those radio emitters are optically unknown in the SDSS survey footprint \citep{ive02,hel15}, mainly due to the shallow optical depth of the SDSS survey down to only $i_\mathrm{AB}=22$~mag. 
Although the combined efforts are conducted in the previous radio surveys, their studies also indicate that wide and deep counterpart search of such radio emitters is still an unexplored frontier of such ``known unknown'' sources.

Recent Subaru/Hyper Suprime-Cam \citep[HSC; ][]{miy18} strategic survey program (called HSC-SSP) opens the wide field optical photometric view with unprecedented depth down to $i_\mathrm{AB}\sim26$ for a wide area ($\sim1100$~deg$^2$ as of January 2022).
We conducted a search for optically-faint radio-galaxies (RGs) using the Subaru HSC survey catalog \citep{aih18a} and the VLA/FIRST 1.4~GHz radio one, and we have found a large number ($>3\times10^3$ sources) of RGs at $z \sim 0$--$6$ \citep{yam18,yam20,uch21}. 
The project is called \textbf{W}ide and deep \textbf{E}xploration of \textbf{R}adio \textbf{G}alaxies with \textbf{S}ubaru/HSC
\citep[\textbf{WERGS};][]{yam18}.
They also demonstrated that over 60\% of VLA/FIRST radio populations now have reliable optical counterparts thanks to deep HSC/optical imaging\footnote{The optical counterpart fraction reaches $\sim95$\% if the sample is limited to FIRST compact sources, where the cases on the lobe contamination become negligible \citep{yam18}.}.
\cite{tob19a} compiled multi-wavelength data covering the optical and IR of the WERGS sample.
By utilizing the multi-wavelength data, \cite{ich21} showed that some WERGS radio galaxies with high radio-loudness would harbor a unique type of AGN that they are prominent candidates of very rapidly growing black holes reaching Eddington-limited accretion, and such accretion phase might accompany with the powerful jet activity \citep[e.g.,][]{tch11}.

X-ray observations provide another important view of such radio galaxies.
The X-ray emission is thought to arise from a hot electron corona above the accretion disk \citep{haa91} and is predominantly
produced by the inverse Compton scattering of photons from the accretion disk, that is, a tracer of 
the AGN accretion disk luminosity. 
This argument is largely true also for most of the radio galaxies, except blazars that produce a strong contribution of jets in the X-ray band \citep[e.g.,][]{ino09,ghi17}.
Previous X-ray surveys, primarily with \textit{Chandra} and \textit{XMM-Newton}, have made remarkable progress in the study of accretion onto SMBHs across the cosmic epoch in the universe \citep{mer14,ued14,bra15,buc15}.
However, the X-ray properties of radio galaxies across the universe are still poorly known, especially in the high-$z$ universe at $z>1$.
This is because of the combination of the two reasons; the first one is that the number density of radio galaxies is one order of magnitude smaller than radio-quiet AGN \citep[e.g.,][]{ive02}, so they are very rare and therefore X-ray surveys with large field-of-view are needed to uncover this population, and the second is the lack of such X-ray survey achieving both wide-area ($>100$~deg$^{2}$) and medium deep X-ray sensitivity \citep[$\fsoft< 10^{-14}~\fsoftunit$; e.g., see][]{mer12}. 
Therefore, most of the WERGS sample solely relied on the shallow WISE mid-IR bands for the estimation of the radiation power from the AGN  \citep{tob19a,ich21}.

eROSITA \citep[extended ROentgen Survey with an Imaging Telescope Array;][]{pre21} onboard the Spektrum-Roentgen-Gamma mission \citep[SRG;][]{sun21}, provides 
unprecedented very wide field view of X-ray imaging and spectroscopy thanks to its large collecting area
and high grasp (collecting area $\times$ FoV).
The eROSITA surveys all-sky \citep[the eROSITA All-Sky Survey: eRASS;][]{pre21} in the two energy bands; the soft (0.2--2.3~keV) and hard (2.3--8~keV) bands.
In the soft X-ray band, the eROSITA survey will achieve about 25 times more sensitive than the
ROSAT all-sky survey \citep{tru93}, while in the hard band it will provide the first ever true imaging survey of all-sky at those energies.

To demonstrate the ground-breaking survey capabilities of eROSITA as well as maximize the science exploitation of the uninterrupted eRASS program,
the eROSITA team observed the contiguous 140 deg$^2$ field called eROSITA Final Equatorial-Depth Survey \citep[eFEDS; ][]{bru21} during the SRG and performance verification phase, between 2019 November 3 and 7, reaching
a flux limit of $\fsoft = 6.5\times 10^{-15}~\fsoftunit$, which is 50\% deeper 
than the final integration of the planned four years program (eRASS8) in the ecliptic equatorial region \citep[$\fsoft = 1.1 \times 10^{-14}~\fsoftunit$][]{pre21}, 
therefore the eFEDS is considered to be a representation of the final eROSITA all-sky survey. 
Also, the 0.5--2~keV flux limit of eFEDS is significantly deeper than the expected one estimated from the WISE W3 (12~$\mu$m) band flux density limit, where $f_{\nu,\mathrm{lim}} = 1$~mJy corresponds to $f_\mathrm{0.5-2keV} = 8.3 \times 10^{-14}$~erg~s$^{-1}$~cm$^{-2}$, by assuming the local X-ray--MIR luminosity correlation of AGN \citep{gan09,asm15,ich12,ich17a,ich19a} and by assuming the X-ray photon index of $\Gamma=1.8$ \citep[e.g.,][]{ric17}.
Another advantage of the eFEDS field is its location of GAMA09 field, where multi-wavelength surveys are already conducted, including the DESI Legacy Survey \citep{dey19} and the Subaru/HSC SSP optical survey \citep{aih18a}.

In this study, we explore the AGN and host galaxy properties of the eROSITA-detected X-ray bright radio galaxies in the eFEDS field, selected by the combination of eROSITA/eFEDS, Subaru/HSC optical, and VLA/FIRST radio survey.
The joint effort between the eROSITA and Subaru/HSC teams enable us to explore
the HSC five band optical imaging and photometries for most of the eFEDS field.
Throughout this paper, we use the AB magnitude system in the optical bands and we adopt the same cosmological parameters 
as \cite{yam18}; $H_0 = 70$~km~s$^{-1}$~Mpc$^{-1}$, $\Omega_\mathrm{M}=0.27$, and $\Omega_\Lambda=0.73$.
This paper is not only part of the eFEDS paper series, but it is also labeled as the eighth paper of the WERGS paper series.

\section{Sample Selection and Properties}\label{sec:sample}

We here describe the eFEDS detected radio galaxy catalog (eFEDS-WERGS) based on the cross-matching between the VLA/FIRST radio surveys and the optical counterparts of the eFEDS X-ray sources.
The HSC-SSP currently covers $>10^3$~deg$^2$ region
including most of the eFEDS area and
the reader should refer to \cite{yam18} for the original S16A-WERGS catalog and \cite{tob19a} and \cite{ich21} for the IR catalog of the WERGS sample.

\subsection{VLA/FIRST}\label{sec:sample:first}
The VLA/FIRST survey contains radio continuum imaging
data at 1.4~GHz with a spatial resolution of 5.4~arcsec \citep[or $\lesssim 40$~kpc at $z\sim1$--$4$;][]{bec95,whi97}, 
which completely covers the footprint of the HSC-SSP Wide-layer (see the text below).
Following the same manner with \cite{yam18}, who utilized the final release catalog of FIRST \citep{hel15} with the flux limit of $>1$~mJy,
we extracted 12375 FIRST sources in the eFEDS footprint.

\begin{figure}
\begin{center}
\includegraphics[width=0.48\textwidth]{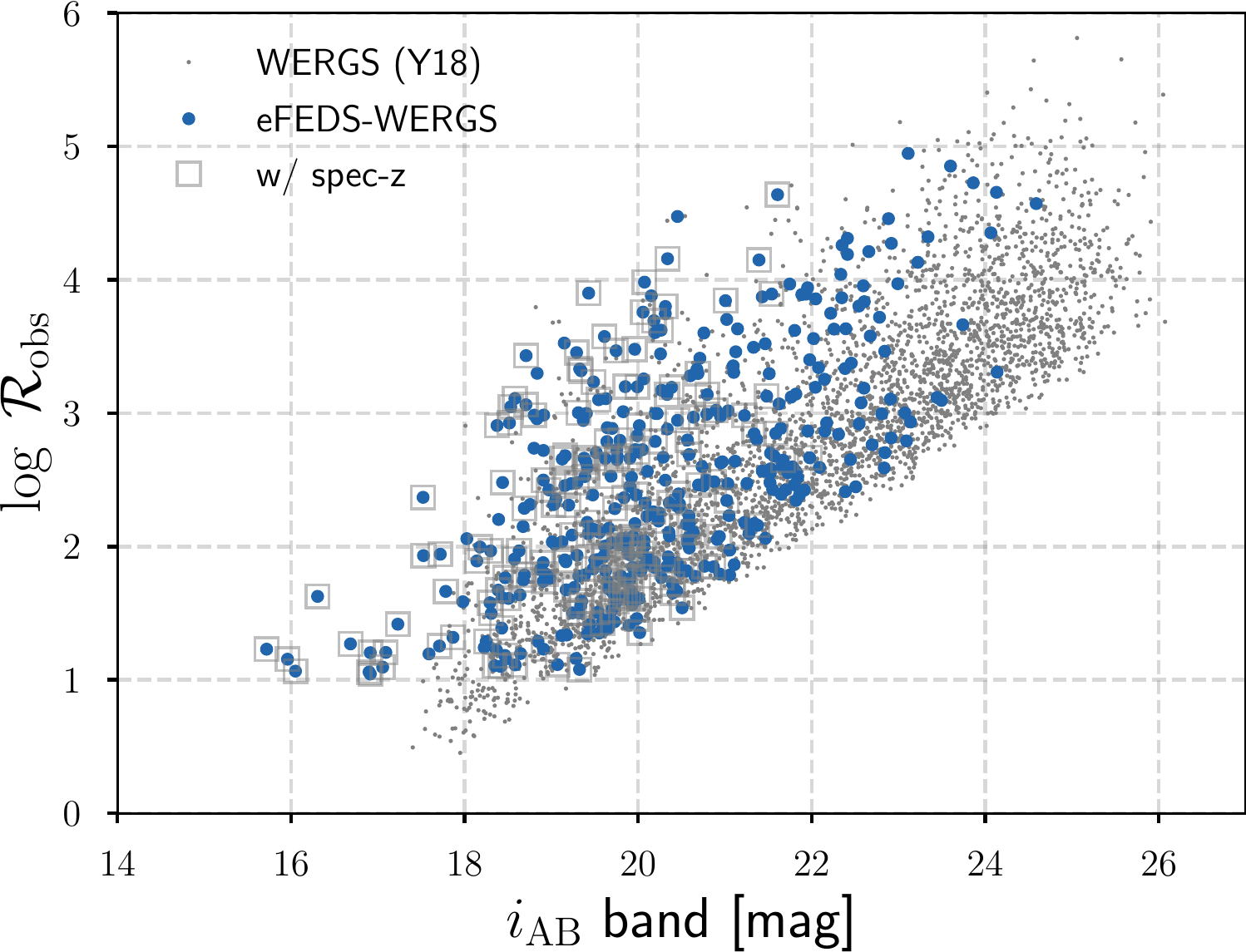}
\caption{
Logarithmic value of radio-loudness 
($\log\Robs=\log(\Fint/F_\mathrm{i band})$) versus the observed $i_\mathrm{AB}$ band magnitude of our sample. See Section~\ref{sec:sample:crossmatch} for the definition.
The original WERGS sample (3579 sources) from \cite{yam18} is shown with gray circles.
The finally selected 393~eFEDS-WERGS sources are shown with blue circles. The eFEDS-WERGS sources with spec-$z$ are encircled with gray open
squares.
The statistical errors of both values are vanishingly small and therefore are not shown.
}\label{fig:R_vs_imag}
\end{center}
\end{figure}

\subsection{eFEDS}

\subsubsection{eFEDS X-ray and Optical counterpart Catalog}\label{sec:sample:efeds_Xopt}

eROSITA team compiled the eFEDS X-ray source catalog \citep{bru21},
which contains the 27910 X-ray point sources with a detection likelihood
at 0.2--2.3~keV of \verb|DET_LIKE>6|, which corresponds to the significant detection with limiting flux of $\fsoft = 6.5 \times 10^{-15} \fsoftunit$. Out of those sources, 27369 X-ray point
sources show no extended signature with \verb|EXT_LIKE=0|.
This catalog gives a secured signal-to-noise as the detection and the estimated spurious detection rate is around $3$\%.

The counterparts have been presented in \cite{sal21} and they have been identified using the DESI Legacy Survey Data Release 8 \citep[LS8;][]{dey19}, which covers the northern hemisphere, as well as the whole eFEDS field, in three optical bands ($g$, $r$, and $z$) using telescopes at the Kitt Peak National Observatory and the Cerro Tololo Inter-American Observatory. The main reason for using LS8 in \cite{sal21} is because LS8 covers the field homogeneously and has sufficient depth, based on the expected optical properties of the X-ray population.

Two independent methods have been used and compared for the counterpart identifications: {\sc{NWAY}} \citep{sal18}  and {\sc{astromatch}} \citep{rui18}. 
Both methods made use of a large training sample of X-ray sources with secure counterparts to determine different complex priors.
Both methods provide high completeness and purity, but {\sc{NWAY}} provides a larger fraction of correct counterparts and a smaller fraction of sources for which the identification of the counterparts is not unique \citep[see][for details]{sal18}. 
The identification of the counterparts was
finalized by comparing the solutions provided by the two methods, especially based on the respective probability values of the right counterparts. 
Then, to each source, a flag was assigned with the ranking of reliability  \citep[see][for more details]{sal21}.
In this study, we used only the sources that have 
reliable optical counterparts with \verb|CTP_quality>=2|, where both {\sc{NWAY}} and {\sc{astromatch}} methods give the same counterparts, and securely extra-galactic sources with \verb|CTP_CLASS>=2|  \citep[see ][]{sal21}.
This reduces the sample size to 21952.

In the catalog, \cite{sal21} also compiled the high spatial resolution Subaru/HSC photometries obtained by the Subaru/HSC SSP survey.
The Subaru/ HSC-SSP is an ongoing wide and deep imaging survey covering five broadband filters \citep[$g$-, $r$-, $i$-, $z$-, and $y$-band;][]{aih18a,bos18,fur18,kaw18,kom18,hua18},
consisting of three layers (Wide, Deep, and UltraDeep).
\cite{sal21} utilized the Wide-layer S19A data release, which covers the almost entire eFEDS footprint. The forced photometry of $5\sigma$ limiting magnitude reaches down to 
$26.8, 26.4, 26.4, 25.5,$ and $24.7$ for $g, r, i, z,$ and $y$-band,  respectively \citep{aih18b}.
The average seeing in the $i$-band is $0.6$ arcsec, 
and the astrometric root mean squared uncertainty is about 40~mas. After removing spurious sources flagged by the pipeline, \cite{sal21} cross-matched the HSC sources with a search radius of 1~arcsec.
For more details on the HSC photometric catalog in the eFEDS field, please see \cite{sal21} and \cite{tob21} as well.

\cite{sal21} also gathered the optical spectroscopic data for obtaining the spec-$z$.
The eFEDS field has previously been observed by several spectroscopic surveys, mostly by SDSS I-IV \citep{ahu20}, GAMA \citep[DR3; ][]{bal18b}, WiggleZ \citep[final data release; ][]{dri18}, 2SLAQ \citep{cro09}, 6dFGS \citep[final data release;][]{jon09}, LAMOST \citep[DR5; ][]{luo15}, and Gaia RVS \citep[DR2][]{gai18}.
Most of the existing spectra are of high quality enough for the spec-$z$ measurements and source classification for separating the stars, quasars, and/or galaxies. The sources with good quality spec-$z$ are classified as
\verb|CTP_REDSHIFT_GRADE=5| in the optical counterpart catalog \citep[for more details, please refer][]{sal21}.

The catalog also contains photo-$z$ information. 
The computation of the photo-$z$ follows the same manner of \cite{sal09,sal11}
using \verb|LePhare| code \citep{arn99,ilb06},
with the aid of the multiwavelength photometries covering from optical to IR and fed by the comparison of the independent photometric redshift method \verb|DNNz| \citep{nis20} for the photometric redshift grade evaluation.
We utilized the redshift quality of \verb|CTP_REDSHIFT_GRADE>=3|,
whose sources have consistent photo-$z$ values between \verb|LePhare| and \verb|DNNz| codes or have the high photo-$z$ reliability.
This reduces the total number of the sample to 20850. Out of the sample, 5284 sources have spec-$z$ and the remaining 15566 sources have photo-$z$.
Please refer \cite{sal21} for more details.

\subsubsection{eFEDS X-ray spectral fitting Catalog}\label{sec:sample:efeds_Xspec}
After the identification of the optical counterpart
and the obtained redshift information, 
eROSITA team further analyzed their X-ray spectra properties. 
\cite{liu21} compiled the X-ray fitting catalog for 
21952 sources with reliable counterparts and with good signal-to-noise,
and then performed the simple spectral X-ray fitting with \verb|XSPEC| terminology of \verb|TBabs*zTBabs*powerlaw| \citep{liu21}
where power law (\verb|powerlaw|) index ($\Gamma$) set as free with the prior distribution centered at $\Gamma=2.0$ with $\sigma=0.5$,
and log-uniform prior for the column density $\NH$ (\verb|zTBabs|) with $4\times 10^{19} < \NHunit < 4 \times 10^{24}$.
The galactic absorption (\verb|TBabs|) is also applied using the total $N_\mathrm{H,Gal}$ measured by the neutral HI observations through the H4PI collaboration \citep{ben16} in the direction of the eFEDS field.
Utilizing the obtained redshift information by \cite{sal21}, \cite{liu21} derived key AGN properties, including the absorption corrected rest-frame 0.5--2.0~keV ($\lxsoftabs$) and the 2--10~keV X-ray luminosities ($\lxhardabs$), the column density ($\NH$), and X-ray spectral index ($\Gamma$).
In this study, we limit the sample only to reliable spec-$z$ and photo-$z$
measurements (as discussed in Section~\ref{sec:sample:efeds_Xopt}) and reliable spectral fitting with $N_\mathrm{H}$ measurements
of \verb|NHclass>=2|.
This reduces the sample to 17603 sources.
This is the ``parent eFEDS catalog'' used for the
cross-matching in Section~\ref{sec:sample:crossmatch}.

\subsection{Cross-matching with parent eFEDS catalog}\label{sec:sample:crossmatch}

The parent eFEDS catalog already contains both LS8 and HSC optical counterpart information,
therefore we apply a simpler cross-matching between the FIRST radio sources and the eFEDS optical counterpart, instead of the cross-matching between the WERGS catalog in the eFEDS footprint and eFEDS optical counterparts. 
Although the original WERGS catalog should follow the latter approach, this method misses optically bright sources $i_\mathrm{AB}<18$ where Subaru/HSC optical bands show a saturation \citep{tan18,yam18}.
On the other hand, the former approach has a great advantage without losing the optically bright sources since LS8 could cover reliable photometries instead.

We cross-matched the FIRST radio sample with the parent eFEDS catalog with the nearest matching  within the 1~arcsec cross-matching radius.
We used the VLA/FIRST coordinates (\verb|RA| and \verb|DEC|) and optical LS8 best counterpart coordinates (\verb|BEST_LS8_RA| and \verb|BEST_LS8_Dec|) for the parent eFEDS catalog. 
After the cross-matching, 558 sources were left.
We confirm that the probability of wrong identification with unassociated FIRST sources is negligible based on the current nearest cross-matching.
The estimated number of false identification in the sample is 0.37\footnote{In the eFEDS field, there are $N_\mathrm{eFEDS}=17603$ circles with the area of $S = \pi R^2$~arcsec$^2$, where $R=1$~arcsec. The surface number density of the FIRST sources in the eFEDS field (140~deg$^2$) is $\Omega_\mathrm{FIRST} = N_\mathrm{FIRST}/(140\times3600\times3600) = 6.8\times 10^{-6}$ counts/arcsec$^2$ by using $N_\mathrm{FIRST} = 12375$. 
 The expected coincident matching in the eFEDS is $n= N_\mathrm{eFEDS} \times S \times \Omega_\mathrm{FIRST} = 0.37$.}
 within the radius of the 1~arcsec for the searched $\sim10^4$ FIRST sources in the eFEDS footprint.

We then applied additional cuts to this sample.
We first applied removing possible
side-lobe artifacts by utilizing
\verb|SIDEPROB<0.05| \citep{hel15}. This reduces the sample size to 504.
We then set the radio-loudness cut since our target is radio galaxies whose radio emission should originate from AGN/jet activity, not from the starformation in the host galaxies.
Radio-loudness is a useful tool to distinguish the radio-quiet and -loud AGN, and
the observed radio-loudness $\Robs$ is defined as $\log \mathcal{R}_\mathrm{obs} = \log (\Fint/F_\mathrm{i band})$, where $\Fint$ is the total integrated flux density in the FIRST band, $F_\mathrm{i band}$ is the cmodel flux density in the Subaru/HSC band. 
In this study, we set $\Robs > 10$ \citep{ive02,kel89}.
This cut further reduces the sample to 429.
Note that the radio-loudness cut does not fully exclude the radio-quiet AGN population \citep[e.g.,][]{sik07,ho08,chi11}, whose radio-loudness sometimes reach $\Robs \simeq100$. Thus our sample possibly contains the radio-quiet AGN population, especially in the low radio luminosity end. On the other hand, as shown in Section~\ref{sec:results:Lzplane}, our selection cut provides that all of our sources have $\lfirst>10^{23}$~W~Hz$^{-1}$, whose radio emission is difficult for the host galaxy starformation alone to reach \citep[e.g.,][]{kim11,pad16,tad16}.

Next, we limit our sample to compact sources in the radio band. This is important in order to reduce false optical identification by mismatching the optical
sources with the locations of spatially 
extended radio-lobe emission.
We follow the same manner as the previous WERGS series \citep{yam18,ich21} using the ratio of the total integrated radio flux density to the peak radio flux density $f_\mathrm{int}/f_\mathrm{peak}$.
They treated the source as compact if the ratio fulfills  either of the two following equations,
\begin{align}
&f_\mathrm{int}/f_\mathrm{peak} < 1 + 6.5 \times (f_\mathrm{peak}/\mathrm{rms})^{-1}\\
&\log (f_\mathrm{int}/f_\mathrm{peak}) < 0.1,
\end{align}
where $f_\mathrm{peak}/\mathrm{rms}$ is the S/N, and rms is a local rms noise in the FIRST catalog, and the above two equations are obtained from the study by \cite{sch07} \citep[see also the original criterion: ][]{ive02}.
This reduces the final sample to 393 sources.
The sources are shown as blue-filled circles in the 
$\Robs$--$i_\mathrm{AB}$ plane in Figure~\ref{fig:R_vs_imag}.

We also investigated how much our radio compact source selection would affect the radio luminosity bias in the sample. In general, the selection of the smaller radio sizes tends to be radio fainter (and therefore could be potentially younger than other radio AGN) as reported in the FR0 sources \citep[e.g.,][]{bal15a,bal18c} and some radio compact sources in the local universe \citep[e.g.,][]{cap17,jim19}. 
In our case, the sample shows that the mean radio luminosities 
are almost comparable between the two subsamples of extended and compact, with 
$ \langle \log (L_\mathrm{1.4GHz}/\mathrm{W~Hz{^{-1}}}) \rangle = 25.6 \pm 1.0$ for the extended and 
$ \langle \log (L_\mathrm{1.4GHz}/\mathrm{W~Hz{^{-1}}}) \rangle = 25.5 \pm 1.1$ for the compact sources.
Therefore, we conclude that our sample selection limited to the radio compact sources generally does not affect the main results hereafter.

\subsection{Additional Analysis and Physical Quantities}\label{sec:sample:additional}

\subsubsection{Luminosities}\label{sec:sample:luminosity}

While the X-ray luminosities are already calculated for all sources in this study \citep{liu21},
we here summarize how other luminosities are calculated for other energy bands.

The 1.4~GHz radio luminosities ($\lfirst$) are calculated by following the same manner of WERGS catalog \citep{yam18,ich21}, in which the integrated flux density ($f_\mathrm{int}$) in the VLA/FIRST catalog \citep{hel15} is used, and $k$-correction is applied by assuming the power-law radio spectrum with $f_\nu \propto \nu^{\alpha}$, and $\alpha=-0.7$ is used for this study \citep[e.g.,][]{con92}.
The uncertainty of the radio luminosities mainly originates from the assumed $\alpha$, which actually varies source by source.
Although there are very steep spectral sources with $\alpha < -1.0$ \citep{kom94} and flat spectral or very young AGN sources show $\alpha > -0.5$ \citep{blu99}, 
most of the radio galaxies reside within the range of $-1<\alpha<-0.5$ \citep{deg18,tob19a}.
This uncertainty of $\alpha$ produces maximum uncertainty of 0.2~dex in the luminosity, thus we add the conservative (=over-estimated) uncertainty, 0.2~dex, to the obtained 1.4~GHz luminosities.

The IR luminosities are also calculated by using the obtained IR fluxes from the WISE all-sky survey \citep{wri10}.
The eFEDS optical counterpart catalog already
compiles the ``unWISE'' flux with a unit of nanomaggie \citep{sch19}.
We adopted the quality flag of \verb|FRAC_FLUX<0.5|, which is the same manner as \cite{tob21}, for avoiding the
sources with a severe flux contribution from their neighborhoods,
which reduces the total available sample from 
393 into 323, 345, 198, and 46 source
at W1 (3.4~$\mu$m), W2 (4.6~$\mu$m), W3 (12~$\mu$m), and W4 (22~$\mu$m) band, respectively.
The obtained flux is first converted into the flux density at each band, and then applied the $k$-correction to shift to the rest-frame flux densities by assuming a typical AGN IR SED template obtained by \cite{mul11}.
The rest-frame band luminosities ($L_{3.4 \mu\mathrm{m}},L_{4.6 \mu\mathrm{m}}, \ltwelve, L_{22 \mu\mathrm{m}}$) in the unit of erg~s$^{-1}$ are calculated based on those rest-frame flux densities.
The 6~$\mu$m IR luminosity is also calculated by assuming the
SED of \cite{mul11}, with the conversion factor of $L_\mathrm{6\mu m} = L_\mathrm{12\mu m}/1.74$.
Considering the measurement error of WISE band is always vanishingly small, with $\Delta \mathrm{W?}<0.01$~mag ($?=1,2,3,4$) for most of the sources,
the scatter of the obtained IR luminosities are dominated from the uncertainty of the $k$-correction based on the assumed AGN SED, which is around 0.2--0.3~dex \citep{mul11}.
Therefore, we add up 0.3~dex undertainties to the all IR luminosity errors.

The bolometric AGN luminosities ($\lbol$) are
estimated from $\lxhardabs$ in \cite{liu21} by using the luminosity dependent bolometric correction of \cite{mar04}.
For the sources with low SN, $\lxhardabs$ are extrapolated from $\lxsoftabs$ on
the fixed powerlaw and therefore the constant conversion factor of $\lxhardabs/\lxsoftabs=1.16$ \citep[see more details in ][]{liu21}. We also confirmed that the bolometric luminosities are almost same within the scatter of $\sigma=0.5$~dex even if $\ltwelve$ is utilized by assuming the conversion of $\lbol \simeq 2 \liragn$ \citep{del14,ina18} and $\liragn/\ltwelve=2.33$ \citep{mul11}.

\begin{figure}
\begin{center}
\includegraphics[width=0.42\textwidth]{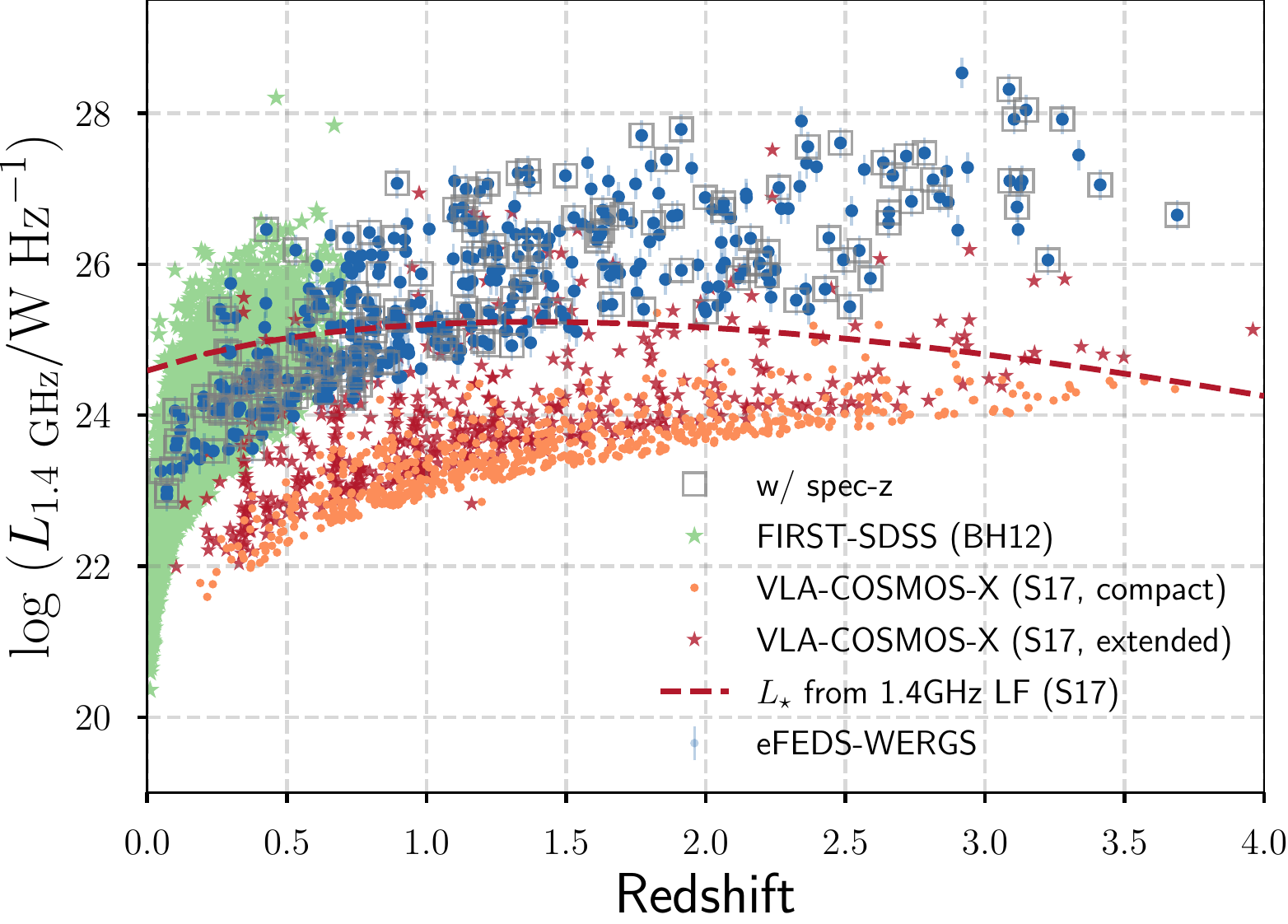}
\\
\includegraphics[width=0.42\textwidth]{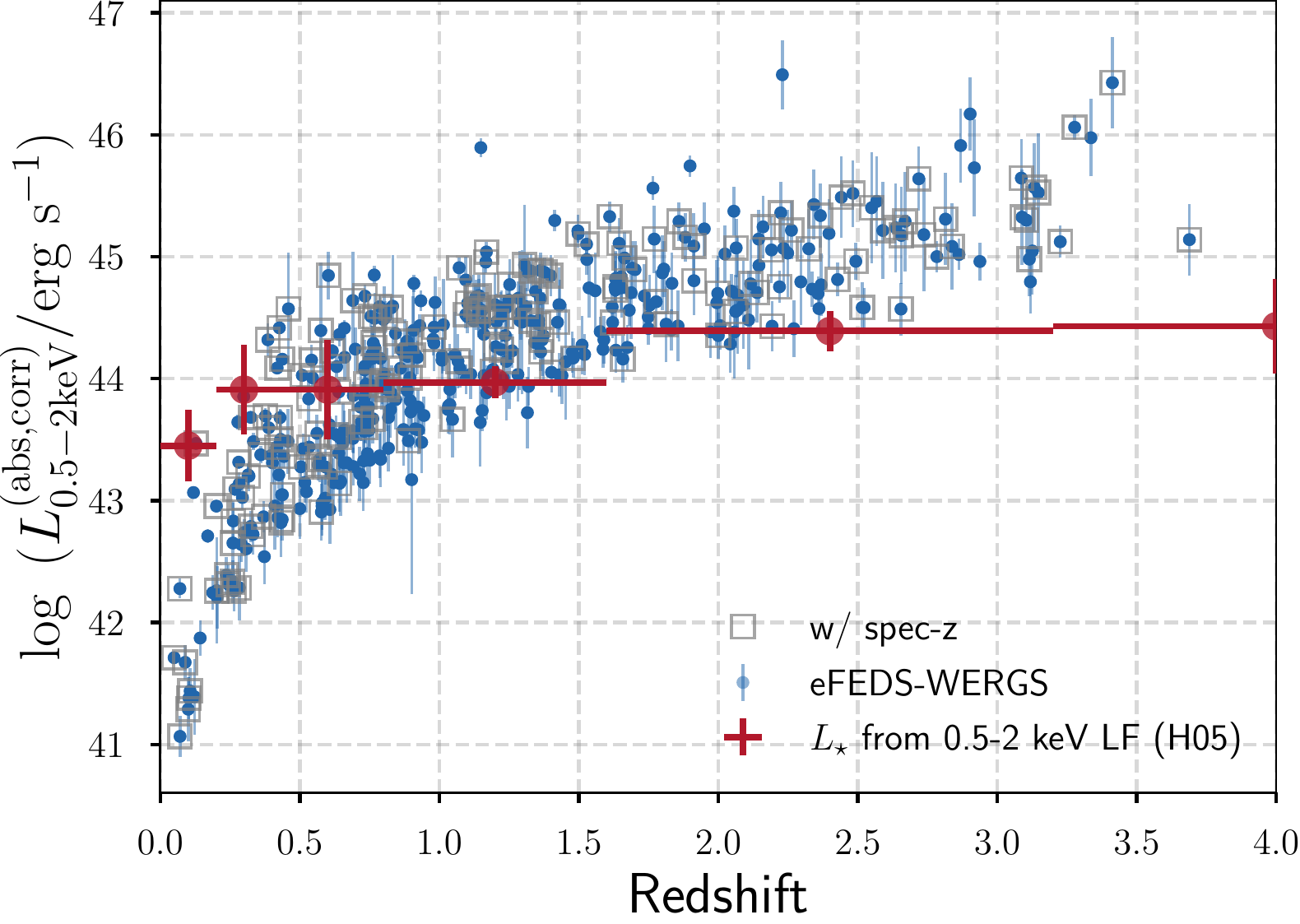}\\
\includegraphics[width=0.42\textwidth]{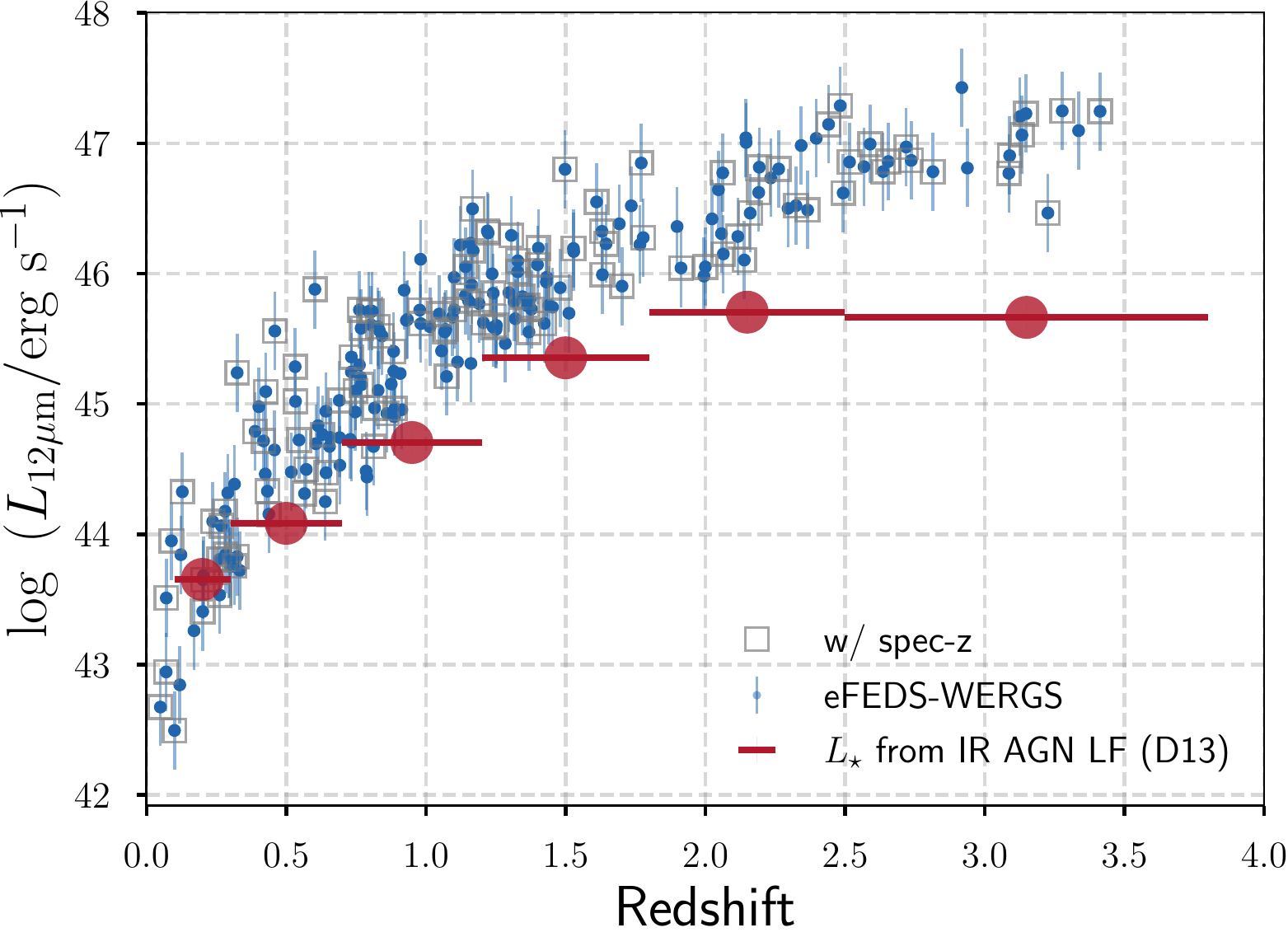}
\caption{
Luminosities of eFEDS-WERGS sample as a function of 
redshift. The points are the same as in Figure~\ref{fig:R_vs_imag}.
(Top) Rest-frame
radio luminosity $\log \lfirst$ ($\whz$) as a function of redshift, where $L_\mathrm{1.4GHz}$ is obtained from VLA/FIRST \citep{hel15},
and the $k$-correction was made by \cite{yam18} based on redshift.
The green stars and orange circle/red star points are data obtained from FIRST-SDSS \citep{bes12} and VLA-COSMOS catalog, respectively.
(Middle) 0.5--2~keV AGN luminosity ($\lxsoftabs$) as a function of redsfhit.
The red-filled circle with error-bar represents the break luminosity at 0.5--2~keV obtained from \cite{has05}.
(Bottom) 12~$\mu$m luminosity $\ltwelve$ (erg~s$^{-1}$) as a function of redshift.
The k-correction is made by assuming the AGN SED of \cite{mul11}. The red filled circle with error-bar represents the break luminosity estimated from the IR AGN studies \citep{del14}.
}\label{fig:L_vs_z}
\end{center}
\end{figure}

\subsubsection{2D image decomposition}\label{sec:sample:2Ddecomp}

To obtain the host galaxy properties such as the stellar-mass
and radial profile, we performed the 2D image decomposition utilizing the high spatial resolution HSC optical images to the final sample.
We followed the same manner as \cite{li21}, who have recently conducted the HSC image decomposition of
the SDSS quasars at $z<0.8$ based on the image modeling tool
originally optimized for the lensing detection \citep{bir15,bir18}.
We briefly summarize the analysis here.
We first chose the HSC $i$-band among the 5 HSC optical bands as a fiducial frame for the image decomposition.
We then performed the 2D brightness-profile fitting using 
one point-spread function (representing the point source)
and one single elliptical Sersic model (representing the host galaxy),
and determined the galactic structural properties including the
Sersic index ($n$) and the half-light radius ($R_\mathrm{e}$).
The final best fit model image is measured by using the 
$\chi^2$ minimization with the particle-swarm optimization algorithm.

We also obtained the host properties by utilizing the 
decomposed host galaxy emission at $grizy$ bands.
We used the SED fitting code \verb|CIGALE| \citep{boq19} to
derive stellar mass $\mstar$ and rest-frame colors of the AGN
host galaxies. 
We applied the model SEDs with a delayed star-formation history,
stellar-sysnthesis model of \cite{bru03}, and a \cite{cha03} initial mass function, and a \cite{cal00} attenuation law.

One limitation of image decomposition with the combination of the HSC
image is that its application returns the reasonable fitting only
up to $z<1$, beyond which the morphology information is hard to obtain
because the surface brightness dimming produces lower values of the host-to-total flux ratios that make the host galaxy challenging to detect even for the deep HSC data \citep{ish20,li21}.
We follow the same manner of \cite{li21}, using those morphological
parameters for objects at $z<1.0$ with reduced $\chi^2 <5$ to ensure
reliable host properties.
This leaves 139 sources.

In summary, our sample contains 393 sources spanning redshift 
range up to $0<z<6$, 391 out of them are at $z<4$.
The sample consists from the reliable redshift information; 180 spec-$z$ confirmed sources and 213 reliable photo-$z$ estimations.
The total sample (393 sources) covers X-ray and radio luminosities as well as the column density $N_\mathrm{H}$ measurements,
supplemented by the IR luminosities obtained by WISE (198 sources at W3).
Especially at low-$z$ sources at $z<1$, 139 sources in the sample
has host galaxy properties such as stellar-mass thanks
to the image decomposition technique by utilizing the high spatial resolution optical images of Subaru/HSC.

\begin{figure*}
\begin{center}
\includegraphics[width=0.43\textwidth]{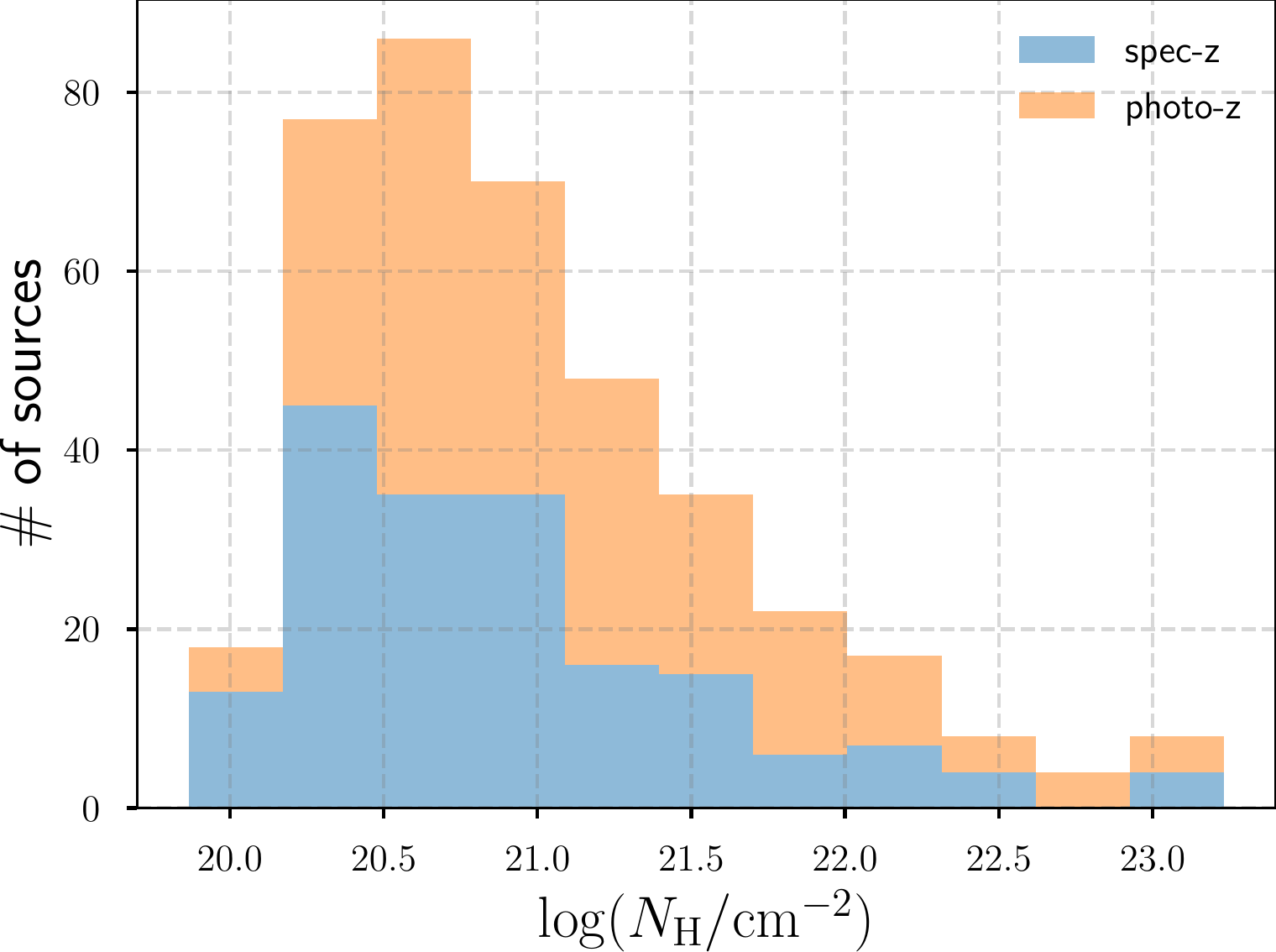}~
\includegraphics[width=0.45\textwidth]{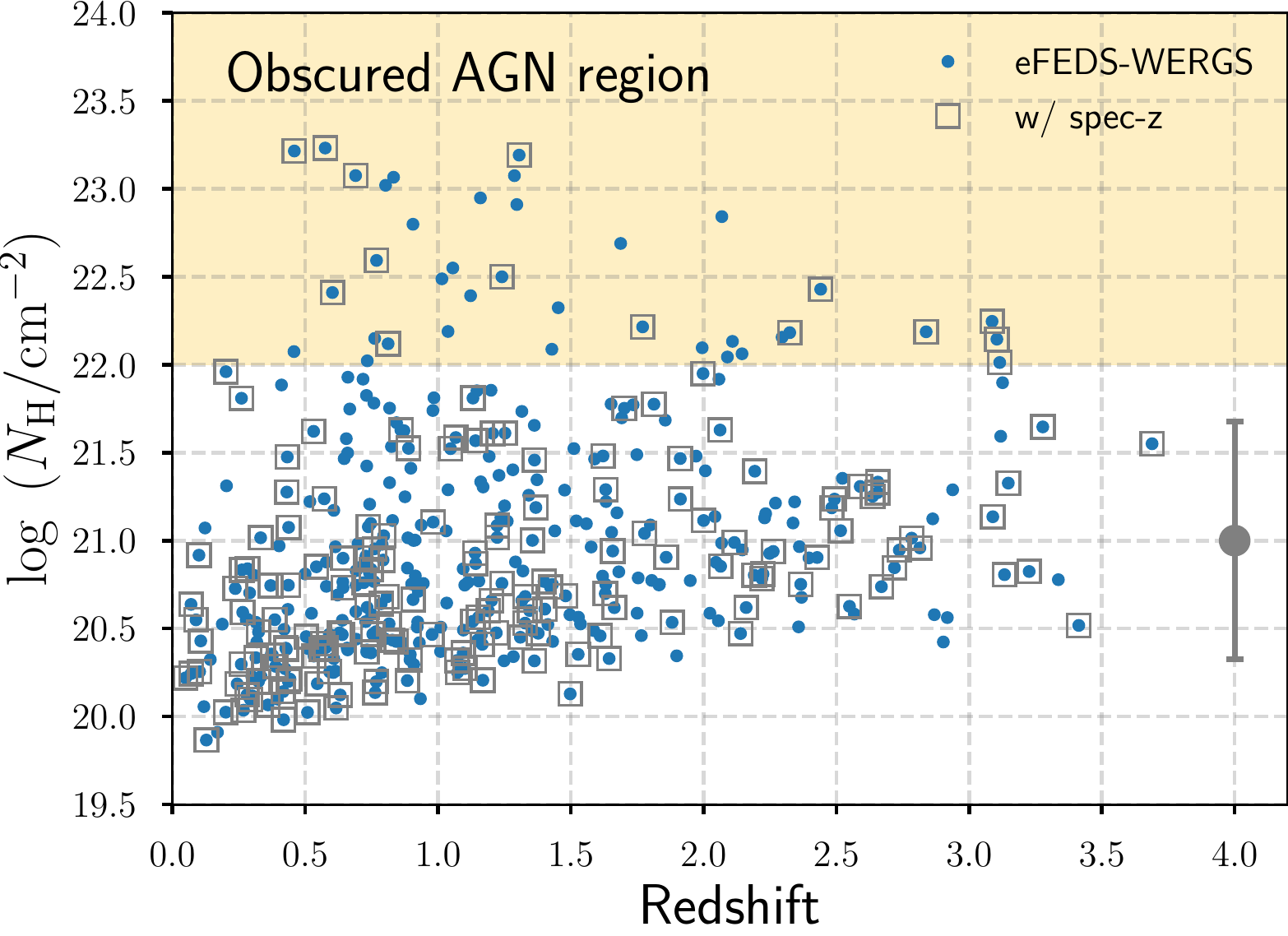}\\
\includegraphics[width=0.43\textwidth]{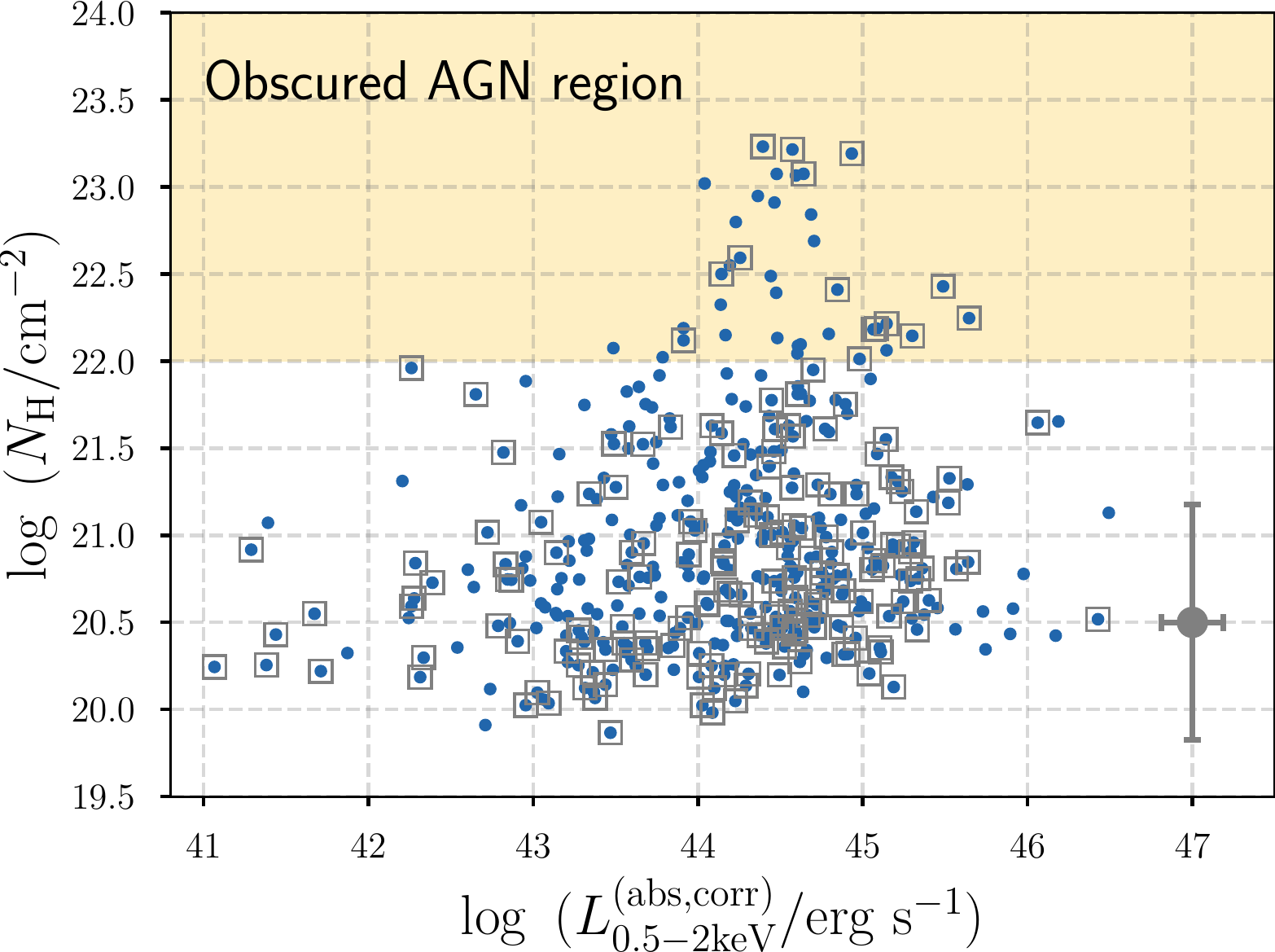}~
\includegraphics[width=0.45\textwidth]{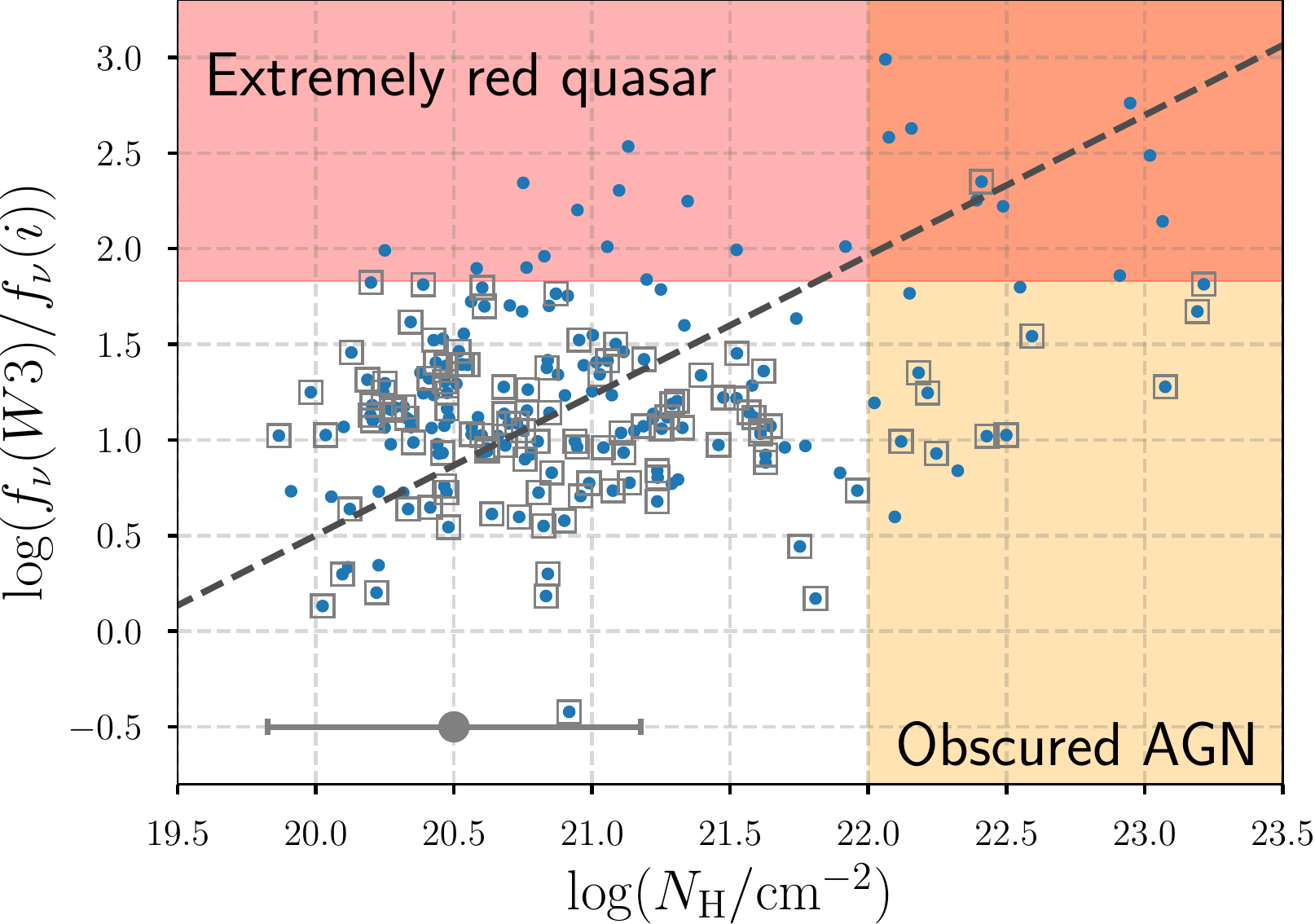}~
\caption{
(Top left) The histogram of $\log (\NH/\mathrm{cm}^{-2})$.
The sources are divided into two subgroups with spec-$z$ (cyan) and phot-$z$ (orange).
(Top right) $\log \NH$ as a function of redshift.
The gray errorbar shows a typical logarithmic error of $\log \NH$.
(Bottom left) $\log \NH$ as a function of 0.5--2~KeV X-ray luminosity
($\lxsoftabs$).
The orange shaded area represents the region of obscured AGN
with $\log (\NHunit) > 22$.
The symbols are same as Figure~\ref{fig:R_vs_imag}
and the gray errorbar shows a typical logarithmic error of $\log \NH$ and $\log \lxsoftabs$.
(Bottom right) 
The relation between the flux density ratio of 12~$\mu$m to optical $i_\mathrm{AB}$ band and the $\log(\NHunit)$.
The dashed black line represents the slope of the fitting.
The red-shaded area represents the region of 
extremely red quasars
which fulfill $f_\nu\mathrm{(12 \mu m)}/f_{\nu,\mathrm{i-band}}>10^{1.84}$ \citep{ham17}. The orange-shaded area represents obscured AGN region.
The other symbols are the same as in Figure~\ref{fig:R_vs_imag}.
}\label{fig:NH}
\end{center}
\end{figure*}

\begin{figure*}
\begin{center}
\includegraphics[width=0.33\textwidth]{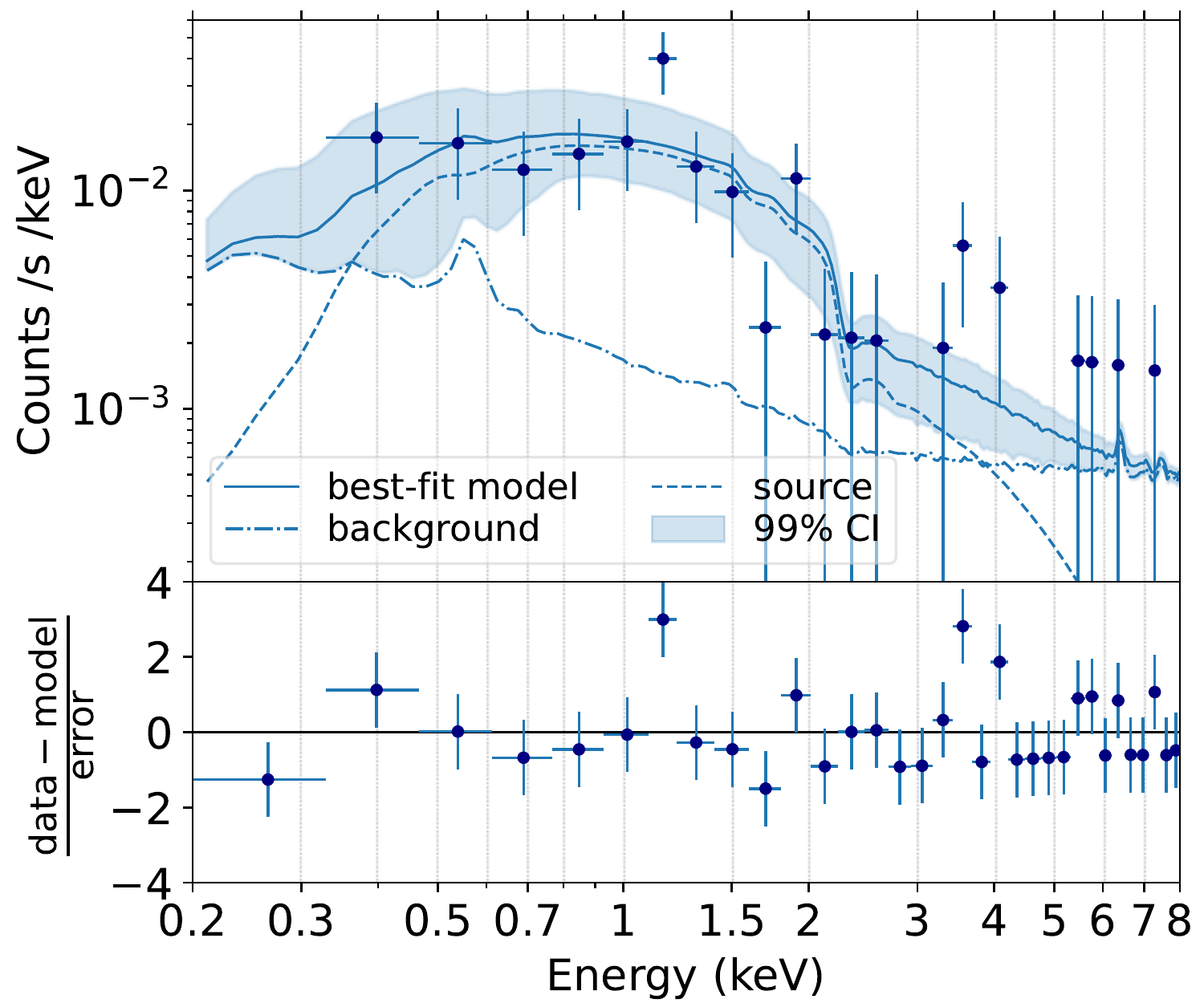}~
\includegraphics[width=0.33\textwidth]{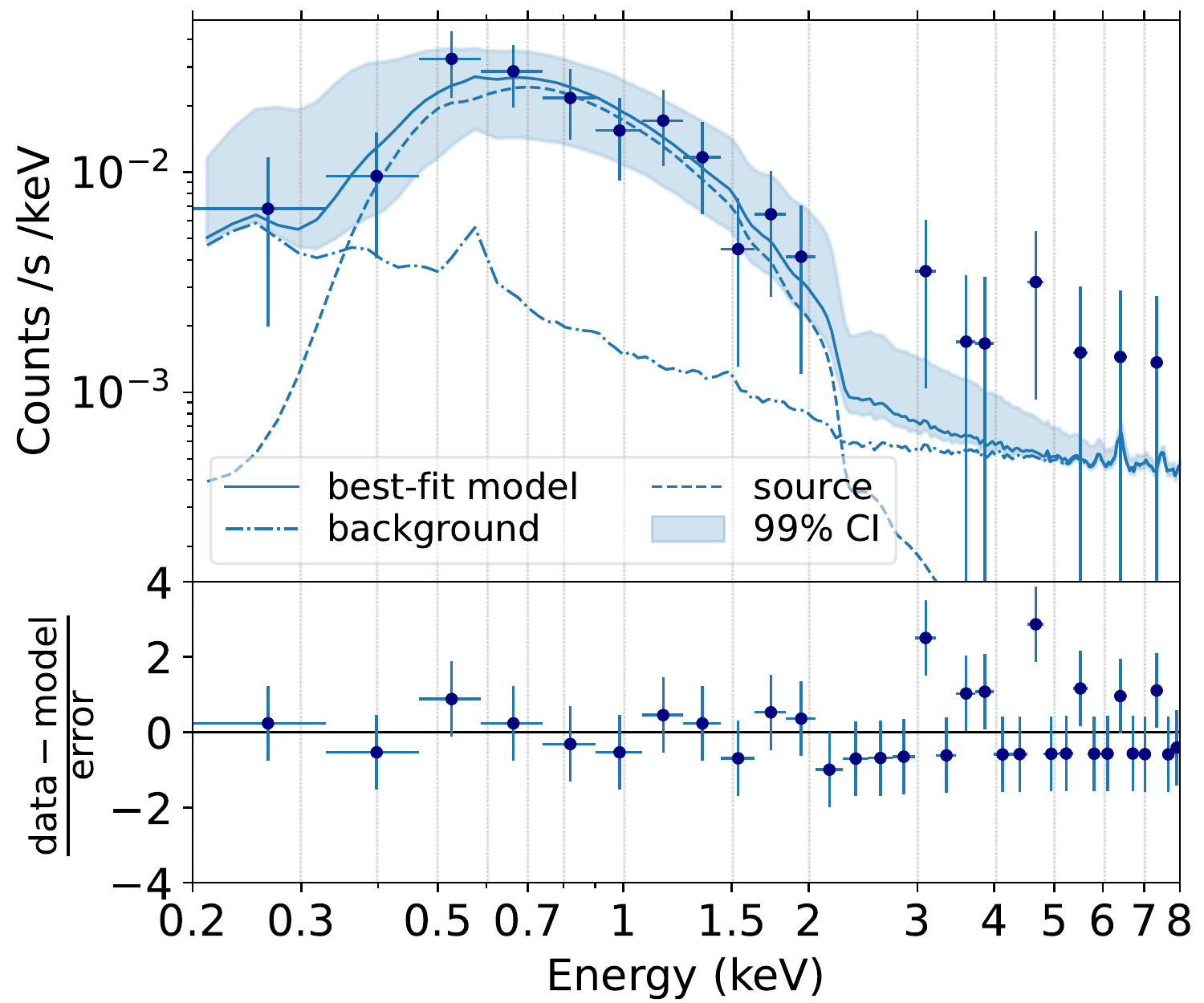}~
\includegraphics[width=0.33\textwidth]
{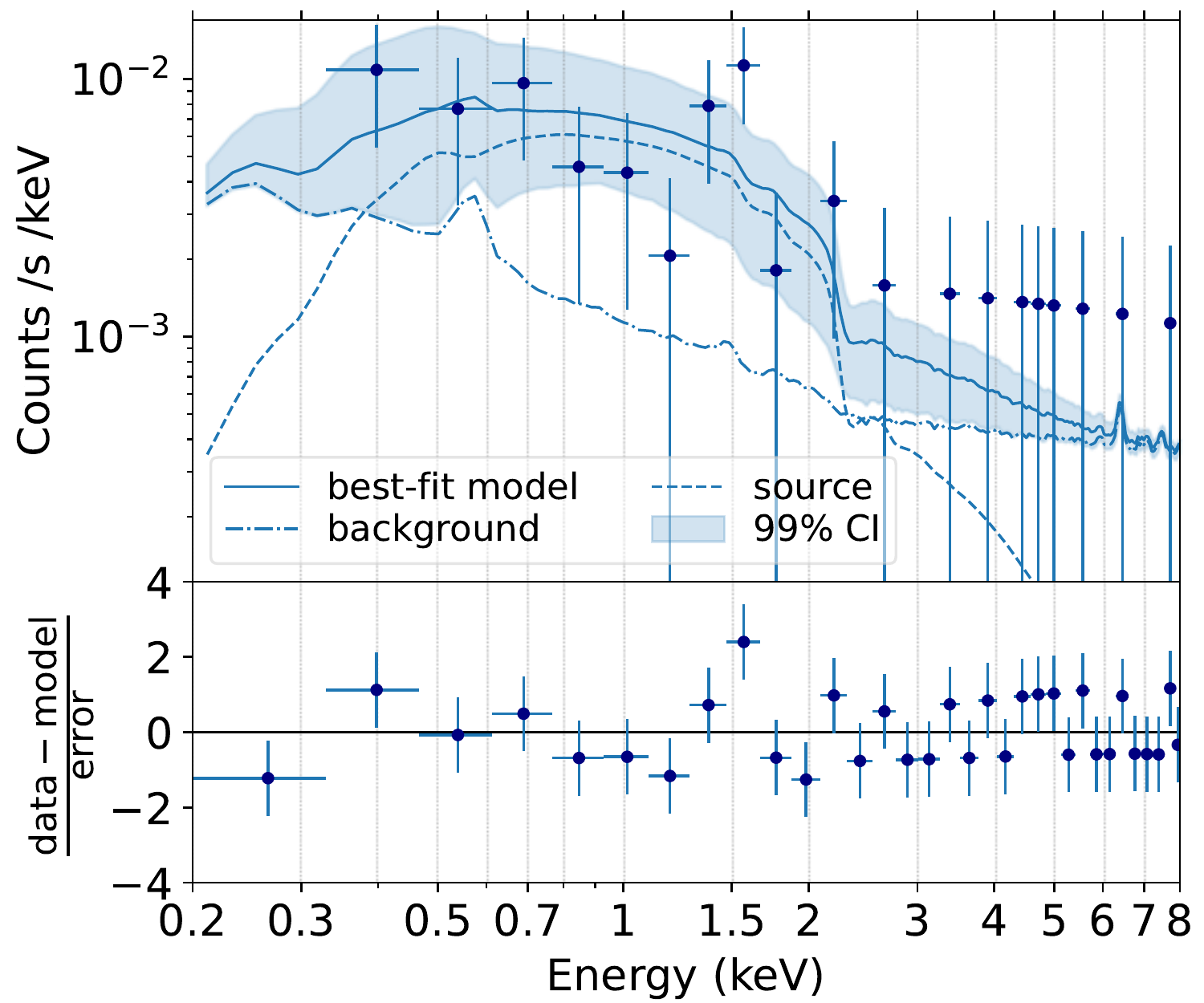}~\\
\includegraphics[width=0.7\textwidth]
{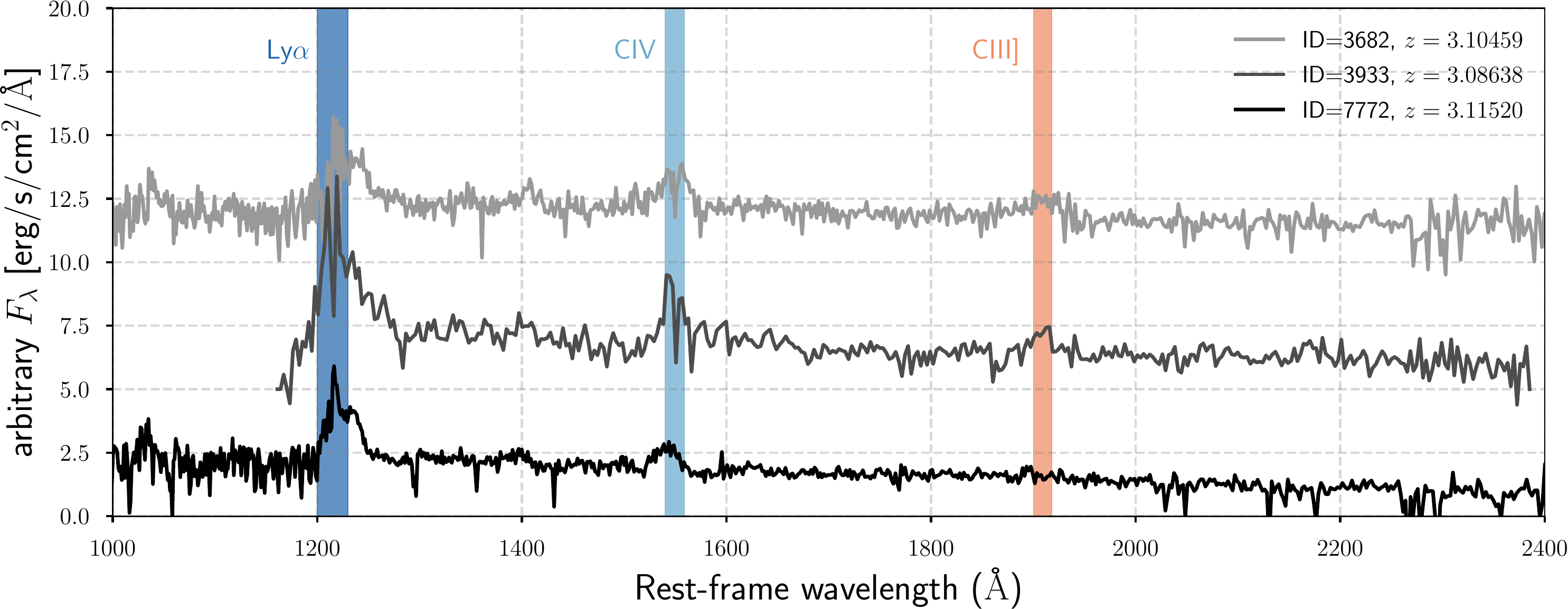}
\caption{
(Top)
X-ray spectra of three obscured radio galaxies at $z>3$ obtained by eROSITA.
All three sources are spec-$z$ confirmed sources.
The left panel is ERO\_ID=3682, $z=3.10$, $\log (\NHunit)=22.1$.
The middle panel is ERO\_ID=3933, $z=3.09$, $\log (\NHunit)=22.1$.
The right panel is ERO\_ID=7772, $z=3.11$, $\log (\NHunit)=22.0$.
The best-fit model and its 99\% confidence interval are displayed as blue-shaded regions.
For each source, the lower panel displays the ratio to the best-fit model.
(Bottom) Optical spectra of the three obscured radio galaxies at $z>3$. The spectra for ERO\_ID=3682 and 7772 are taken from the SDSS survey \citep{abd21}
and the one for ERO\_ID=3933 is taken from the WiggleZ survey, which uses AAT/AAOmega spectrograph \citep{dri18}. }
\label{fig:Xspec_obsAGN}
\end{center}
\end{figure*}

\section{Results and Discussion}\label{sec:results}

\subsection{Basic Sample Properties}\label{sec:results:basic}

\subsubsection{$\Robs$ vs. $i_\mathrm{AB}$-band magnitude}\label{sec:results:Robs_vs_imag}
Figure~\ref{fig:R_vs_imag} shows the 
distribution of $\Robs$ as a function of the
observed $i_\mathrm{AB}$ magnitude.
The gray point shows the WERGS sample \citep[e.g.,][]{yam18}, spanning down to $i_\mathrm{AB}\sim26$, but has a saturation limit at $i_\mathrm{AB}\sim18$ as discussed in Section~\ref{sec:sample:crossmatch}.
The eFEDS detected WERGS sources are mainly
clustered at $16 < i_\mathrm{AB} < 22$ (337 out of 393 sources) with a median
of $\left<i_\mathrm{AB}\right> = 20.1 \pm 1.5$, and with
a small fraction of sources down to $22<i_\mathrm{AB}<24.6$ (56 sources).
The sources enclosed by the open square are
spec-$z$ sources and over half of the sources have spec-$z$
information down to $i_\mathrm{AB}\sim21$ (completeness of 61\%), 
thanks to the power of SDSS IV spectroscopic follow-up (Merloni et al. in prep.,).

\subsubsection{L-z plane}\label{sec:results:Lzplane}

Figure~\ref{fig:L_vs_z} shows 1.4~GHz, absorption corrected 0.5--2~keV, and 12~$\mu$m luminosities
of our sample as a function of redshift
at $0<z<4$\footnote{There are two sources at $z>4$, but are not shown in Figure~\ref{fig:L_vs_z} for the illustrative purpose.
Both are photo-$z$ sources, but each shows a clear $g$-dropout ($z=4.2$) and $r$-dropout feature ($z=5.4$).}.
The top panel of Figure~\ref{fig:L_vs_z} shows how uniquely eFEDS-WERGS covers the plane of $\lfirst$--$z$.
Our eFEDS-WERGS sample covers relatively
high radio luminosities with a median of  $ \langle \log (L_\mathrm{1.4GHz}/\mathrm{W~Hz{^{-1}}}) \rangle = 25.5$, 
which is roughly two orders of magnitude
brighter than the overall VLA-COSMOS sources with significant X-ray detections \citep[VLA-COSMOS-X, $ \langle \log (L_\mathrm{1.4GHz}/\mathrm{W~Hz{^{-1}}}) \rangle = 23.6$;][]{smo17a,smo17b} across the redshift range of $0<z<4$,
indicating that most of our sample is composed of rare and radio-powerful AGN thanks to the wide area coverage of eFEDS-WERGS sample with $>100$~deg$^2$.

We also investigate the impact of the radio observation baseline differences on the luminosity discrepancy between our
eFEDS-WERGS sample and the VLA-COSMOS-X. 
While the VLA-FIRST (and therefore the eFEDS-WERGS) sources are based on observations from the VLA in its B-configuration, the VLA-COSMOS(-X) sample is constructed from VLA observations in the longer A configurations and supplementary observations in the shorter C configurations. 
As demonstrated by \cite{smo17a}, the VLA-COSMOS(-X) sources are classified as radio-extended or compact based on the flux ratio of total to peak radio flux, and its total flux is complimented by their C-configuration observations.
The top panel of Figure~\ref{fig:L_vs_z} shows that although the radio-extended or VLA-COSMOS-X sources (red stars) have higher radio-luminosities of $ \langle \log (L_\mathrm{1.4GHz}/\mathrm{W~Hz{^{-1}}}) \rangle = 23.9$ compared to the radio-compact VLA-COSMOS sources (orange stars) with $ \langle \log (L_\mathrm{1.4GHz}/\mathrm{W~Hz{^{-1}}}) \rangle = 23.5$. However, this difference does not bridge the gap in radio luminosity between the eFEDS-WERGS and VLA-COSMOS-X.

Figure~\ref{fig:L_vs_z} (top panel) also shows the comparison from the sample by the FIRST-SDSS survey \citep{bes12}. 
Our sample covers higher-$z$ sources
($\langle z \rangle = 1.1$ for eFEDS-WERGS, and
$\langle z \rangle = 0.2$ for the FIRST-SDSS sources) thanks to relatively deeper X-ray follow-ups and the SDSS IV spectroscopic survey, as well as the power of deep photometries of Subaru/HSC.
The red dashed line represents
the 1.4~GHz break luminosity of lower-power radio AGN, which is obtained from the 1.4~GHz luminosity function studies by \cite{smo17a}.
More than half/most of the sources are located above the
line of the break luminosity at $z>1.0$/$z>1.5$, respectively. This indicates that eFEDS-WERGS are relatively rare and radio luminous AGN,
and their luminosity range is similar level with the high radio luminosity AGN populations such as FRI ($10^{25} < \lfirst / \mathrm{W}~\mathrm{Hz}^{-1} < 3\times 10^{26}$) and FR II (mainly with $\lfirst>3\times10^{26}$~W~Hz$^{-1}$) \citep{dun90,wil01}.

The middle panel of Figure~\ref{fig:L_vs_z} shows the absorption corrected 0.5--2~keV X-ray luminosity ($\lxsoftabs$) as a function of $z$. 
The red filled circle represents the
break 0.5--2~keV AGN luminosity
($L_{\star,0.5-2 \mathrm{keV}}$)
at each redshift obtained from \cite{has05}.
This clearly shows that our sample covers the relatively rare and luminous end of the X-ray AGN at $z>1$ and cover the sources around the shoulder at $0.2<z<1$, thanks
to the wide coverage of eFEDS field.
We discuss the possible jet contamination in the X-ray band later in Section~\ref{sec:results:Lx_vs_L6}.

One notable trend in the middle panel of Figure~\ref{fig:L_vs_z} is that
eFEDS-WERGS catalog contains the most X-ray luminous radio AGN exceeding $\lxsoftabs > 10^{45.5}$~erg~s$^{-1}$. 
Even limiting the sample with the spec-$z$ sources,
there are 7 sources above $\lxsoftabs > 10^{45.5}$~erg~s$^{-1}$
and 2 out of them are above $\lxsoftabs > 10^{46}$~erg~s$^{-1}$,
whose expected bolometric luminosity reaches $\lbol \simeq 10^{48}$~erg~s$^{-1}$,
which is equivalent with the Eddington limit luminosity of maximum mass SMBHs with $\mbh = 10^{10}\msun$ \citep[e.g.,][]{kor13}. This suggests that they are a promising candidate population for the super-Eddington phase, and the association of strong 1.4~GHz emission indicates that the super-Eddington phase is somehow linked to the jet emission.
Although we cannot discard the possibilities that they are also blazars and therefore both radio and X-ray emission is boosted by jet, those sources are not listed in the eFEDS-blazar catalog (Collmer et al. in prep.) based on the cross-matching to the optical counterparts of previously known blazar catalogs such as BZCAT \citep{mas15} and gamma-ray detected blazars.
In addition, the archival SDSS spectra show that at least three sources show quasar-like spectra, showing a broad Ly$\alpha$
ad CIV emission with a blue continuum, suggesting that the accretion disk emission is dominated at least in the optical band.

The bottom panel of Figure~\ref{fig:L_vs_z} shows the rest-frame 12~$\mu$m luminosity as a function of $z$, which is considered to be
mainly dominated by AGN dust emission \citep[e.g.,][]{gan09,asm15,ich17a}.
The red-filled circles represent
the break 12~$\mu$m AGN luminosity ($L_{\star,12\mu\mathrm{m}}$) at each redshift obtained from \cite{del14}, who compiled the break AGN bolometric luminosities from the IR SED fitting, with the combination of the bolometric correction of $L_\mathrm{AGN,bol}/L_\mathrm{AGN,TIR} \simeq 3$ \citep{del14} and the 12~$\mu$m to total IR AGN conversion factor of $L_\star(\mathrm{TIR})/L_\star(12 \mu\mathrm{m}) = 2.3$ \citep{mul11,ich19a}.
Again, our sample is located above the break luminosity of IR AGN luminosity function \citep{del14}, indicating that our eFEDS-WERGS sample covers the relatively rare and luminous end of AGN population at $z>0.5$ and the sample locates around the break luminosity at $z<0.5$.
This illustrates a slight difference in the location of AGN luminosities between the X-ray and MIR bands at $z<1$, where most of the 12~$\mu$m sources are located on or above the shoulder at $0.5<z<1$. This seemingly discrepant location might be a result of the contamination from the host galaxies into the 12~$\mu$m band,
which is sometimes significant especially at the lower luminosity end
at $\ltwelve < 10^{44}$~erg~s$^{-1}$ \citep{ram09,alo11,ich15}. 
We will discuss this point later in Section~\ref{sec:results:Lx_vs_L6}.

To summarize, Figure~\ref{fig:L_vs_z} illustrates two important properties.
One is that the eFEDS-WERGS sample at $z<1$
covers the sources around the break luminosity
of each luminosity function in each wavelength band.
The second is that the sample at $1<z<4$ covers the sources 
above the break luminosities, indicating that they are
rare and luminous AGN populations in radio, X-ray, and IR bands at each redshift range in $1<z<4$.

\subsection{Obscuration Properties}\label{sec:results:obscuration}

The X-ray spectral analysis provides $\NH$, which is an indicator of obscuring gas properties around AGN.
eROSIA already discovered statistically large number of obscured ($\NH > 10^{22}$~cm$^{-2}$) AGN at $z>0.5$ in the eFEDS field \citep{bru21a,tob21}.
As already discussed in \cite{liu21},
we note that the uncertainty of $\NH$ is not small, reaching $\Delta \log (\NHunit) = 0.67$ for the sample (see the top right panel of Figure~\ref{fig:NH}).
In this study, we use the best-fitted $\NH$
as a face value, but note that the
large uncertainty always harbors
the possibility that unobscured AGN might be obscured one, and vice versa
if the source is around $\log (\NHunit) \sim 22$.

The top left panel of Figure~\ref{fig:NH} shows the histogram of $\NH$ of our sample, divided by the spec-$z$ and photo-$z$ sample, and it illustrates most of the eFEDS-WERGS sources can be classified as unobscured AGN with $\log (\NHunit)<22$, comprising 91\% (356 out of 393 sources) of the sources.
This is a natural outcome considering that eROSITA is more sensitive in the soft 0.5--2~keV band, which is weak against the strong absorption with $\log (\NHunit)>22$ \citep[e.g.,][]{mer12,ric15,liu21}.
This means that 91\% of our eFEDS-WERGS sources are unobscured radio-loud AGN. Considering that most of them at $z>1$ have the luminosity range of $\log (\lxsoftabs/\mathrm{erg}~\mathrm{s}^{-1})>44$ (93\% or 123 out of the 132 unobscured AGN at $z>1$), they are likely radio-loud quasars.

While they are rare, 9\% (37 out of the 393 sources) of the population shows a strong absorption reaching $\log (\NHunit)>22.0$.
The top right panel of Figure~\ref{fig:NH} shows that most of those obscured AGN are distributed widely at $0.4<z<3.2$, which is slightly biased to higher-$z$.
It is a natural outcome since the observed 0.5--2~keV covers more higher energy band that is stronger against the gas absorption at higher redshift. This produces a negative $k$-correction and thus the observed energy coverage enables eROSITA to detect relatively obscured AGN.

The bottom left panel of Figure~\ref{fig:NH} shows 
the distribution of $\NH$ as a function of $\lxsoftabs$.
All of the obscured AGN are X-ray luminous,
with the median of 
$\left< \log \left( \lxsoftabs/\mathrm{erg}~\mathrm{s}^{-1} \right) \right> = 44.6$, 
or $\left< \log \left( \lbol/\mathrm{erg}~\mathrm{s}^{-1} \right) \right> = 46.6$,
which is similar luminosity with the optical SDSS quasar \citep[e.g.,][]{she11,rak20}. This indicates that they are X-ray and radio-luminous obscured AGN.
Considering that those sources are located at $0.4<z<3.2$,
they are equivalent of radio SDSS quasars, but with strong absorption, and those populations are widely missed in the previous SDSS optical band studies, showing the power of
wide and medium-depth X-ray surveys by eROSITA.

Out of the 37 obscured radio AGN, 15 sources are spec-$z$ confirmed sources,
and the SDSS spectra of 9 sources out of them are publicly available.
The three sources show clear quasar spectra, 
and other three sources show a broad emission component either in the H$\beta$, MgII, or Ly$\alpha$ but a redder continuum.
The remaining three show type-2 AGN-like spectra whose continuum is dominated by the host galaxies.
Interestingly, three sources are high-$z$ sources at $z>3$.
Figure~\ref{fig:Xspec_obsAGN} shows both the X-ray (top panel) and optical spectra (bottom panel) of all three spec-$z$ confirmed obscured radio AGN at $z>3$ found in this study.
Three sources are highly X-ray and radio luminous AGN with $45.0< \log (\lxsoftabs/\mathrm{erg}~\mathrm{s}^{-1})<45.6$ and 
$42.9<\log (\lvfirst/\mathrm{erg}~\mathrm{s}^{-1})<44.5$.
We will explore in more detail the X-ray and multi-wavelength properties of those high-$z$ obscured radio AGN in the forthcoming paper.

Since some of our samples also cover optical to mid-IR bands,
flux density ratio of $f_\nu(12\mu \mathrm{m})/f_{\nu,\mathrm{i-band}}$ is
calculated as an indicator of dust extinction \citep{ros18}.
Here, observed WISE W3 (12~$\mu$m) flux density and Subaru/HSC $i$-band flux densities are used in this study. 
The bottom right panel of Figure~\ref{fig:NH} shows
the relation between the $f_\nu(12\mu\mathrm{m})/f_{\nu,\mathrm{i-band}}$ and
$\NH$, each of them is a tracer of the dust extinction and the gas absorption,
respectively. 
The scatter is huge, but the two parameters show a statistically significant correlation with each other,
which is supported by Pearson's test with a coefficient value of 
0.34 and p-value of $7.2 \times 10^{-7}$.
The linear fitting result is also shown with the black dashed line,
with the equation of 
$\log \left( f_\nu(12\mu\mathrm{m})/f_{\nu,\mathrm{i-band}} \right) = -14.1 + 0.73 \log (\NHunit)$.

The bottom right panel of Figure~\ref{fig:NH} also shows the sources which fulfill the color criterion of extremely red quasars, which is defined as optically identified quasars with an observed flux density ratio of
$f_\nu\mathrm{(12\mu m)}/f_{\nu,\mathrm{i-band}}>10^{1.84}$ \citep{ros15,ham17}.
Thanks to the deep photometries of Subaru/HSC, 16 out of the 23 sources have faint $i_\mathrm{AB}>22$, which is not covered by the previous studies \citep[e.g.,][]{ros15}.
As shown in the Figure, there is one spec-$z$ source in the region of the extremely red quasar color.
The source is \verb|ERO_ID=608|,
which \cite{bru21a} already identified as an
X-ray luminous obscured AGN with strong [OIII]$\lambda5007$ outflow feature, undergoing strong AGN feedback.
This is consistent with one of the notable features of extremely red quasars \citep{zak16}.
Although the other 22 sources of our candidates should wait for further spectroscopic confirmation of whether they are genuinely extremely red quasars or not, our sample contains interesting candidates of such radio-loud extremely red quasars.
\cite{hwa18} proposed that the origins of radio-emission for extremely red quasars would be a shocked induced emission by the strong outflows \citep[e.g.,][]{zak14} for the moderate radio luminosities of $\lfirstunit = 7\times 10^{22}$--$10^{25}$, where 
6 of our extremely red quasar candidates fulfill this criterion. On the other hand, we also cover 17 radio luminous extremely red quasar candidates at $10^{25} < \left(\lfirstunit\right) < 10^{27.3}$, which might have not been investigated before.

\begin{figure}
\begin{center}
\includegraphics[width=0.47\textwidth]{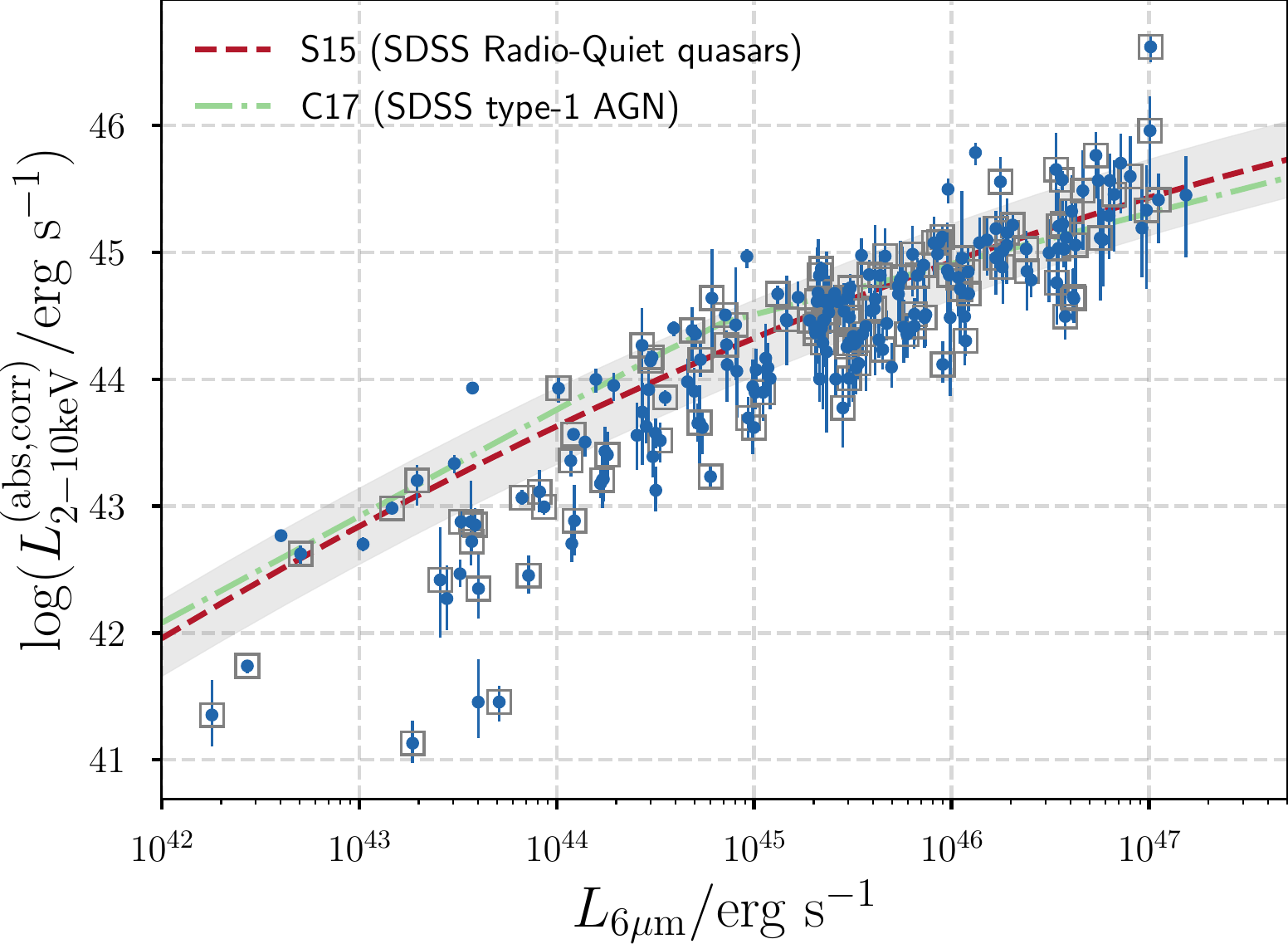}~
\caption{
The luminosity correlation between the
luminosities at absorption corrected 2--10~keV  ($\lxhardabs$) and 6~$\mu$m ($\lsix$).
The symbols are the same as in Figure~\ref{fig:R_vs_imag}.
The red dashed- and green dot-dashed line represents the slope in the study of optically selected radio-quiet quasars at $1.5<z<5$ \citep{ste15}
and \cite{che17}, respectively. 
}\label{fig:LvsLrelation}
\end{center}
\end{figure}

\begin{figure*}
\begin{center}
\includegraphics[width=0.48\textwidth]{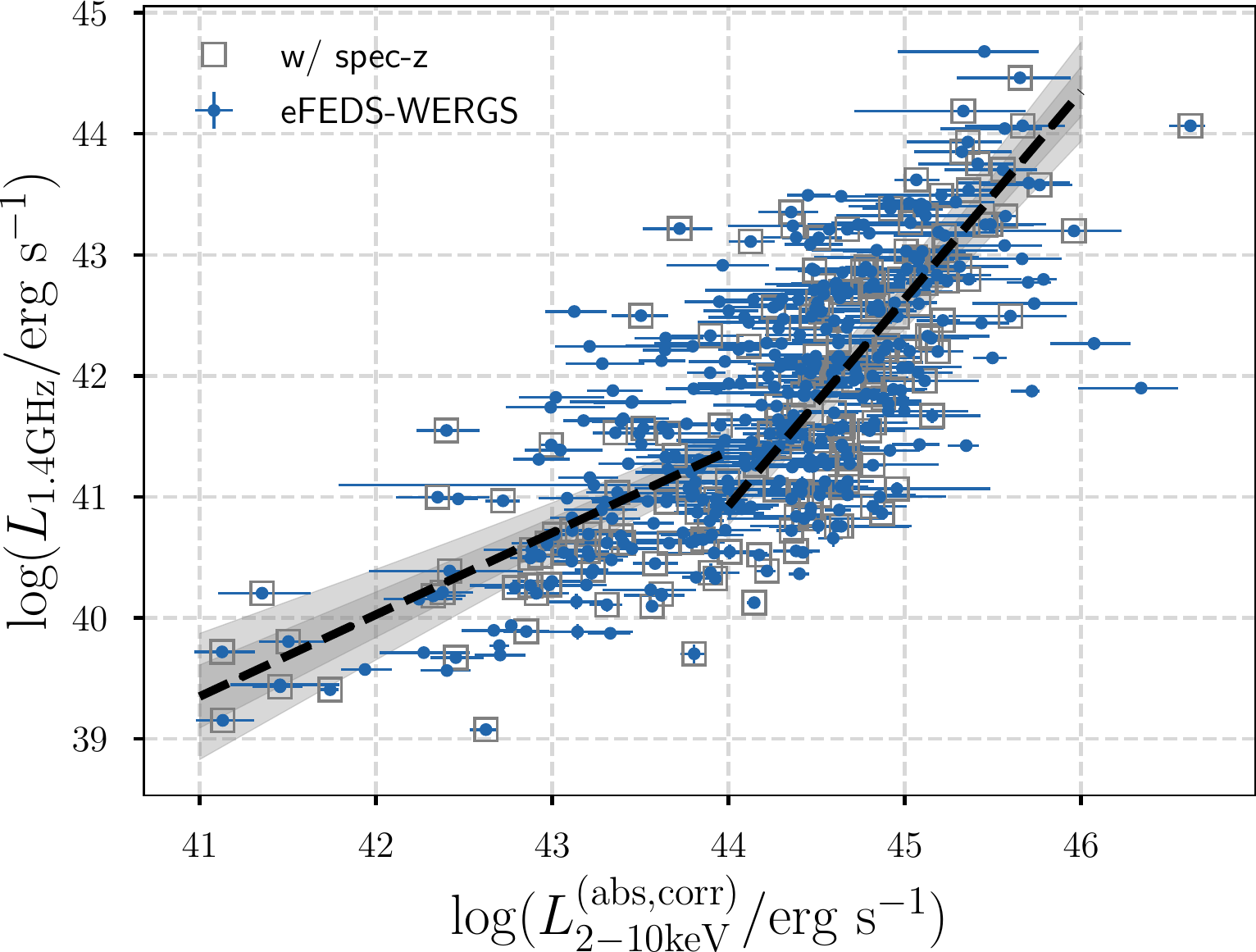}~
\includegraphics[width=0.48\textwidth]{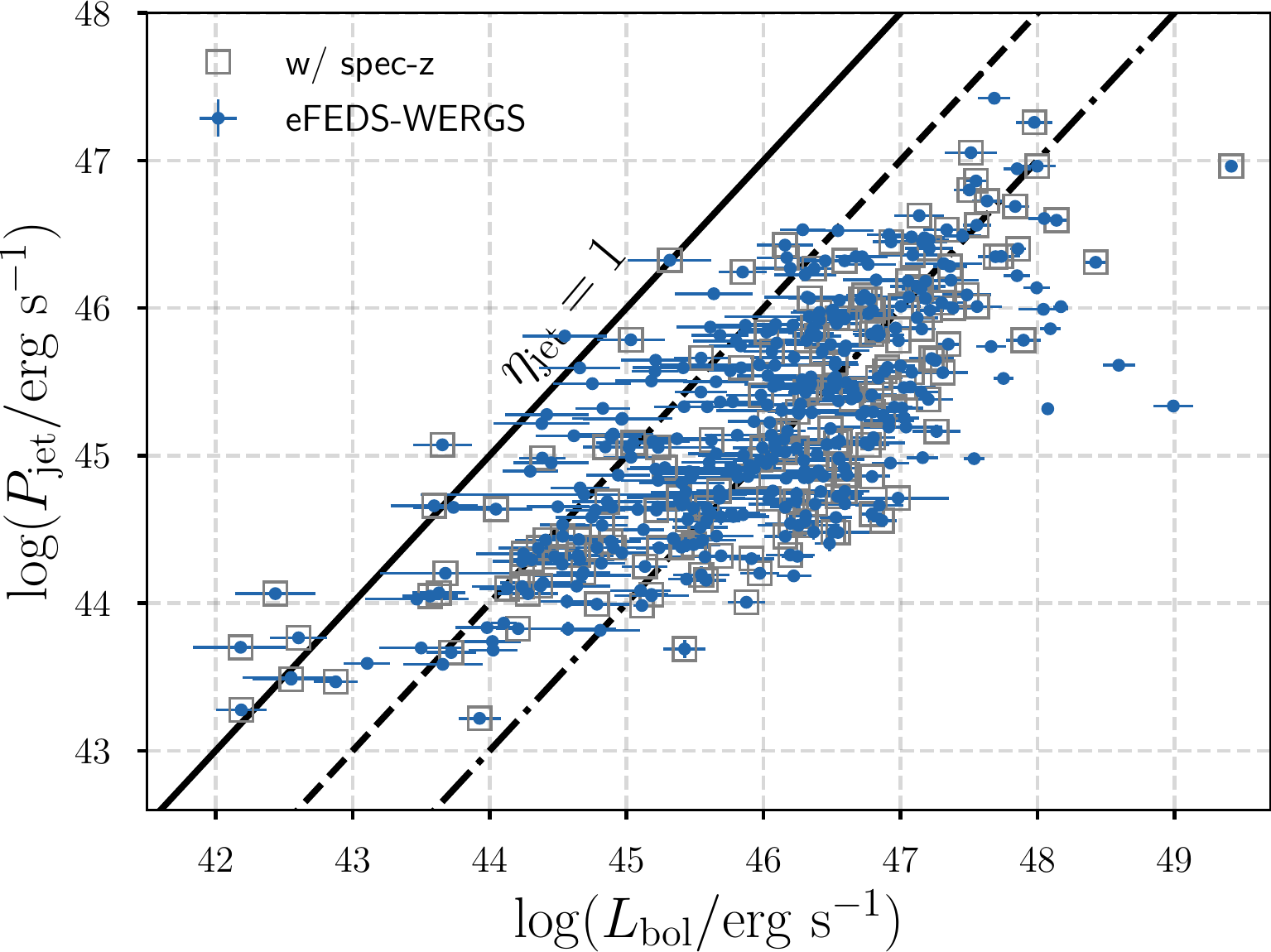}
\caption{
(Left)
The luminosity correlation between the
luminosities at 1.4~GHz ($\lfirst$ [erg~s$^{-1}$])
and absorption corrected 2--10~keV ($\lxhardabs$).
The two black dashed lines in the lower and higher $\lxhardabs$ with the boundary of $\lxhardabs = 10^{44}$~erg~s$^{-1}$ represents the fitting lines of Eq.~\ref{eq:Lr_vs_Lx_lowL} and ~\ref{eq:Lr_vs_Lx_highL}, respectively.
(Right) The correlation between the jet power ($\ljet$) and bolometric AGN luminosity ($\lbol$). The black solid/dashed/dot-dashed line shows the case with the jet production efficiency $\etaj=1$, $\etaj=0.1$, and $\etaj=0.01$, respectively, by assuming the radiation efficiency of $\etar=0.1$ \citep{sol82}.
The other symbols are the same as in Figure~\ref{fig:R_vs_imag}.
}\label{fig:LxvsLradio}
\end{center}
\end{figure*}

\begin{figure*}
\begin{center}
\includegraphics[width=0.43\textwidth]{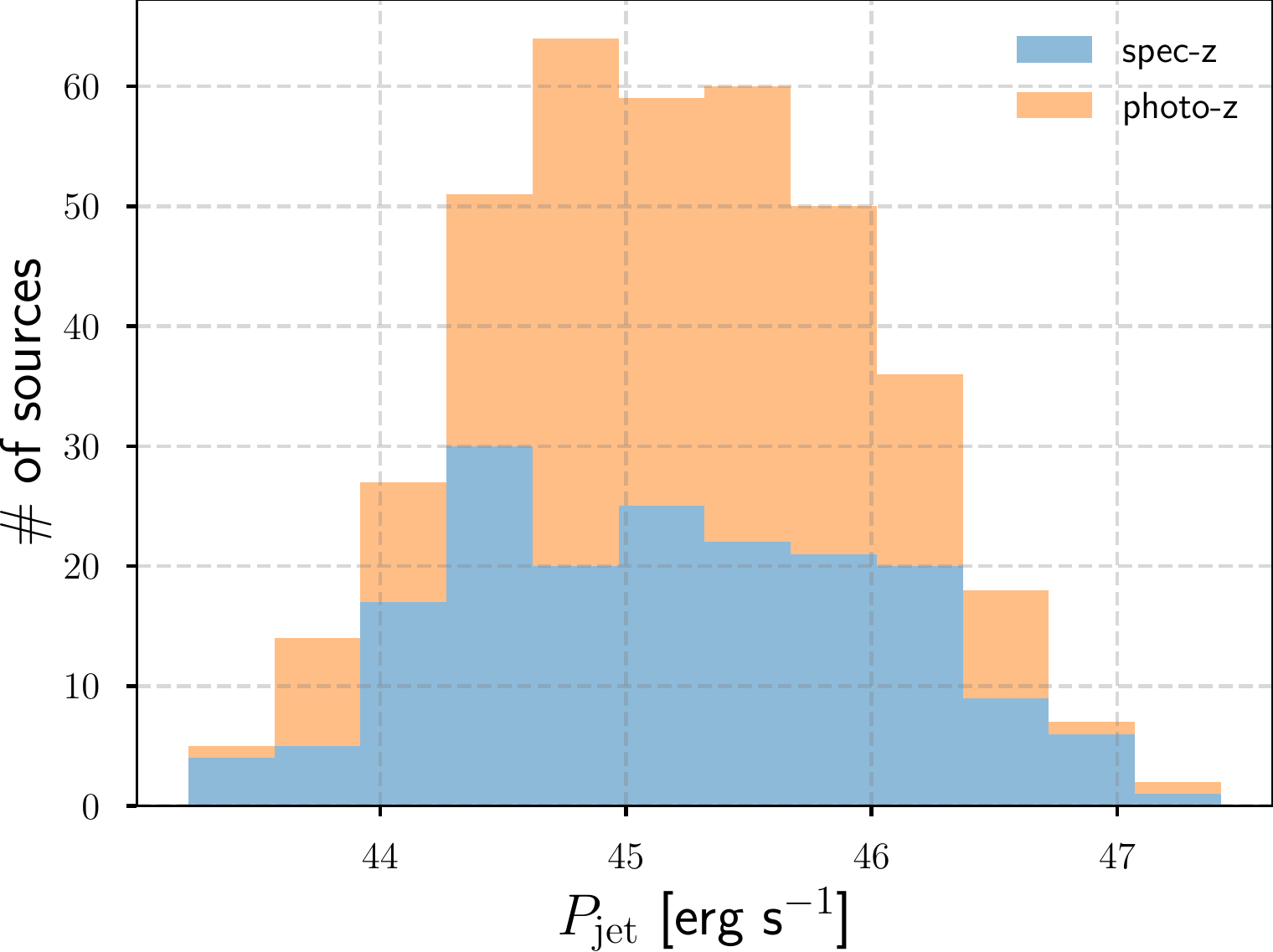}~
\includegraphics[width=0.43\textwidth]{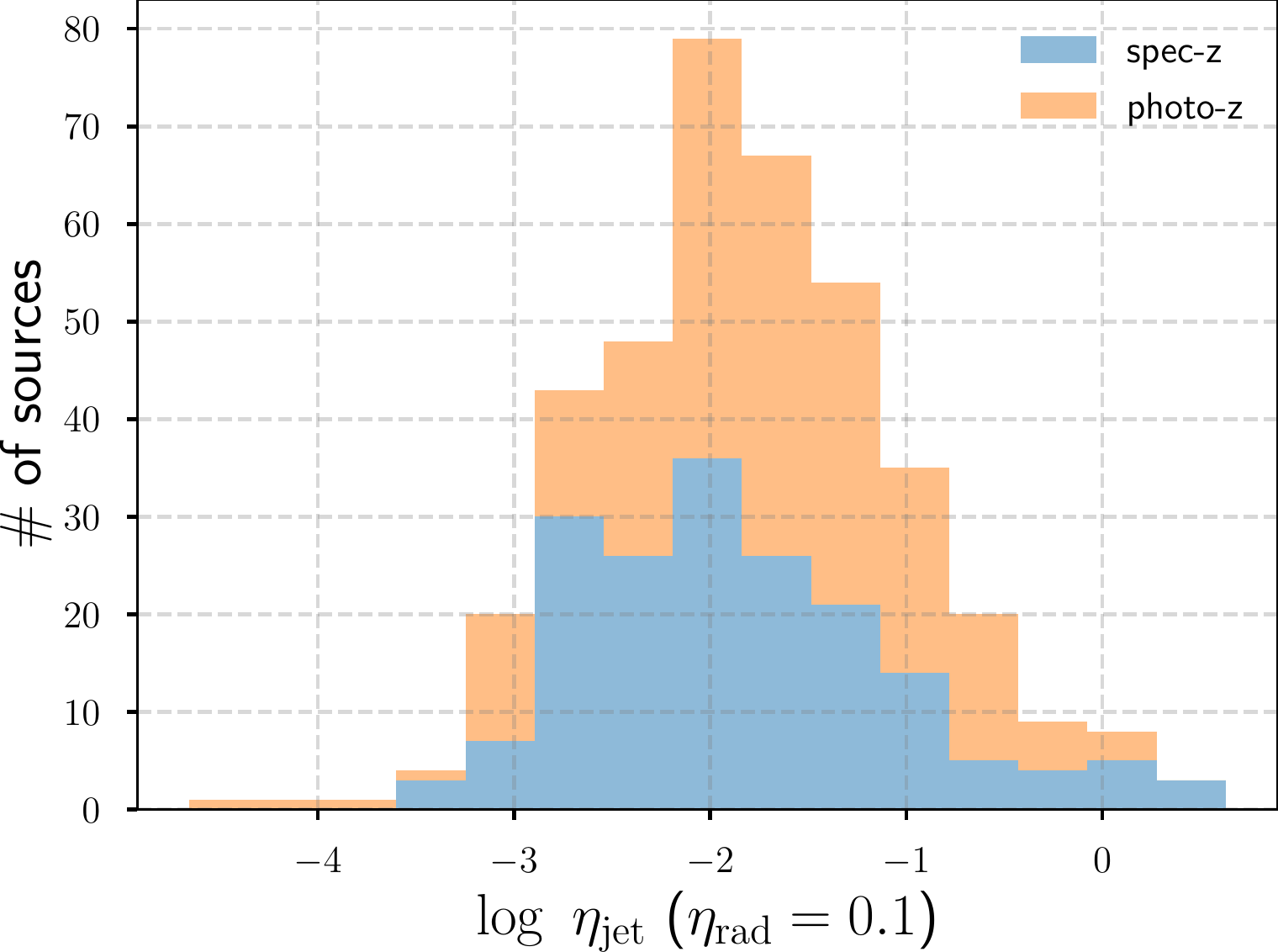}
\caption{
The distribution of $\ljetunit$ (left)
 and $\etaj$ assuming the radiative efficiency of 0.1 (right).
 The sample is divided into subgroups of
 spec-$z$ sources (cyan) and photo-$z$ one (orange).
}\label{fig:hist_Rint}
\end{center}
\end{figure*}

\subsection{X-ray and 6~$\mu$m Luminosity Correlation}\label{sec:results:Lx_vs_L6}

It is not trivial for radio AGN
to assume that the X-ray emission is dominated by the radiation from the corona around the accretion disk because of the possible contamination of the emission from the jet as shown in blazars \citep[e.g.,][]{ghi17}.
To investigate this jet contribution in the X-ray bands, 
we utilize the MIR bands which are generally dominated by the thermal dust emission heated by the accretion disk of AGN, and 
investigate the luminosity relation between the X-ray and MIR band of our sample and compare it to the radio-quiet AGN, which shows a tight correlation each other both in the low-redshift ($z<0.1$) and low-luminosity AGN at $\lxhardabs < 10^{44}$~erg~s$^{-1}$ \citep{gan09,mat12a,asm15,ich12,ich17a,ich19a}
and high-redshift ($z>1$) and high-luminosity AGN or quasars \citep{ste15,mat15,che17}.

Figure~\ref{fig:LvsLrelation} shows a luminosity correlation between the rest-frame 6~$\mu$m ($\lsix$) and absorption corrected 2--10~keV band ($\lxhardabs$) extrapolated from the absorption corrected 0.5--2~keV luminosities ($\lxsoftabs$). 
The reason on the usage of $\lxhardabs$ here instead of $\lxsoftabs$ is for the comparison of the previous luminosity correlation studies that mostly utilize absorption corrected 2--10~keV luminosities
\citep[e.g.,][]{ste15,che17}.
Overall, most sources of our sample follow nicely the luminosity correlation of radio-quiet QSOs of the similar redshift \citep{ste15,che17}, which shows a saturation of X-ray emission in the high luminosity end.
This trend is also consistent with the results of obscured AGN sample in the eFEDS field \citep{tob21}.
On the other hand, while the sample size is small, blazar populations are known to follow the 1:1 relation without showing the saturation of $\lxhardabs$ \citep[e.g.,][]{mat12a}.
This indicates that both X-ray and 6~$\mu$m luminosity
of our eFEDS-WERGS sample is dominated by the
same radiation origins for radio-quiet populations and 
the contribution from the jet emission would be not a significant one on average.

We also note that Figure~\ref{fig:LvsLrelation} shows a deviation of the sample from the slopes of \cite{ste15} and \cite{che17} in the low-luminosity end, notably at $\lsix < 10^{44}$~erg~s$^{-1}$.
Such a trend is already discussed in Section~\ref{sec:results:Lzplane} and Figure~\ref{fig:L_vs_z}, and possible origins are the contamination from the host galaxies into the MIR bands where the AGN dust does not always outshine the host galaxy scale dust emission anymore \citep{ich17a}.
The other possibility is the contamination from the Synchrotron radiation of the jet. Several low-luminosity AGN are known to show such Synchrotron emission contamination even in the MIR bands \citep{mas12,pri12,lop14,lop18}. As we will discuss in Section~\ref{fig:LxvsLradio}, those low-luminosity AGN tend to show high jet production efficiency and therefore the relative contribution from the jet might be strong compared to the AGN dust emission.
Nonetheless, in either case, our main conclusion does not change that the X-ray emission for the majority of the eFEDS-WERGS sample is likely dominated by the radiation from the corona, not by the jet.

\begin{figure}
\begin{center}
\includegraphics[width=0.45\textwidth]{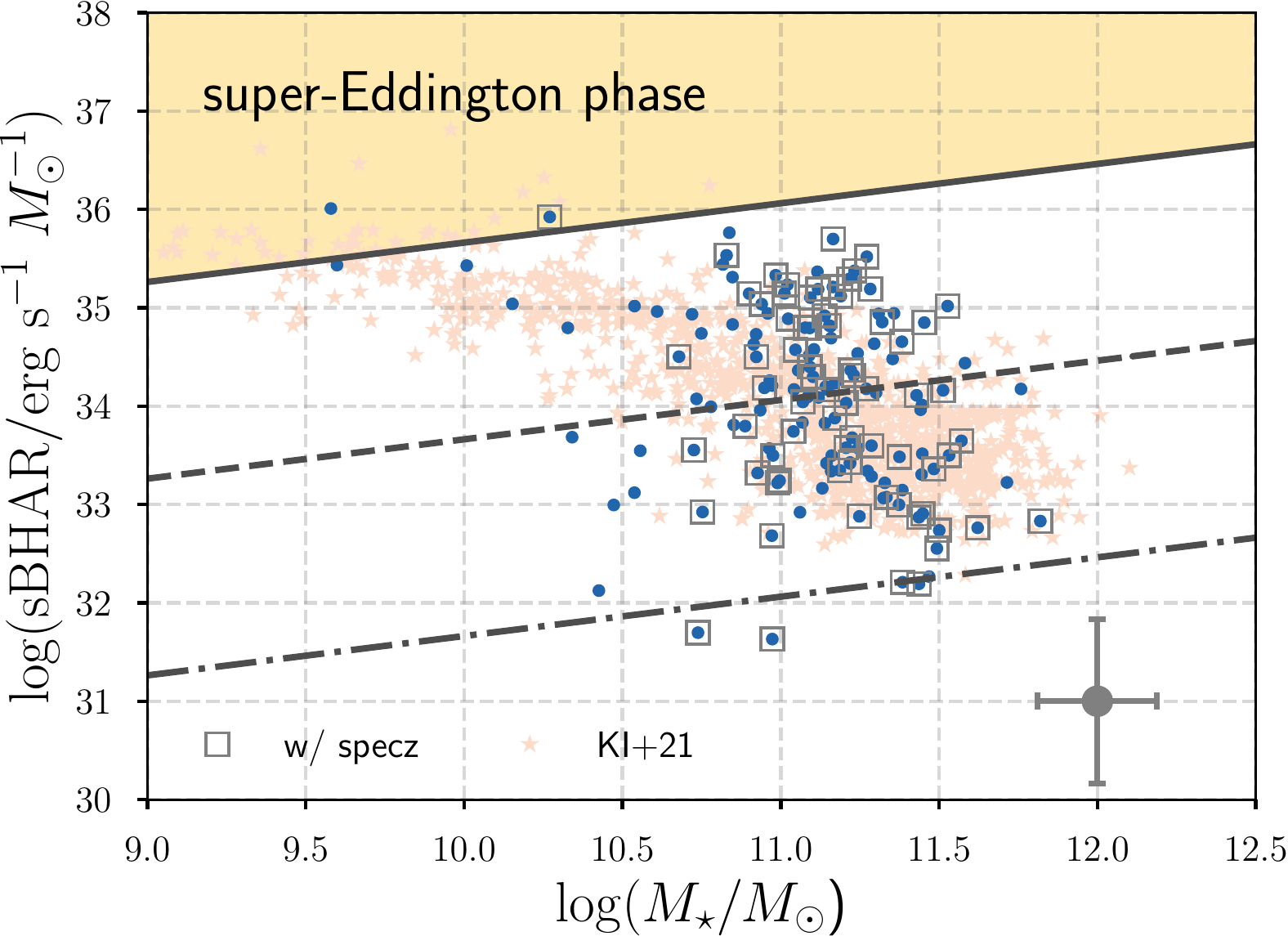}~
\caption{
The relation between sBHAR ($=\lbol/\mstar$~erg~s$^{-1}$~$\msun^{-1}$) and $\mstar$. 
The gray cross represents the typical errorbar of the two values.
The orange shaded area represents the region of super-Eddington of $\lambdaedd>1$.
The three straight lines from the top to bottom are the expected Eddington ratio of $\lambdaedd=1$ (Eddington limit; black solid line), $\lambdaedd=0.01$ (black dashed line), and $\lambdaedd=10^{-4}$ (black dot-dashed line) using Equation~\ref{eq:sBHAR_KH13}.
The pale pink stars in the background represent the sources from the original WERGS sample with IR detections \citep[KI+21;][]{ich21}.
The other symbols are the same as Figure~\ref{fig:R_vs_imag}.}
\label{fig:sBHARvsMstar}
\end{center}
\end{figure}

\subsection{Radio 1.4~GHz and X-ray luminosity correlation}\label{sec:results:Lr_vs_Lx}

The combination of the radio and X-ray data offers the opportunity to investigate the connection between the disk-corona (traced by the X-ray) and disk-jet (traced by the radio).
The slope of the luminosity correlation between the radio and X-ray bands is argued to depend on the accretion disk properties.
Several authors argue that the slope of $L_\mathrm{R}$--$L_\mathrm{X}$ is different among the accretion disk states; in low-accretion rate, radiatively inefficient accretion flow
\citep[RIAF; $b\sim0.5$--$0.7$; see][]{cor03,gal03}
and in high accretion rate, standard disk state \citep[$b\sim1.4$;][]{cor12}. 

The left panel of Figure~\ref{fig:LxvsLradio} shows the luminosity correlation between rest-frame 1.4~GHz luminosity and absorption corrected 2--10~keV luminosity.
While the scatter is large, there is a sign of the luminosity slope change at $\log (\lxhardabs/\mathrm{erg}~\mathrm{s}^{-1})\simeq 44$, below/above which the accretion state is likely the RIAF/standard disk state.
This is partially supported from the fact that $\log (\lxhardabs/\mathrm{erg}~\mathrm{s}^{-1})\simeq 44$
corresponds to $\log (\lbol/\mathrm{erg}~\mathrm{s}^{-1})\simeq 45.7$, which is the Eddington luminosity of the 
sources with $\log (\mbh/\msun) = 7.7$, and 
therefore $\lambdaedd > 0.01$ boundary for the sources with
$\log (\mbh/\msun) < 9.7$.
In other words, most of the sources above the boundary
$\log (\lxhardabs/\mathrm{erg}~\mathrm{s}^{-1})\simeq 44$
should be in the efficient accretion phase of $\lambdaedd > 0.01$.

We apply ordinary least-squares Bisector fits, which minimize the 
perpendicular distance from the slope line to data points \citep{sch85,iso90}.
This gives the correlation at low luminosity end ($\lxhardabs<10^{44}$~erg~s$^{-1}$)
\begin{align}
\log \frac{\lfirst}{10^{41}~\mathrm{erg}~\mathrm{s}^{-1}}= (0.38 \pm 0.08) + (0.69 \pm 0.07)
\log \frac{\lxhardabs}{10^{44}~\mathrm{erg}~\mathrm{s}^{-1}},
\label{eq:Lr_vs_Lx_lowL}
\end{align}
and at high luminosity end ($\lxhardabs \geq 10^{44}$~erg~s$^{-1}$) with
\begin{align}
\log \frac{\lfirst}{10^{41}~\mathrm{erg}~\mathrm{s}^{-1}}= (-0.07 \pm 0.10) + (1.73 \pm 0.12)
\log \frac{\lxhardabs}{10^{44}~\mathrm{erg}~\mathrm{s}^{-1}}.
\label{eq:Lr_vs_Lx_highL}
\end{align}

The obtained slope value of $b=0.69$, which is defined by 
$\log \lfirst \propto b \log \lxhardabs$, is consistent with those of RIAF for the low-luminosity end and the values expected from the fundamental planes \citep[$b\sim0.7$, e.g.,][]{mer03}.

On the other hand, the slope has a steeper value of $b=1.73$ at the high $\lxhardabs$ end.
This high slope value is also reported in the ROSAT-FIRST radio quasar sample, which shows $b=2.1$
 \citep{bri00}.
This suggests that our sample contains radio jet efficient sources in higher luminosity end, especially when X-ray emission saturates at $\lxhardabs\simeq 10^{45-46}$~erg~s$^{-1}$.
Based on the results above, thanks to the broad coverage of $\log (\lxhardabs/\mathrm{erg}~\mathrm{s}^{-1})=$41--46
including both low and high accretion sources, 
we find the two important trends of 1) the luminosity correlation shows a break around $\lxhardabs \sim 10^{44}$~erg~s$^{-1}$ above which $\lxhardabs$ shows a saturation compared to $\lfirst$, which suggests that X-ray corona and the radio-jet connection is fundamentally different below/above the break luminosity of 
$\lxhardabs\simeq 10^{44}$~erg~s$^{-1}$.
We will discuss this point later in Section~\ref{sec:results:Ljet_vs_Lbol} and ~\ref{sec:discussion:FP}.

\subsection{$\ljet$ and $\lbol$ Relation}\label{sec:results:Ljet_vs_Lbol}

To investigate the jet and disk connection more quantitatively, 
we estimate the jet power and the bolometric AGN luminosities,
both of which are the total power of the jet and radiation from the accretion disk.
For the bolometric luminosity, we use the one estimated from $\lxhardabs$ as discussed in Section~\ref{sec:sample:luminosity}. 

There are several studies measuring the relation between jet power and radio luminosities, and two approaches are commonly used. One is based on the analytic models of the sources to predict the radio luminosity for a given radio jet power \citep[e.g.,][]{wil99}. The other is based on the estimation of the jet power from the size of the X-ray cavities in the hot gas inflated by the radio lobes, and the X-ray cavity powers can be empirically related to the radio luminosity \citep[e.g.,][]{bir08,cav10}. In this study, we use the relation based on the latter approach and
the total jet power of the AGN ($\ljet$) is estimated from the VLA/FIRST radio luminosity.
 We adopt the relation between $\ljet$ and $\lfirst$ of \cite{cav10},
 \begin{align}
  \ljet = 1.05 \times 10^{44}~(\lfirst/10^{24}~\whz)^{0.75}~\mathrm{erg}~\mathrm{s}^{-1} \label{eq:ljet_lfirst}.
 \end{align}
 The results based on other relations are discussed in Appendix~\ref{sec:AppendixA}.
 Note that the jet power is known to depend not only on the observed radio luminosity but also its radio source size, which is roughly an indicator of the source age \citep[so called the power-linear size plane or $P_\mathrm{jet}$--$D$ diagram, eg.,][]{bal82,kai97,blu99,an12}, and its redshift due to the Compton loss at higher-$z$ at $z \geq 2$ \citep[e.g.,][]{har18,har19}.
 \cite{sha13} recently investigated the effect of the radio source size in the jet power estimation,
 and found the dependence on the source size
 to be $\propto D^{0.58}$, suggesting that
 the effect of the source size is small.
 Since our sample is selected only for radio-compact emission in VLA/FIRST 
 with a spatial resolution of $\sim 5$~arcsec
  (or $\lesssim 20$~kpc at $z\sim1$)
this again mitigates the size dependence; otherwise the sources are ultra-compact sources.
We will discuss the redshift dependence below.

 The left panel of Figure~\ref{fig:hist_Rint} shows the
 distribution of $\ljet$, spanning wide luminosity range of $43.2 < \log (\ljetunit) < 47.4$.
Especially, the high $\ljet$ sources with $\log (\ljetunit) > 45.0$ is almost equivalent to the radio-loud quasar level \citep[e.g.,][]{bes05,ino17}.
The eFEDS-WERGS sample covers high $\ljet$ sources which are powerful enough to produce an expanding jet with $\gtrsim 10$~kpc that can disrupt the interstellar medium \citep[e.g.,][]{nes17} and
larger scale intergalactic environment possibly causing to quench star-formation in host galaxies \citep[e.g.,][]{mcn07}.
 
The right panel of Figure~\ref{fig:LxvsLradio} shows 
the relation between the jet power $\ljet$ and 
the bolometric AGN luminosity $\lbol$.
As expected from the correlation between $\lfirst$ and $\lxhardabs$,
the relation shows a flatter trend in the low $\lbol$ and
the slope becomes steeper at higher $\lbol$, notably at $\lbol>10^{45.5}$~erg~s$^{-1}$, which shows a
slope value of $b\approx1.0$.
The slope is shallower than $b=1.73$ in Eq.~\ref{eq:Lr_vs_Lx_highL} because of the
X-ray luminosity dependent bolometric correction by \cite{mar04}.
This 1:1 slope is consistent with the value
previously reported in the SDSS radio quasars by using optical and UV continuum as an indicator of AGN radiation \citep[e.g.,][]{ino17}.

Considering that jet production is correlated to the accretion rate to the BH $\mdotbh$, jet power is limited by the BH accretion rate with a jet production efficiency.
The jet production efficiency is defined as 
 \begin{equation}
 \etaj = \ljet / \mdotbh c^2 = \etar \ljet / \lbol 
 \end{equation}
 where $\etar$ is the radiation efficiency of an AGN accretion disk, $c$ is the speed of light, and
$\mdotbh$ is the mass accretion rate onto the SMBH through the disk.
In the following, we adopt a  canonical value of $\etar=0.1$ based on the Soltan-Paczynski argument \citep{sol82}.
The solid line in the right panel of Figure~\ref{fig:LxvsLradio}
represents $\ljet = \lbol / \etar = \mdotbh c^2$,
corresponding to the maximum efficiency of $\etaj=1$, and the dashed and dotted-dashed line represents the one with $\etaj=0.1$ and $\etaj=0.01$.

The median value of $\etaj$ is $\left< \log \etaj \right> = -1.0 \pm 0.6$ for the sources with $\lbol < 10^{45}$~erg~s$^{-1}$ and
$\left< \log \etaj \right> = -2.0 \pm 0.7$ for high luminosity sources with $\lbol > 10^{45}$~erg~s$^{-1}$,
 respectively. The overall distribution is summarized in the right panel of Figure~\ref{fig:hist_Rint}.
 At the high luminosity end, 
 the jet production efficiency is almost consistent
 with the studies of the SDSS radio quasar population, whose radio jet efficiency lies at $\left< \log \etaj \right> = -2.0 \pm 0.5$ \citep[e.g.,][]{van13b,ino17}.
The origin of the jet emission for those high luminosity sources is still unclear since they would host standard geometrically thin disks, which are not expected to have powerful jets \citep{lub94}. However, recent general relativistic magneto-hydrodynamic (GR-MHD) simulations succeed to reproduce the jet launching even in such geometrically thin disks \citep[e.g.,][]{lis19}.
Finally, we note that our estimation of $\etaj$ might be under-estimated for such high luminosity sources with $\lbol > 10^{45}$~erg~s$^{-1}$, whose median redshift is $\left<z\right>=1.3$. \cite{har18} discussed that the observed radio luminosities in high-$z$ sources tend to show lower values for a given $\ljet$ because of the increased inverse-Compton losses due to the increased CMB backgrounds with redshift.
At $z\sim2$, the luminosity loss reaches by a factor of a few at the jet age of a few Myr, and its radio size is around a few 10~kpc~\citep{har18}, which is the maximum end of the size in our VLA/FIRST radio compact sources.
Thus, a possible underestimation of $\etaj$ could reach around by a factor of a few. Nonetheless, this difference does not affect our main conclusion here.
 
 At the low-luminosity end,
 the jet production efficiency is slightly higher or consistent with the nearby radio AGN population ($\log \etaj \sim -1.5$) whose host galaxies are massive \citep[$\mstar > 10^{11} \msun$;][]{nem15}.
 One notable trend especially at the low AGN luminosity end is that some low-luminosity sources 
 reach extremely high jet production efficiency at $\etaj\sim1$ where 14 sources fulfill the criteria of $\etaj + \Delta \etaj >1$).
 This extremely high value can be achieved if
 the sources are blazar with extremely high BH spin value
 \citep[e.g.,][]{ghi14}.
 However, most of such high $\etaj$ sources are spec-$z$ confirmed
 sources and they are categorized as non-blazar.
 Therefore it is unlikely that blazars are the majority of the  population.
 
 The second possibility is that our assumption of $\etar=0.1$ is no longer valid in the low-luminosity AGN regime at $\lambdaedd<0.01$.
 At sufficiently low accretion rates, the infalling matter
 to the BH is not dense enough to cool anymore, deviating from the standard-disk model. 
 Thus, the flow is essentially adiabatic and
 mass accretion is led by anomalous viscosity produced by the 
magneto-rotational instability \citep[e.g.,][]{bal91}.
 There are several accretion disk models in such low accretion regimes,
 and the radiation efficiency differs among them.
 One RIAF model \citep[][]{ich77,nar94,nar95,sto01,haw02} predicts that
 $\etar \sim 10^{-2}$ and $\etar \sim 10^{-3}$ at $\lambdaedd\sim10^{-2}$ and $\lambdaedd \sim 10^{-3}$, respectively \citep[][]{cio09,xie12,ina19}.
 On the other hand, several authors have shown that rotating accretion flows
 become convectively unstable and becomes convection dominated accretion flow \citep[][]{igu99,sto99,nar00},
which shows higher $\etar$ than the other RIAF models, 
with $\etar \sim 10^{-2}$ and $\etar \sim 10^{-3}$ at $\lambdaedd \sim 10^{-4}$ and 
 $\lambdaedd \sim 10^{-6}$, respectively \citep{xie12,rya17,ina19}.
 As we will discuss later in Section~\ref{sec:results:sBHAR_vs_Mstar}, 
 the expected $\lambdaedd$ range is $-4<\log \lambdaedd<-2$ and therefore 
 $\etar=10^{-3}$ to $10^{-2}$ for the both model, resulting $\etaj$ is alleviated
 from $\etaj\sim1$ into lower values with $\etaj\sim10^{-1}$ to $\sim 10^{-2}$.

 The third possibility is that sources show high $\etaj$ 
 because the central engine is experiencing a quenching of AGN activity.
 It is known that jet emission is quite stable over the timescale of $\sim10$~Myr
 even after the energy injection stops from the central engine \citep[e.g.,][]{kai97,sha20,jur21,mor21}, 
 indicating that the jet luminosity does not fade drastically with this timescale.
Note that the main discussions in the references were based on the 140~MHz band, but the spectral aging method shows that the trend does not change drastically even at 1.4~GHz in a timescale of $<10$~Myr \citep[e.g.,][]{har13}.
 On the other hand, $\lbol$ in this study is
 estimated from the X-ray observation.
 This emission can be used as the indicator of the current AGN luminosities \citep[e.g.,][]{ich19c}
since it originates from the X-ray corona with the physical scale of $\sim10$ gravitational radius \citep[e.g.,][]{mor10}.
 Once the galaxies experience the AGN feedback 
 and prompt gas feeding to the BH decreases within $\sim10$~Myr timescale, we would witness that decreased $\lxsoftabs$, and therefore the decrease of $\lbol$, while $\lfirst$ remains stable. This different response time of the two luminosities might increase $\etaj$ higher, sometimes reaching nearly the maximum limit of $\etaj$, and it also allows even high than the limit of $\etaj=1$, as shown in the right panel of Figure~\ref{fig:LxvsLradio}. For the sources at $\etaj\sim1$, considering that averaged $\etaj$ is higher at one order of magnitude compared to typical radio galaxies, and two orders of magnitude higher than the typical radio-loud quasars,
 the central engine might have experienced a drastic luminosity decline by a factor of $\sim$10--100 within the
 constant jet regime, i.e., the order of $\sim10$~Myr.
 
 Some might wonder how likely such a drastic AGN luminosity decline could happen. Actually, recent multi-wavelength observations have
enable us to probe different AGN signatures with different physical scales covering up to $\sim10$~kpc, and they have discovered an AGN population with drastic luminosity declines by a factor of 10--$10^3$ in the last $10^{4-5}$~yr. 
They are called fading AGN and nearly 100 sources have been reported so far
\citep{sch10,kee15,ich16,sch16,kee17,sar18a,vil18,wyl18,ich19a,ich19c,che20,pfl22,saa22,fin22}, and some of the sources with $\etaj\sim1$ might be one of them. In order to judge whether they are such fading AGN populations, additional multiwavelength information is necessary to obtain the AGN signature tracing the difference physical scale, and therefore the probes of AGN luminosities (such as [OIII]$\lambda5007$ emissions as a tracer of an extended narrow line region) in the past $10^{3-4}$~yr \citep[e.g.,][]{kee17,ich19a}.

\subsection{SMBH properties: Relation between sBHAR and $\mstar$}
\label{sec:results:sBHAR_vs_Mstar}

Considering that some eFEDS-WERGS are faint in optical, as shown in Figure~\ref{fig:R_vs_imag}, sometimes it is difficult to obtain the BH mass through the optical spectroscopy (but see Merloni et al. in prep.).
In addition, direct BH mass estimation is not available for optically type-2 AGN although several indirect BH mass estimations are presented \citep[e.g.,][]{bar19}.
Instead, we investigate the SMBH properties through the specific black hole accretion rate (hereafter, sBHAR), which is considered to be a good proxy for the Eddington ratio.
The sBHAR is conventionally defines as sBHAR$=\lbol/\mstar$~erg~s$^{-1}$~$\msun^{-1}$ \citep{air12,mul12,del18,air19}, and the relation between the sBHAR and the Eddginton ratio $\lambdaedd$
is summarized by \cite{ich21} and it can be written as a function of $\lambdaedd$ and $\mstar$ by
\begin{align}
\mathrm{sBHAR}= 4.6\times 10^{35} \lambda_\mathrm{Edd} \left( \mstar / 10^{10}~\msun \right)^{0.4}, \label{eq:sBHAR_KH13}
\end{align}
here we apply stellar-mass dependent values for $\mbh$ with $\mbh \propto M_\star^\beta$, where 
$\beta=1.4$ \citep[][]{kor13} and $\mstar$ is available at $0<z<1$ through the image decomposition of HSC optical images of the targets.

Figure~\ref{fig:sBHARvsMstar} shows sBHAR as a function of $\mstar$
for eFEDS-WERGS sources at $z<1$.
This exhibits that 58\% of eFEDS-WERGS sources (81 out of 139 sources) are above the line of $\lambdaedd=0.01$ (black dashed line), indicating that roughly half of them are highly gas accreting sources even though they are radio-bright. 
Especially, some fraction of sources is located at the vicinity of the Eddington limit of $\lambdaedd=1$.
On the other hand, the remaining 42\% of the population (58 sources) are below the line of $\lambdaedd=0.01$ and are clustered at $10^{-4} < \lambdaedd < 0.01$.
While the sample in Figure~\ref{fig:sBHARvsMstar} is limited for the sources at $z<1$, which misses a large population of high-luminosity sources,
this trend is consistent with the result in Figure~\ref{fig:LxvsLradio},
which shows a steep slope trend in higher luminosity trend, indicating those sources are in standard disk or even in the slim disk state
while the low-luminosity sources are in a radiatively inefficient state.

The background red points are WERGS radio galaxies at $0.3<z<1.6$ obtained from \cite{ich21}, which
covers optically very faint sources down to $i_\mathrm{AB}\sim26$.
Their AGN luminosities and stellar-masses are obtained from the SED fitting using \verb|CIGALE| to the photometric data from the optical to mid-IR bands.
Thanks to the deep optical photometries and wide coverage of wavelength bands which enable them to covers low stellar-mass regimes down to $\log (\mstar/\msun) \simeq 9.5$.
On the other hand, our eFEDS-WERGS sample is biased into relatively brighter sources
with $i_\mathrm{AB}<23$, and only for the reliable stellar-mass measurements
from \cite{li21}, which produces a clear stellar-mass cut at $\mstar > 10^{10.5} \msun$. Once more multiwavelength information is assembled,
the expansion of the sample into higher redshift and lower stellar-mass might uncover more high sBHAR sources.

\begin{figure}
\begin{center}
\includegraphics[width=0.45\textwidth]{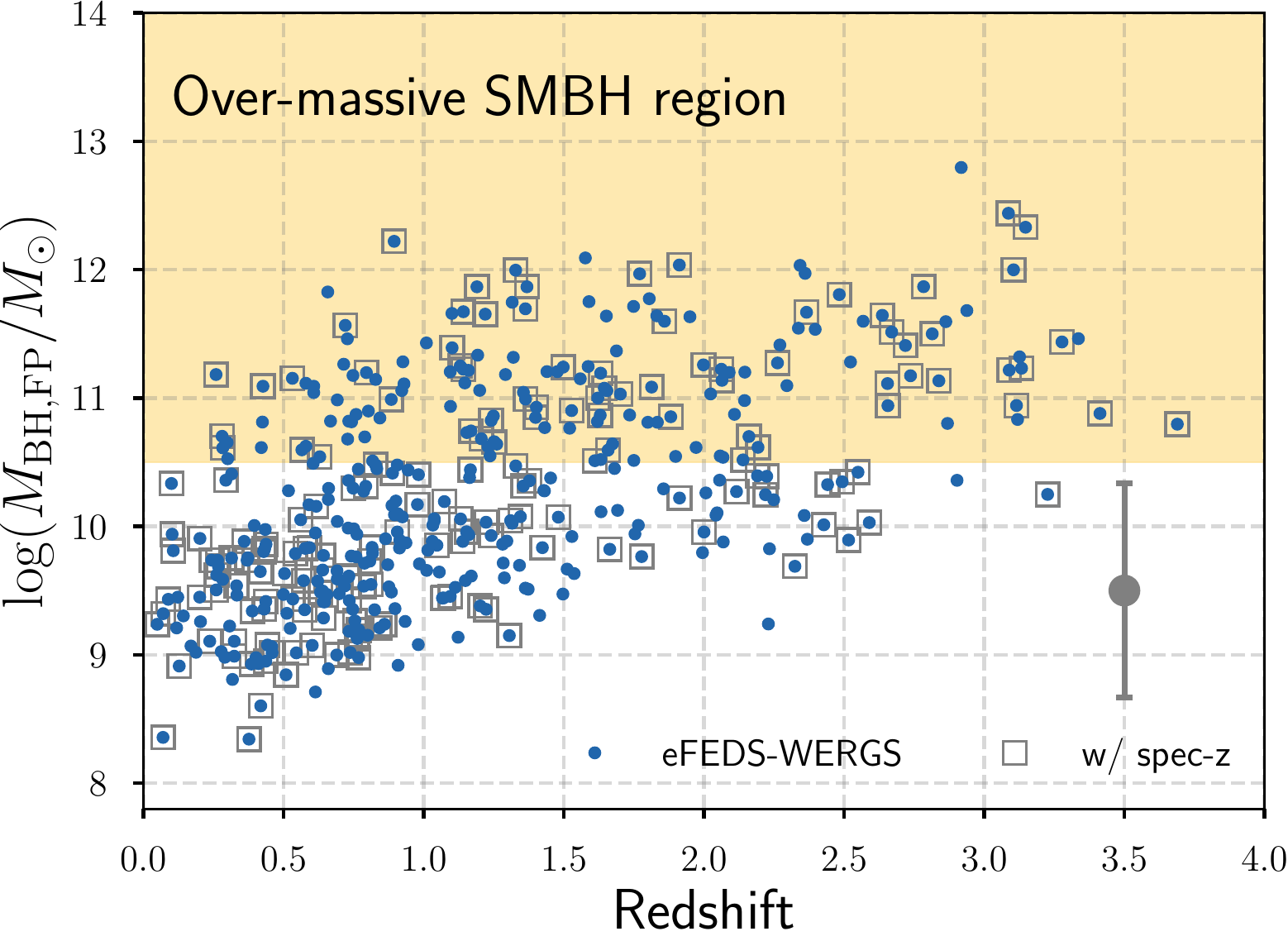}
\caption{
The BH mass estimated from the fundamental plane as a function of redshift.
The orange-shaded area represents
the over-massive SMBH region
above the maximum mass limit of known SMBHs \citep[e.g.,][]{kor13}.
}\label{fig:MBHFP_vs_z}
\end{center}
\end{figure}

\begin{figure*}
\begin{center}
\includegraphics[width=0.45\textwidth]{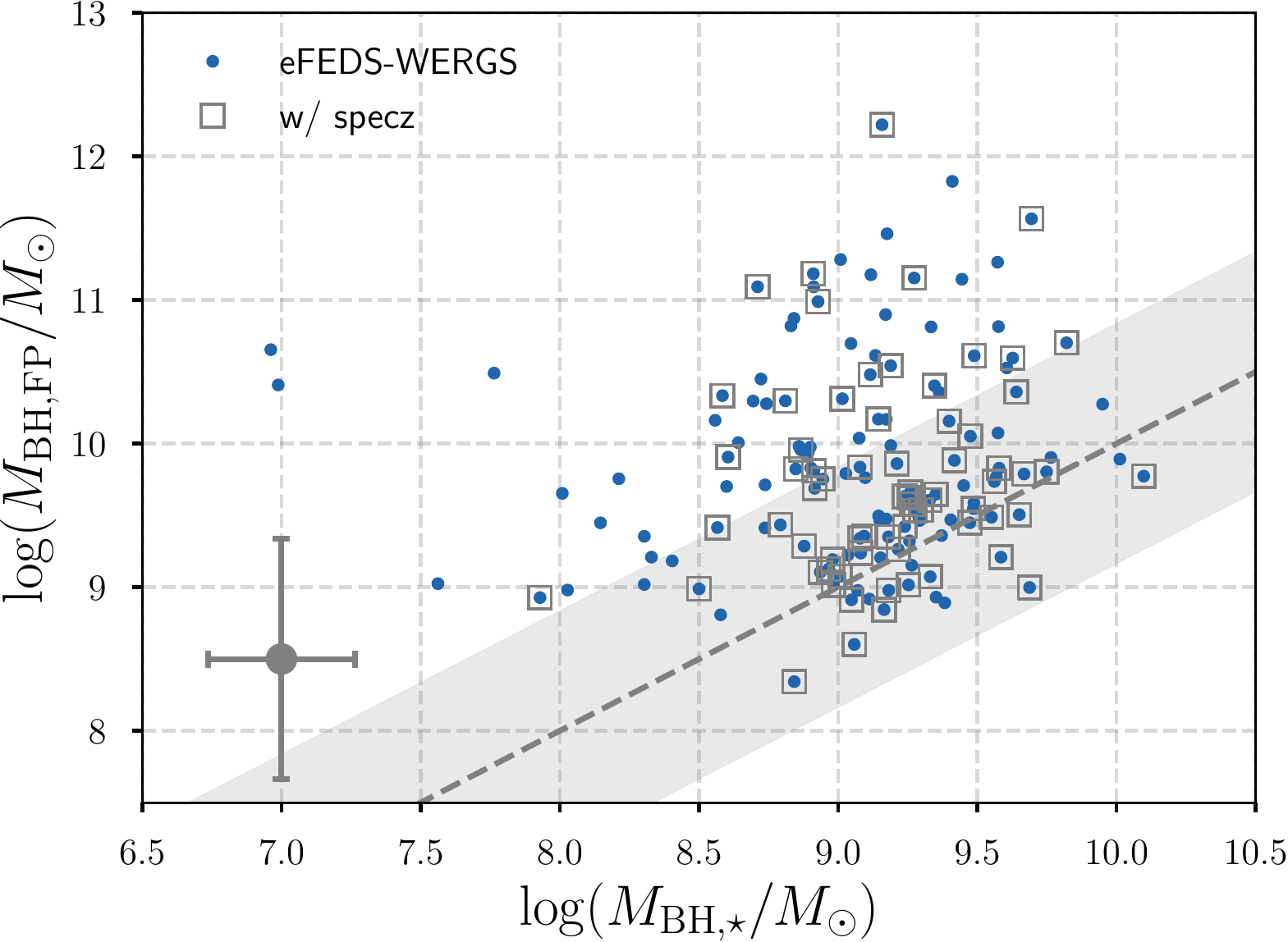}~
\includegraphics[width=0.45\textwidth]{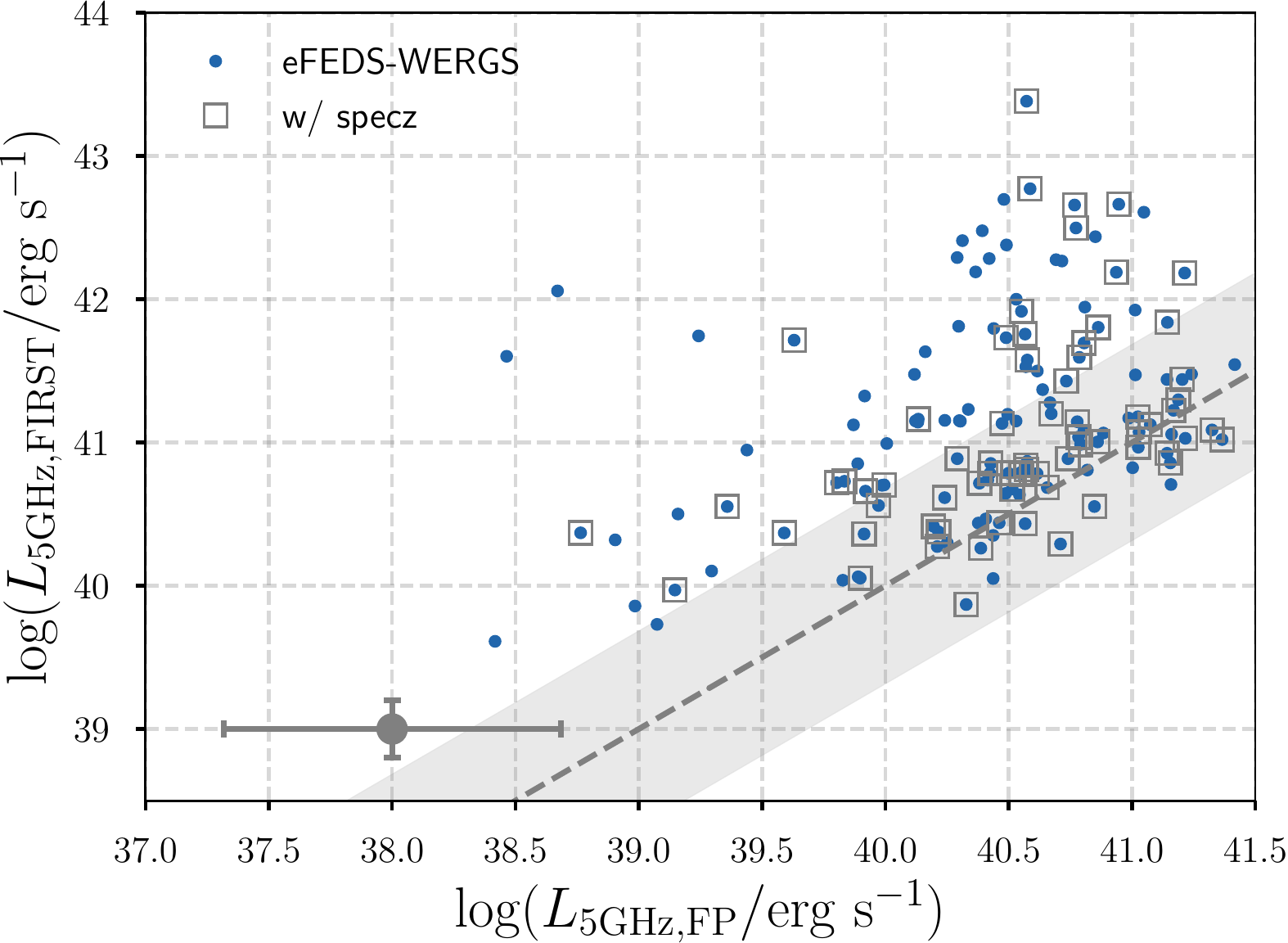}\\
\includegraphics[width=0.45\textwidth]{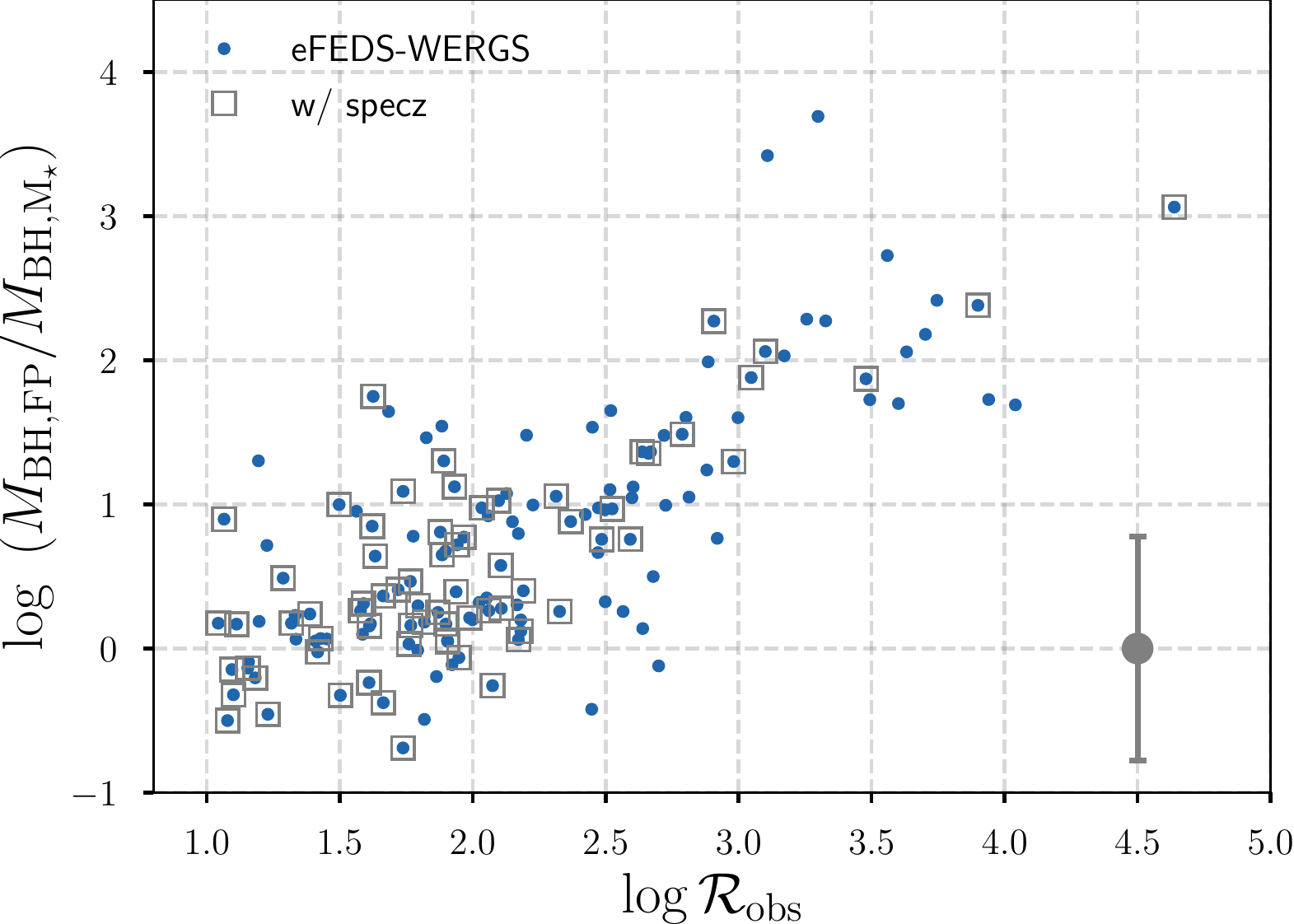}~
\includegraphics[width=0.45\textwidth]{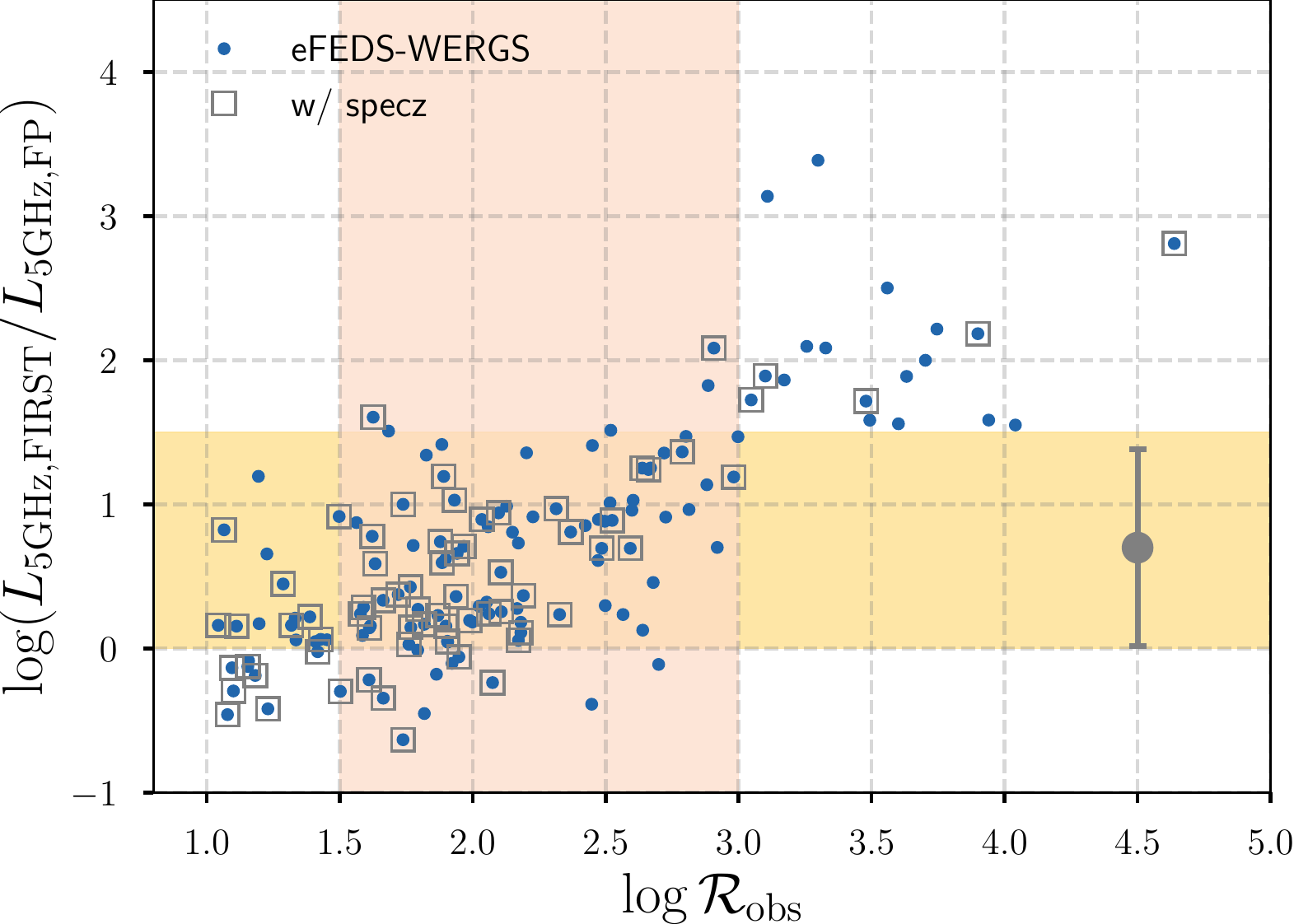}
\caption{
(Top left) 
The relation between the two different BH mass estimations. 
$\mbhfp$ is obtained from the fundamental plane of \cite{gul19} and by using the obtained X-ray and radio luminosities in this study.
$\mbhmstar$ is the BH mass estimation estimated from the scaling relation of \cite{kor13} and by using the obtained stellar mass.
The gray cross represents the typical errorbar of the two parameters.
The gray dashed line represents the 1:1 line with the scatter area of the typical errorbar.
(Top right)
The relation between the $L_\mathrm{5GHz,FIRST}$ and $L_\mathrm{5GHz,FP}$, which is estimated from the fundamental plane relation of \cite{gul19} with the values of $\mbhmstar$ and $\lxhardabs$.
The gray cross represents the typical errorbar of the two parameters.
The gray dashed line represents the 1:1 line, with the scatter area of the typical errorbar.
(Bottom left)
The BH mass ratio of $\mbhfp/\mbhmstar$ as a function of $\Robs$.
(Bottom right)
The radio-luminosity ratio of $L_\mathrm{5GHz,FIRST}/L_\mathrm{5GHz,FP}$ as a function of $\Robs$.
The yellow/pink shaded area represents the possible Doppler boost area by assuming a typical radio-loud AGN/blazars.
The red shaded area represents the possible range of the $\Robs$ for blazars by assuming a blazar sequence at $0<z<5$.
The gray cross represents the typical errorbar of the two values in each panel.
}\label{fig:FP}
\end{center}
\end{figure*}

\subsection{Fundamental Plane of eFEDS-WERGS AGN}\label{sec:discussion:FP}

The fundamental plane
of BH activity is the relation among three parameters of $M_\mathrm{BH}$, $L_\mathrm{R}$, and $L_\mathrm{X}$, where $L_\mathrm{R}=\nu L_\nu (\mathrm{5GHz})$, $L_\mathrm{X} = L_\mathrm{2-10keV}$.  
The empirical relation of the fundamental plane has been reported since the late 1990s \citep{han98}
and this relation gives the important observational connection related to accretion-jet phenomena. The studies of fundamental planes
are mainly focused on the low-accretion rate phase such as RIAF state with $< 0.01 \dot{M}_\mathrm{BH}$,
where the radiation from a BH is non-thermally dominated \citep[e.g.,][]{mer03,hei03,fal04}.
On the other hand, at the phase of high accretion rate, the radiation is more thermally dominated, and it is of great importance whether
the universal trend of the fundamental plane maintains even including the luminous radio galaxies, such as our sample. If so, the fundamental plane hosts the powerful ability to use radio and X-ray observations to estimate black hole mass,
which would be a good alternative method where other methods are not available.

Since both of the nuclear radio and X-ray luminosities are already obtained in this study, we estimate the BH from the fundamental plane relation by adapting
the studies of \cite{gul19}, which is written as:
\begin{align}
\log (\mbhfp / 10^8 M_\odot) = & \mu_0 + \xi_\mathrm{5GHz} \log (L_\mathrm{5GHz}/10^{38} \mathrm{erg/s}) \nonumber\\
&+ \xi_\mathrm{2-10keV} \log (L_\mathrm{2-10keV} / 10^{40} \mathrm{erg/s})
\end{align}
where $\mu_0 = 0.55 \pm 0.22$, $\xi_\mathrm{5GHz}=1.09 \pm 0.10$, $\xi_\mathrm{2-10keV} = -0.59 \pm 0.16$.
Note that this relation is based on the sample of X-ray binaries and AGN whose Eddington ratios are mostly $\lambdaedd<0.01$.
The 5~GHz radio luminosity ($L_\mathrm{5GHz}$) is extrapolated from $\lfirst$ by assuming the power-law radio spectrum with $f_\nu \propto \nu^\alpha$ and $\alpha=-0.7$.

Figure~\ref{fig:MBHFP_vs_z} shows the BH masses estimated from the fundamental plane ($\mbhfp$) as a function of redshift.
Some of the obtained $\mbhfp$ exceeds the observationally known maximum BH mass limit of $\mbhfp > 10^{10.5} \msun$ across the Universe at $0<z<7.5$ \citep{mcc11,kor13,tra14,jun15,wu15,ban18,yan20}. This suggests that $\mbhfp$~is significantly over-estimated, sometimes by one-to-three orders of magnitude.
This huge discrepancy is well beyond the intrinsically large scatter of the original fundamental plane of $\sigma \approx 0.8$~dex.

To investigate this point, we compare the obtained $\mbhfp$ with the BH masses estimated from the stellar mass ($\mbhmstar$) estimated from the 2D image decomposition at $z<1$. Here we apply the $\mbh$-$\mstar$ relation in the local universe \citep{kor13}.
The top left panel of Figure~\ref{fig:FP} shows the relation between the two BH mass estimates of $\mbhfp$ and $\mbhmstar$.
The obtained $\mbhmstar$ is in the range of the known SMBH mass range of $\mbh < 10^{10.5}\msun$.
While 78 out of the 139 sources are within the scatter of 
$\sigma \approx 0.8$~dex between the two BH mass values,
two BH mass values do not follow the expected 1:1 relation
for the remaining 61 sources, but rather shows a clear excess in $\mbhfp/\mbhmstar$.
The result above indicates that our eFEDS-WERGS sources do not follow the fundamental plane for some sources. 
That might be a natural outcome considering that most of the fundamental plane is based on the low-accreting black holes, while roughly half of our sources are likely highly accreting sources, as shown in Section~\ref{sec:results:sBHAR_vs_Mstar}.

The top right panel of Figure~\ref{fig:FP} shows the relation between the $L_\mathrm{5GH,FIRST}$ and the estimated 5~GHz radio luminosity from the fundamental plane by using the obtained $\mbhmstar$ and $\lxhardabs$ ($L_\mathrm{5GH,FP}$). This shows that our eFEDS-WERGS shows a strong radio emission excess compared to the expected one from the fundamental plane.
%
The obtained radio emission excess ranges up to three orders of magnitude and cannot be solely explained by differences in the radio slope $\alpha$ (where $f_\nu \propto \nu^{\alpha}$), which only results in a 0.2 dex difference for $-1.0 < \alpha < -0.5$, as demonstrated by the typical error bar in the panel, and at most a 0.5 dex difference even for the steepest value of $\alpha=-1.7$.

The bottom left panel of Figure~\ref{fig:FP} shows a relation between the ratio of $\mbhfp/\mbhmstar$ as a function of the observed radio-loudness ($\Robs$). The Figure shows a good linear relation between the two values, supporting that radio-emission excess compared to the optical radiation produces the excess of the $\mbhfp$ estimation.

One possible origin of such radio emission excess might be from the beaming effect of the jet emission in the radio band \citep{wan06,li08,don15}. 
\cite{li08} discussed how much the boosting factor can produce the observed excess of the radio emission as a function of the Lorentz factor ($\Gamma_\mathrm{L}$) and the inclination angle $\theta$, where $\theta=0$ means face-on view \citep[see also the discussion of ][]{hei04}.
They demonstrate that radio emission excess is estimated to be less than 30 for most of the cases by assuming the typical Lorentz factor of $\Gamma_\mathrm{L}=5$ \citep{orr82} and an expected inclination angle of $\theta>10$~degree \citep{mar94}.
The bottom right panel of Figure~\ref{fig:FP} shows the radio luminosity ratio of $L_\mathrm{5GHz,FIRST}/L_\mathrm{5GHz,FP}$ as a function of $\Robs$. The yellow-shaded region is the expected range of the radio emission excess for typical radio-loud AGN and quasars as discussed above. Some of the sources with $\log \Robs < 3$ are in this range, suggesting that radio emission excess might be a boost origin. On the other hand, for the sources with $\log \Robs >3$, the radio emission excess cannot be explained with the Lorentz boost alone, with a typical Lorentz factor and the inclination angle. 

\cite{li08} also demonstrated that radio emission can boost at maximum of
$L_\mathrm{5GHz,FIRST}/L_\mathrm{5GHz,FP} \lesssim 10^3$ assuming the extreme case with $\Gamma_\mathrm{L}>10$ and $\theta \ll 10$~degree, which would be achievable for blazar cases.
On the other hand, \cite{ich21} discussed the possible $\Robs$ range for blazars based on the known blazar sequence
in the radio and optical bands \citep{fos98,don01,ino09,ghi17}.
The observed radio-loudness of blazars is in the range of $\log \mathcal{R}_\mathrm{obs} \sim 1.5-3$. This range is obtained 
by using the luminosity range of $\lfirst=10^{41-43}$~erg~s$^{-1}$ and shifting redshift to the range of $z\sim0$--$5$. This 
area is shown in pink colored area in the bottom panel of Figure~\ref{fig:FP}.
The combination of them show that although 
some of our eFEDS-WERGS sources might have a boosted radio emission in the FIRST band, still there are sources whose radio emission excess cannot be explained by the Lorentz boost alone. 
This is more prominent if sources have high radio-loudness of $\log \Robs>3$.

\begin{figure}
\begin{center}
\includegraphics[width=0.5\textwidth]{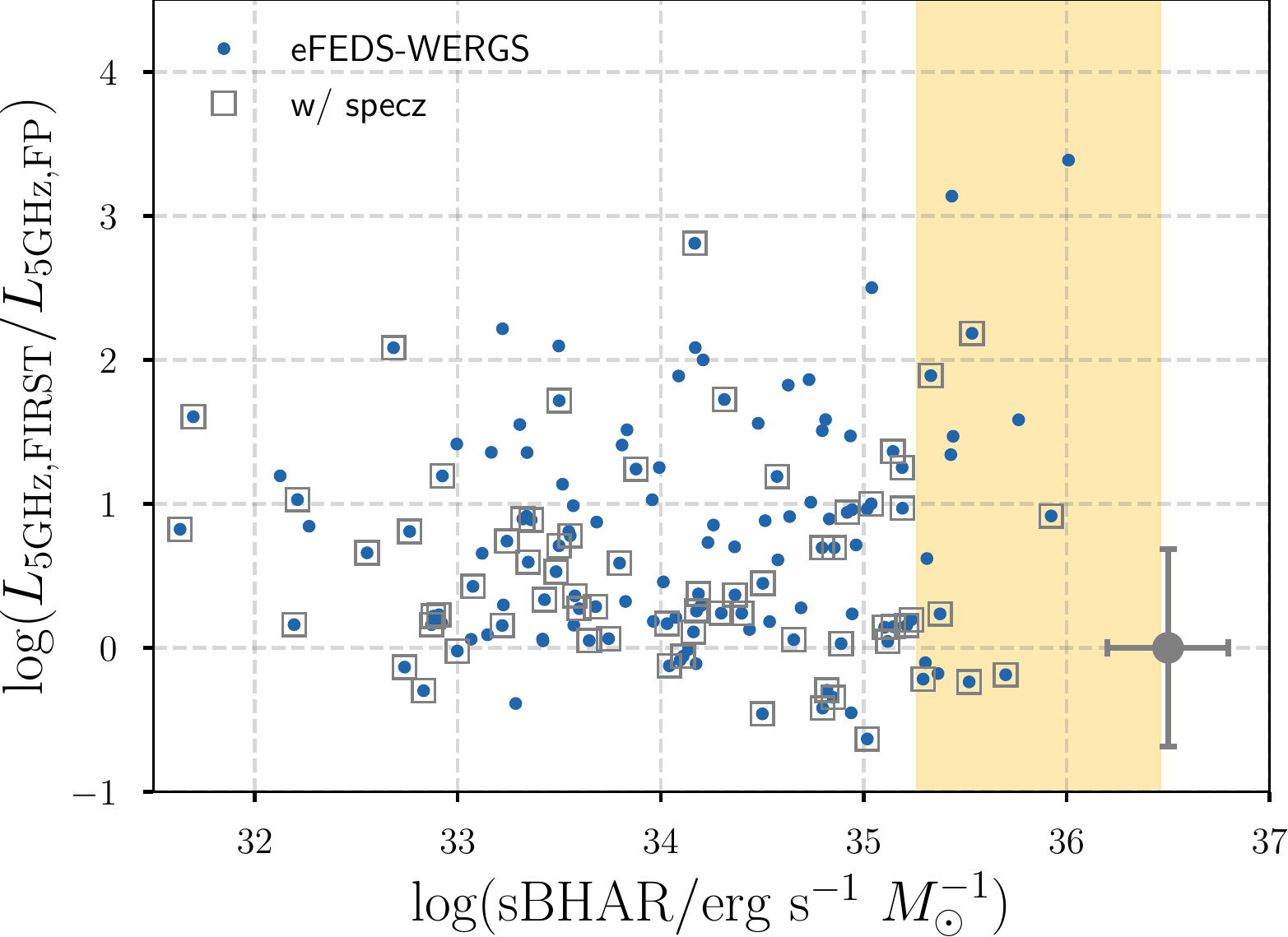}~
\caption{
The radio-luminosity ratio of $L_\mathrm{5GHz,FIRST}/L_\mathrm{5GHz,FP}$ as a function of sBHAR.
The yellow shaded area represents the corresponding Eddington limit accretion ($\lambdaedd=1$) region assuming the local scaling relation of $\mbh$--$\mstar$ \citep{kor13} for the range of
$9<\log (\mstar/\msun)<12$ where our sources reside as shown in Figure~\ref{fig:sBHARvsMstar}.
The gray cross represents the typical errorbar of the two values.
}\label{fig:Lradioratio_vs_sBHAR}
\end{center}
\end{figure}

The additional origins of strong radio emission for those sources with $\log \Robs > 3$ is unclear, but one key aspect is that most of them are likely highly accreting sources.
Such high jet emission efficiency can be achieved if the accretion disk of such sources is actually in the ``radiatively inefficient'' state, but the physical origin of radiative inefficiency is different from the disks in the local radio galaxies. 
They may be undergoing more rapid mass accretion \citep[so called the Slim disk model;][]{abr88}. 
 Recent radiation hydrodynamical simulations suggest that when the mass accretion rate significantly exceeds the Eddington rate,
radiation is effectively trapped within the accreting matter and advected to the central BH before escaping by radiative diffusion. 
As a result, the emergent radiation luminosity is saturated at the order of $\ledd$ (i.e., $\etar \lesssim 0.1$ at $\gg \dot{M}_\mathrm{Edd}$) 
and the accretion flow turns into a radiatively inefficient state even with a super-Eddington accretion rate \citep[e.g.,][]{ohs05,ohs09,mck14,ina16b,tak20}.

This  would produce the suppressed X-ray emission and more prominent radio emission, which produces a larger discrepancy between the $L_\mathrm{5GHz,FIRST}$ and $L_\mathrm{5GH,FP}$ and leading larger
$L_\mathrm{5GHz,FIRST}/L_\mathrm{5GHz,FP}$.
This is prominent in Figure~\ref{fig:Lradioratio_vs_sBHAR}, which shows the eFEDS-WERGS sources in the plane of $L_\mathrm{5GHz,FIRST}/L_\mathrm{5GHz,FP}$ and sBHAR.
While the sample size is limited at $z<1$ because of the availability of $\mstar$ in this study, the sources with high sBHAR, and around the expected Eddington limit region (yellow shaded area) might have a tentative sign of an excess of $L_\mathrm{5GHz,FIRST}/L_\mathrm{5GHz,FP}$. 
However, this trend should be confirmed by obtaining the stellar mass or BH masses of our high-$z$ ($z>1$) eFEDS-WERGS sample, which would contain more Eddington-limit sources rather than the sources discussed here at $z<1$.
 This phase is considered to be a key process of the BH seed growth in the high-$z$ ($z>6$) universe to describe the already known massive high-$z$ SMBHs \citep[e.g.,][]{mor11,wu15,ban18,ono19}
 and such high radio-loudness sources with $\log \Robs>3$ would be prominent candidates for such a high accreting phase.

\section{Conclusion}

The Subaru/HSC SSP survey enables us to find optical counterparts missed in the previous surveys.
WERGS catalog \citep{yam18} provides the widest ($>100$~deg$^2$) and deepest ($i_\mathrm{AB}\sim26$) optical counterparts of the radio bright VLA/FIRST population
whose optical counterparts are missing before.
We have constructed the first X-ray catalog of WERGS radio galaxies in the 140~deg$^2$ eFEDS field.
Thanks to the wide and medium depth X-ray survey down to $f_\mathrm{0.5-2keV} = 6.5 \times 10^{-15}$~erg~s~cm$^{-2}$ by eROSITA onboard the Spektrum-Roentgen-Gamma (SRG) mission in the eFEDS field, 
we have found 393 (180 spec-$z$ and 213 reliable photo-$z$) eROSITA detected WERGS radio galaxies (hereafter, eFEDS-WERGS catalog)
spanning the wide redshift range at $0<z<6$,
where most of them reside in $0<z<4$.
By combining with the eFEDS X-ray catalog by \cite{liu21} and the multi-wavelength catalog by \cite{sal21}, the eFEDS-WERGS catalog contains a wealth of information crucial for both of BHs and the  host galaxies, including X-ray, radio, and mid-infrared luminosities and the column density $\NH$, as well as the host galaxy properties such as stellar-mass for the 139 sources at $z<1$ thanks to the high spatial resolution imaging by Subaru/HSC \citep{li21}.

We have investigated the properties of the eFEDS-WERGS radio galaxies and found the following key results:

\begin{enumerate}

\item Thanks to the wide and medium depth X-ray surveys, eFEDS-WERGS contains the rare, and most X-ray and mid-IR luminous radio galaxies above the knee of the X-ray luminosity function at $1<z<4$, spanning $44 < (\lxsoftabs/\mathrm{erg}~\mathrm{s}^{-1}) < 46.5$ and/or $45 < (\ltwelve/\mathrm{erg}~\mathrm{s}^{-1}) < 47$. On the other hand, the sample covers the sources around and below the knee for the sources at $z<1$.

\item The 91\% of the sample (356 out of 393) are unobscured AGN with $\log (\NHunit) < 22$, while rare, 9\% of the sample (37 sources) show obscured AGN signature with $\log (\NHunit) > 22$. Most of those obscured AGN are relatively high-$z$ ($0.4<z<3.2$) and they are X-ray luminous with $(\lxsoftabs/\mathrm{erg}~\mathrm{s}^{-1})>44$, which is similar luminosities with the SDSS optical quasars. Thus, they are obscured counterparts of the radio-loud quasar, which are largely missed in the previous studies and one of the important populations to study the quasar evolution at $z=1$--$3$.

\item The majority of the eFEDS-WERGS sources follow the almost similar luminosity correlations of $\lxhardabs$--$\lsix$ of radio-quiet AGN/quasars. This indicates that X-ray and mid-IR bands trace the AGN corona and dust emissions respectively without the contamination by the jet emission, even for radio galaxies and the majority of the eFEDS-WERGS are likely to be non-blazars.

\item $\lfirst$--$\lxhardabs$ relation shows a different slope trend between the low- and high luminosity end with the boundary of $(\lxhardabs/\mathrm{erg}~\mathrm{s}^{-1})\sim44$. $\ljet$--$\lbol$ relation revealed that there are sources with high jet production efficiency at $\etaj\approx1$. Such a high $\etaj$ originates either from the lower radiative efficiency for those low-luminosity AGN, or originates from the result of the experienced AGN feedback by jet within $\sim$Myr, which produces the declining of $\lbol$ by a factor of 10--100 by keeping $\ljet$ constant in the previous Myr,
leading the boost of the obtained $\etaj$.

\item The expected BH mass ($\mbhfp$) estimated from the fundamental plane and $L_\mathrm{5GHz}$ and $\lxhardabs$ obtained by the eFEDS-WERGS sources show unreasonably high values at $\mbhfp >10^{10} \msun$ for some sources, indicating that eFEDS-WERGS show a deviation from the conventional fundamental plane relation. This deviation stems from high $\Robs$, the excess of radio luminosities. Such radio emission excess cannot be explained by the Doppler booming alone, and therefore disk-jet connection for some sources of the eFEDS-WERGS sample might be fundamentally different from the conventional plane that mainly traces the low accretion rate based sources.

\end{enumerate}

\begin{acknowledgements}

We appreciate the referee for the constructive and fruitful comments which deepen our understanding of the sample.
We thank Yoshiyuki Inoue for providing us with the blazar sequence SED templates.
This work is supported by Program for Establishing a Consortium for the Development of Human Resources in Science
and Technology, Japan Science and Technology Agency (JST) and is partially supported by Japan Society for the Promotion of Science (JSPS) KAKENHI (20H01939; K.~Ichikawa, T.~Kawamruo; JP20K14529).
K.~Ichikawa also thanks Max Planck Institute for hosting me and most of the studies are conducted at MPE.
K.I. acknowledges support from the National Natural Science Foundation of China (12073003, 12003003, 11721303, 11991052, 11950410493).
T.K acknowledges support from FONDECYT Postdoctral Fellowship (3200470). 

This work is based on data from eROSITA, the soft X-ray instrument aboard SRG, a joint Russian-German science mission supported by the Russian Space Agency (Roskosmos), in the interests of the Russian Academy of Sciences represented by its Space Research Institute (IKI), and the Deutsches Zentrum f\"ur Luft- und Raumfahrt (DLR). The SRG spacecraft was built by Lavochkin Association (NPOL) and its subcontractors, and is operated by NPOL with support from the Max Planck Institute for Extraterrestrial Physics (MPE).

The development and construction of the eROSITA X-ray instrument was led by MPE, with contributions from the Dr. Karl Remeis Observatory Bamberg \& ECAP (FAU Erlangen-Nuernberg), the University of Hamburg Observatory, the Leibniz Institute for Astrophysics Potsdam (AIP), and the Institute for Astronomy and Astrophysics of the University of T\"ubingen, with the support of DLR and the Max Planck Society. The Argelander Institute for Astronomy of the University of Bonn and the Ludwig Maximilians Universit\"at Munich also participated in the science preparation for eROSITA.

The Hyper Suprime-Cam (HSC) collaboration includes the astronomical communities of Japan and Taiwan, and Princeton University.  The HSC instrumentation and software were developed by the National Astronomical Observatory of Japan(NAOJ), the Kavli Institute for the Physics and Mathematics of the Universe (Kavli IPMU), the University of Tokyo, the High Energy Accelerator Research Organization (KEK), the Academia Sinica Institute for Astronomy and Astrophysics in Taiwan (ASIAA), and Princeton University.  Funding was contributed by the FIRST program from Japanese Cabinet Office, the Ministry of Education, Culture, Sports, Science and Technology (MEXT), the Japan Society for the Promotion of Science (JSPS), Japan Science and Technology Agency (JST),the Toray Science Foundation, NAOJ, Kavli IPMU, KEK,ASIAA, and Princeton University.
Funding for the Sloan Digital Sky Survey (SDSS) has been provided by the Alfred P. Sloan Foundation, the Participating Institutions, the National Aeronautics and Space Administration, the National Science Foundation, the US Department of Energy, the Japanese Monbukagakusho, and the Max Planck Society. The SDSS Web site is http://www.sdss.org/. The SDSS is managed by the Astrophysical Research Consortium (ARC) for the Participating Institutions. The Participating Institutions are The University of Chicago, Fermilab, the Institute for Advanced Study, the Japan Participation Group, The Johns Hopkins University, Los Alamos National Laboratory, the Max-Planck-Institute for Astronomy (MPIA), the Max-Planck-Institute for Astrophysics (MPA), New Mexico State University, University of Pittsburgh, Princeton University, the United States Naval Observatory, and the University of Washington.

\end{acknowledgements}

\begin{appendix}

\section{$P_\mathrm{jet}$ estimation based on a different $P_\mathrm{jet}$--$L_\mathrm{radio}$ relation}\label{sec:AppendixA}

There are several studies providing the relation between $\ljet$ and $L_\mathrm{radio}$ \citep[e.g.,][]{cav10,dal12,hec14,ine17}. Each relation has a different slope and normalization, which mainly originates from the different samples with different radio luminosity ranges, radio emission sizes, and environments.
Recently, \cite{har19} compared several relations and found that most of their sample in the $\ljet$--$L_\mathrm{radio}$ plane is located between the relations of \cite{ine17} and \cite{cav10}, and the latter relation is used in this study. Motivated by these findings, we investigate the effect of the different relation by \cite{ine17} on our $\ljet$ distribution and the subsequent $\etaj$ distribution.

The relation by \cite{ine17} shows a steeper slope of $b=0.89$ compared to $b=0.75$ by \cite{cav10}, and the relation is as follows:
\begin{align}
\ljet = 2.6 \times 10^{46}~(\lfirst/10^{27}~\whz)^{0.89}~\mathrm{erg}~\mathrm{s}^{-1} \label{eq:ljet_lfirst_Ine17}.
\end{align}
Note that the original equation by \cite{ine17} is for $L_\mathrm{151MHz}$ and the above equation is derived assuming that most radio sources have a spectral index of $\alpha=0.7$.

\begin{figure*}
\begin{center}
\includegraphics[width=0.48\textwidth]{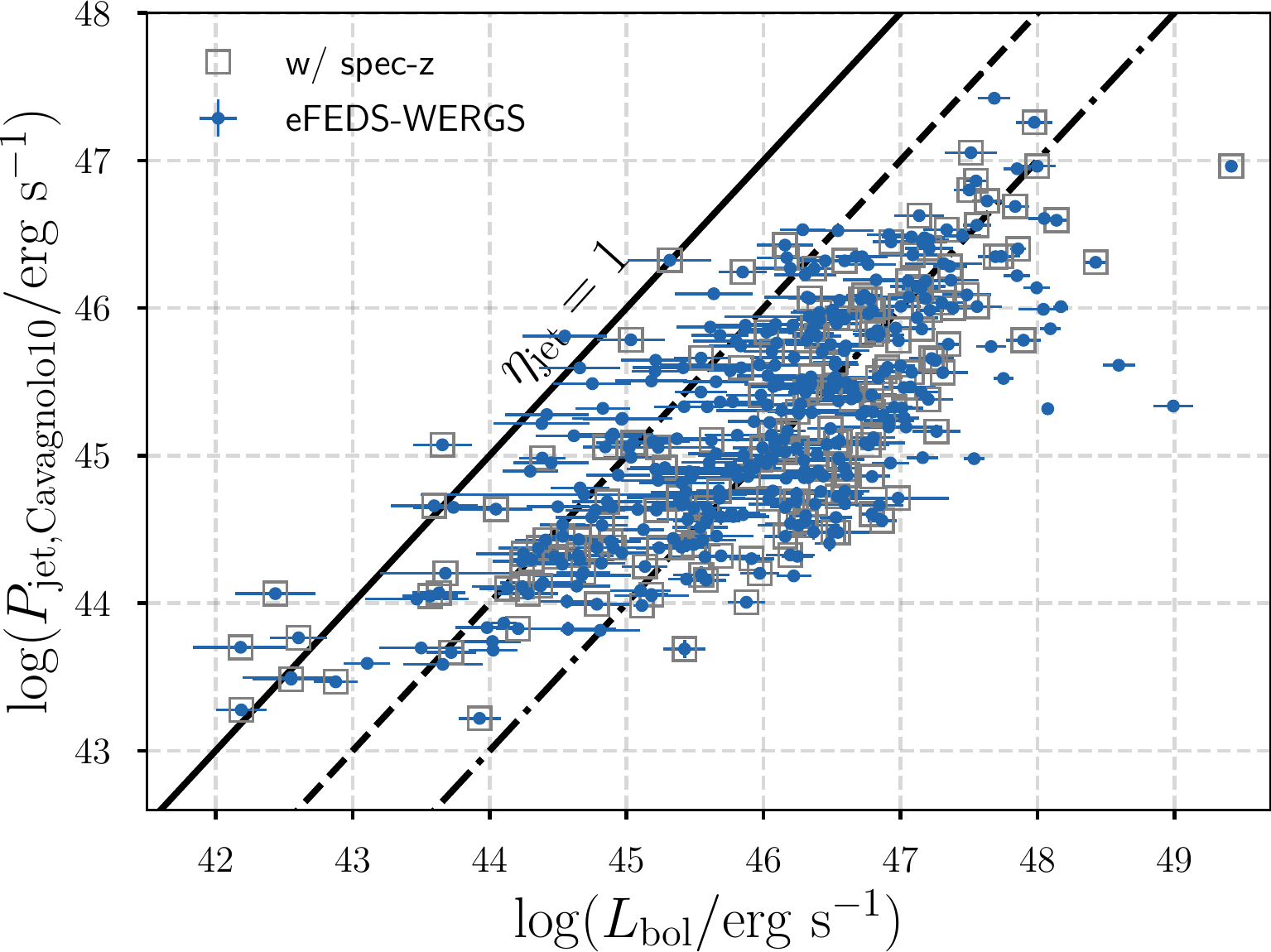}
\includegraphics[width=0.48\textwidth]{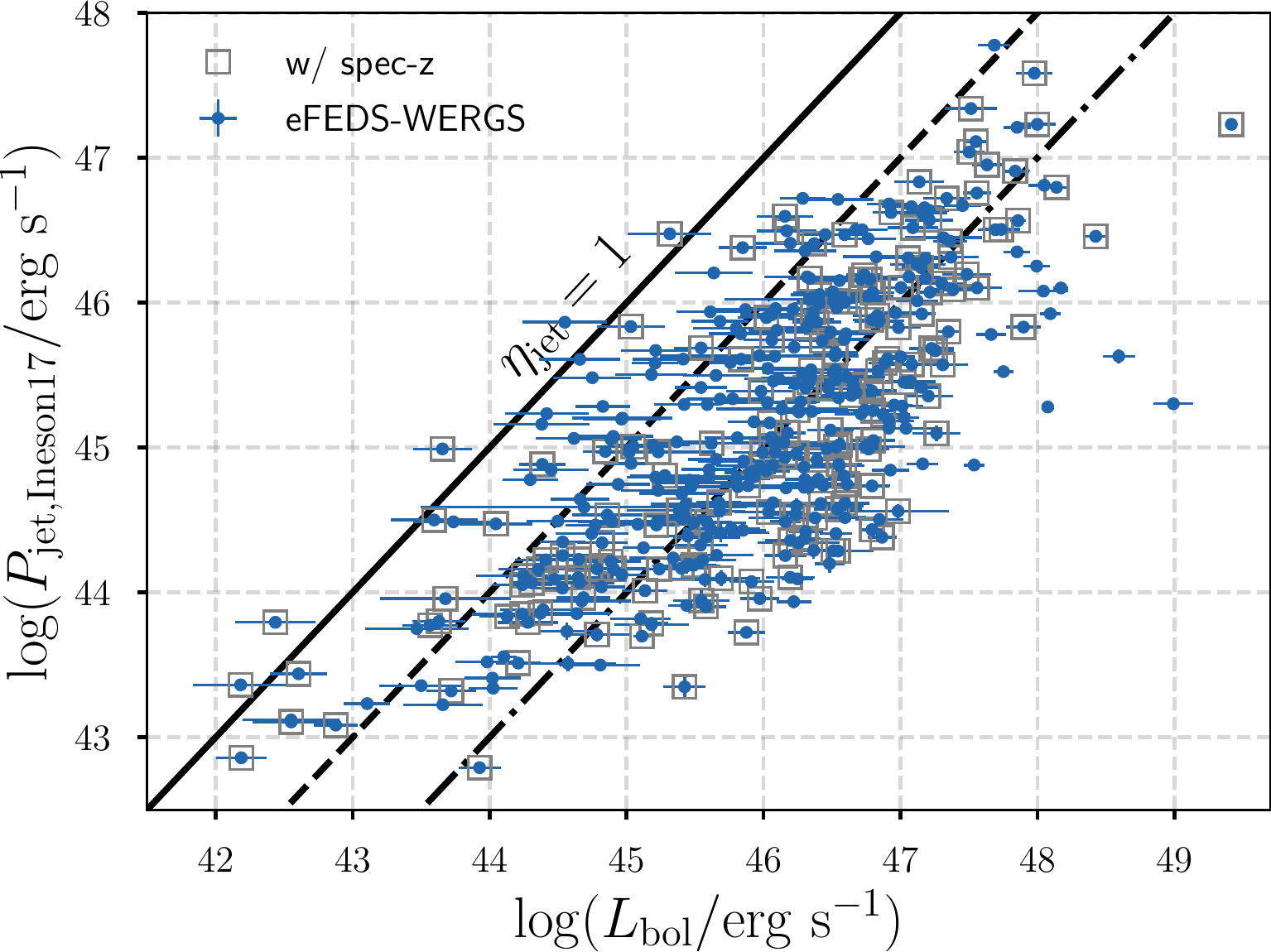}\\
\includegraphics[width=0.48\textwidth]{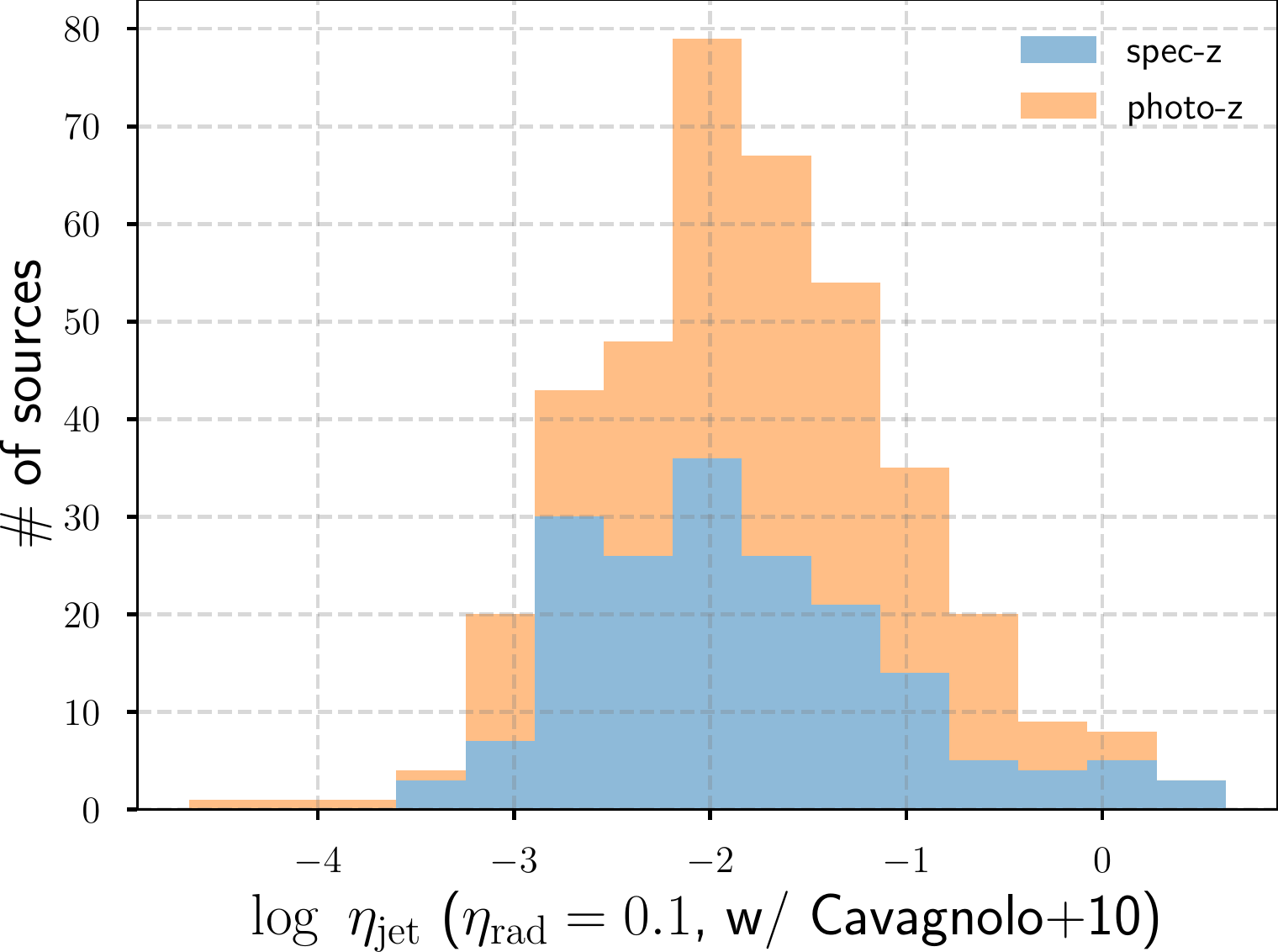}
\includegraphics[width=0.48\textwidth]{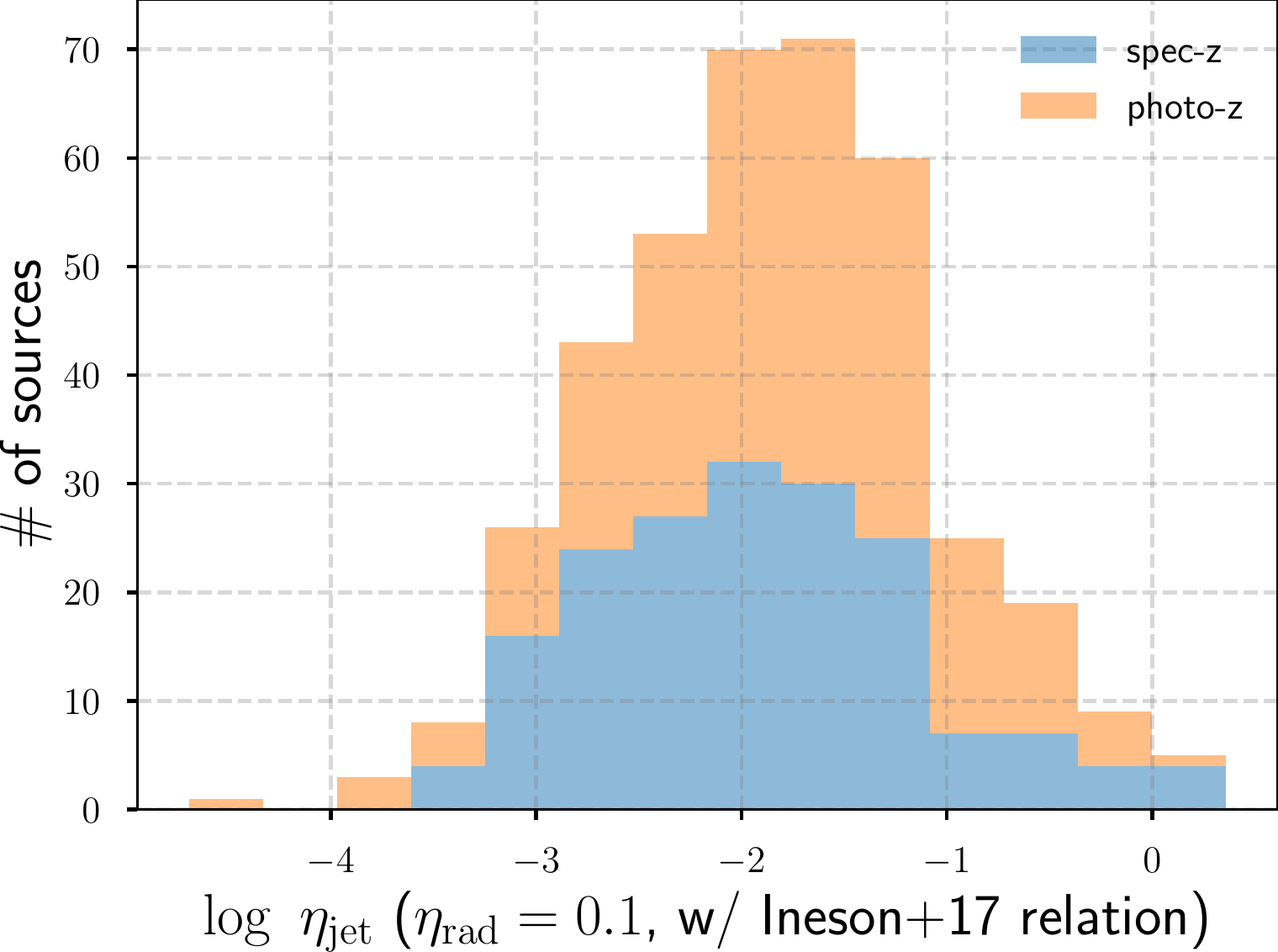}
\caption{
(Top) The correlation between the jet power ($\ljet$) and bolometric AGN luminosity ($\lbol$). The black solid/dashed/dot-dashed line shows the case with the jet production efficiency $\etaj=1$, $\etaj=0.1$, and $\etaj=0.01$, respectively, by assuming the radiation efficiency of $\etar=0.1$ \citep{sol82}.
The other symbols are the same as in Figure~\ref{fig:R_vs_imag}. 
(Bottom) The distribution of $\etaj$ assuming the radiative efficiency of 0.1.
The left panel shows the result using the relation by \cite{cav10}, and right panel shows the one using the relation by \cite{ine17}.}
\label{fig:AppendixA}
\end{center}
\end{figure*}

The top panels of Figure~\ref{fig:AppendixA} show the $\ljet$--$\lfirst$ relation based on the equations by \cite{cav10} (left) and by \cite{ine17} (right), respectively. As expected from the steeper slope of $b=0.89$ by \cite{ine17}, the obtained $\ljet$ values have a more extended distribution than those calculated using the relation by \cite{cav10}. However, the median values of $\ljet$ are comparable, with $\left< \log \ljet \right> = 45.15 \pm 0.79$ for \cite{cav10} and $\left< \log \ljet \right> = 45.08 \pm 0.79$ for \cite{ine17}.

For $\etaj$, the median value for sources with $\lbol < 10^{45}$~erg~s$^{-1}$ is $\left< \log \etaj \right> = -1.0 \pm 0.6$ \citep{cav10} and $\left< \log \etaj \right> = -1.2 \pm 0.6$ \citep{ine17}. For high luminosity sources with $\lbol > 10^{45}$~erg~s$^{-1}$, the median value is $\left< \log \etaj \right> = -2.0 \pm 0.7$ \citep{cav10} and $\left< \log \etaj \right> = -2.0 \pm 0.7$ \citep{ine17}, respectively.

The bottom panels of Figure~\ref{fig:AppendixA} show the distribution of $\etaj$ derived from the two relations. As shown in the Figures, $\etaj \simeq1$ sources still exist even when using the relation of \cite{ine17}, while the number of $\etaj\simeq1$ sources decreases from 14 sources to 10.
Thus, we conclude that the different $\ljet$--$L_\mathrm{radio}$ relation does not affect our main results and discussions.

\end{appendix}

\bibliographystyle{aa} 
\bibliography{eFEDSWERGS} 

\begin{thebibliography}{217}
\expandafter\ifx\csname natexlab\endcsname\relax\def\natexlab#1{#1}\fi

\bibitem[{{Abdurro'uf} {et~al.}(2021){Abdurro'uf}, {Accetta}, {Aerts}, {Silva
  Aguirre}, {Ahumada}, {Ajgaonkar}, {Filiz Ak}, {Alam}, {Allende Prieto},
  {Almeida}, \& et~al.}]{abd21}
{Abdurro'uf}, {Accetta}, K., {Aerts}, C., {et~al.} 2021, arXiv e-prints,
  arXiv:2112.02026

\bibitem[{{Abramowicz} {et~al.}(1988){Abramowicz}, {Czerny}, {Lasota}, \&
  {Szuszkiewicz}}]{abr88}
{Abramowicz}, M.~A., {Czerny}, B., {Lasota}, J.~P., \& {Szuszkiewicz}, E. 1988,
  \apj, 332, 646

\bibitem[{{Ahumada} {et~al.}(2020){Ahumada}, {Prieto}, {Almeida}, {Anders},
  {Anderson}, {Andrews}, {Anguiano}, {Arcodia}, {Armengaud}, {Aubert}, \&
  et~al.}]{ahu20}
{Ahumada}, R., {Prieto}, C.~A., {Almeida}, A., {et~al.} 2020, \apjs, 249, 3

\bibitem[{{Aihara} {et~al.}(2018{\natexlab{a}}){Aihara}, {Arimoto},
  {Armstrong}, {Arnouts}, {Bahcall}, {Bickerton}, {Bosch}, {Bundy}, {Capak},
  {Chan}, {Chiba}, {Coupon}, {Egami}, {Enoki}, {Finet}, {Fujimori}, {Fujimoto},
  {Furusawa}, {Furusawa}, {Goto}, {Goulding}, {Greco}, {Greene}, {Gunn},
  {Hamana}, {Harikane}, {Hashimoto}, {Hattori}, {Hayashi}, {Hayashi},
  {He{\l}miniak}, {Higuchi}, {Hikage}, {Ho}, {Hsieh}, {Huang}, {Huang},
  {Ikeda}, {Imanishi}, {Inoue}, {Iwasawa}, {Iwata}, {Jaelani}, {Jian},
  {Kamata}, {Karoji}, {Kashikawa}, {Katayama}, {Kawanomoto}, {Kayo}, {Koda},
  {Koike}, {Kojima}, {Komiyama}, {Konno}, {Koshida}, {Koyama}, {Kusakabe},
  {Leauthaud}, {Lee}, {Lin}, {Lin}, {Lupton}, {Mand elbaum}, {Matsuoka},
  {Medezinski}, {Mineo}, {Miyama}, {Miyatake}, {Miyazaki}, {Momose}, {More},
  {More}, {Moritani}, {Moriya}, {Morokuma}, {Mukae}, {Murata}, {Murayama},
  {Nagao}, {Nakata}, {Niida}, {Niikura}, {Nishizawa}, {Obuchi}, {Oguri},
  {Oishi}, {Okabe}, {Okamoto}, {Okura}, {Ono}, {Onodera}, {Onoue}, {Osato},
  {Ouchi}, {Price}, {Pyo}, {Sako}, {Sawicki}, {Shibuya}, {Shimasaku},
  {Shimono}, {Shirasaki}, {Silverman}, {Simet}, {Speagle}, {Spergel},
  {Strauss}, {Sugahara}, {Sugiyama}, {Suto}, {Suyu}, {Suzuki}, {Tait},
  {Takada}, {Takata}, {Tamura}, {Tanaka}, {Tanaka}, {Tanaka}, {Tanaka},
  {Terai}, {Terashima}, {Toba}, {Tominaga}, {Toshikawa}, {Turner}, {Uchida},
  {Uchiyama}, {Umetsu}, {Uraguchi}, {Urata}, {Usuda}, {Utsumi}, {Wang}, {Wang},
  {Wong}, {Yabe}, {Yamada}, {Yamanoi}, {Yasuda}, {Yeh}, {Yonehara}, \&
  {Yuma}}]{aih18a}
{Aihara}, H., {Arimoto}, N., {Armstrong}, R., {et~al.} 2018{\natexlab{a}},
  \pasj, 70, S4

\bibitem[{{Aihara} {et~al.}(2018{\natexlab{b}}){Aihara}, {Armstrong},
  {Bickerton}, {Bosch}, {Coupon}, {Furusawa}, {Hayashi}, {Ikeda}, {Kamata},
  {Karoji}, {Kawanomoto}, {Koike}, {Komiyama}, {Lang}, {Lupton}, {Mineo},
  {Miyatake}, {Miyazaki}, {Morokuma}, {Obuchi}, {Oishi}, {Okura}, {Price},
  {Takata}, {Tanaka}, {Tanaka}, {Tanaka}, {Uchida}, {Uraguchi}, {Utsumi},
  {Wang}, {Yamada}, {Yamanoi}, {Yasuda}, {Arimoto}, {Chiba}, {Finet},
  {Fujimori}, {Fujimoto}, {Furusawa}, {Goto}, {Goulding}, {Gunn}, {Harikane},
  {Hattori}, {Hayashi}, {He{\l}miniak}, {Higuchi}, {Hikage}, {Ho}, {Hsieh},
  {Huang}, {Huang}, {Imanishi}, {Iwata}, {Jaelani}, {Jian}, {Kashikawa},
  {Katayama}, {Kojima}, {Konno}, {Koshida}, {Kusakabe}, {Leauthaud}, {Lee},
  {Lin}, {Lin}, {Mandelbaum}, {Matsuoka}, {Medezinski}, {Miyama}, {Momose},
  {More}, {More}, {Mukae}, {Murata}, {Murayama}, {Nagao}, {Nakata}, {Niida},
  {Niikura}, {Nishizawa}, {Oguri}, {Okabe}, {Ono}, {Onodera}, {Onoue}, {Ouchi},
  {Pyo}, {Shibuya}, {Shimasaku}, {Simet}, {Speagle}, {Spergel}, {Strauss},
  {Sugahara}, {Sugiyama}, {Suto}, {Suzuki}, {Tait}, {Takada}, {Terai}, {Toba},
  {Turner}, {Uchiyama}, {Umetsu}, {Urata}, {Usuda}, {Yeh}, \& {Yuma}}]{aih18b}
{Aihara}, H., {Armstrong}, R., {Bickerton}, S., {et~al.} 2018{\natexlab{b}},
  \pasj, 70, S8

\bibitem[{{Aird} {et~al.}(2019){Aird}, {Coil}, \& {Georgakakis}}]{air19}
{Aird}, J., {Coil}, A.~L., \& {Georgakakis}, A. 2019, \mnras, 484, 4360

\bibitem[{{Aird} {et~al.}(2012){Aird}, {Coil}, {Moustakas}, {Blanton},
  {Burles}, {Cool}, {Eisenstein}, {Smith}, {Wong}, \& {Zhu}}]{air12}
{Aird}, J., {Coil}, A.~L., {Moustakas}, J., {et~al.} 2012, \apj, 746, 90

\bibitem[{{Alonso-Herrero} {et~al.}(2011){Alonso-Herrero}, {Ramos Almeida},
  {Mason}, {Asensio Ramos}, {Roche}, {Levenson}, {Elitzur}, {Packham},
  {Rodr{\'\i}guez Espinosa}, {Young}, {D{\'\i}az-Santos}, \&
  {P{\'e}rez-Garc{\'\i}a}}]{alo11}
{Alonso-Herrero}, A., {Ramos Almeida}, C., {Mason}, R., {et~al.} 2011, \apj,
  736, 82

\bibitem[{{An} \& {Baan}(2012)}]{an12}
{An}, T. \& {Baan}, W.~A. 2012, \apj, 760, 77

\bibitem[{{Antonucci}(1993)}]{ant93}
{Antonucci}, R. 1993, \araa, 31, 473

\bibitem[{{Arnouts} {et~al.}(1999){Arnouts}, {Cristiani}, {Moscardini},
  {Matarrese}, {Lucchin}, {Fontana}, \& {Giallongo}}]{arn99}
{Arnouts}, S., {Cristiani}, S., {Moscardini}, L., {et~al.} 1999, \mnras, 310,
  540

\bibitem[{{Asmus} {et~al.}(2015){Asmus}, {Gandhi}, {H{\"o}nig}, {Smette}, \&
  {Duschl}}]{asm15}
{Asmus}, D., {Gandhi}, P., {H{\"o}nig}, S.~F., {Smette}, A., \& {Duschl}, W.~J.
  2015, \mnras, 454, 766

\bibitem[{{Ba{\~n}ados} {et~al.}(2018){Ba{\~n}ados}, {Venemans},
  {Mazzucchelli}, {Farina}, {Walter}, {Wang}, {Decarli}, {Stern}, {Fan},
  {Davies}, {Hennawi}, {Simcoe}, {Turner}, {Rix}, {Yang}, {Kelson}, {Rudie}, \&
  {Winters}}]{ban18}
{Ba{\~n}ados}, E., {Venemans}, B.~P., {Mazzucchelli}, C., {et~al.} 2018, \nat,
  553, 473

\bibitem[{{Balbus} \& {Hawley}(1991)}]{bal91}
{Balbus}, S.~A. \& {Hawley}, J.~F. 1991, \apj, 376, 214

\bibitem[{{Baldi} {et~al.}(2015){Baldi}, {Capetti}, \& {Giovannini}}]{bal15a}
{Baldi}, R.~D., {Capetti}, A., \& {Giovannini}, G. 2015, \aap, 576, A38

\bibitem[{{Baldi} {et~al.}(2018){Baldi}, {Capetti}, \& {Massaro}}]{bal18c}
{Baldi}, R.~D., {Capetti}, A., \& {Massaro}, F. 2018, \aap, 609, A1

\bibitem[{{Baldry} {et~al.}(2018){Baldry}, {Liske}, {Brown}, {Robotham},
  {Driver}, {Dunne}, {Alpaslan}, {Brough}, {Cluver}, {Eardley}, {Farrow},
  {Heymans}, {Hildebrandt}, {Hopkins}, {Kelvin}, {Loveday}, {Moffett},
  {Norberg}, {Owers}, {Taylor}, {Wright}, {Bamford}, {Bland-Hawthorn},
  {Bourne}, {Bremer}, {Colless}, {Conselice}, {Croom}, {Davies}, {Foster},
  {Grootes}, {Holwerda}, {Jones}, {Kafle}, {Kuijken}, {Lara-Lopez},
  {L{\'o}pez-S{\'a}nchez}, {Meyer}, {Phillipps}, {Sutherland}, {van Kampen}, \&
  {Wilkins}}]{bal18b}
{Baldry}, I.~K., {Liske}, J., {Brown}, M.~J.~I., {et~al.} 2018, \mnras, 474,
  3875

\bibitem[{{Baldwin}(1982)}]{bal82}
{Baldwin}, J.~E. 1982, in Extragalactic Radio Sources, ed. D.~S. {Heeschen} \&
  C.~M. {Wade}, Vol.~97, 21--24

\bibitem[{{Baron} \& {M{\'e}nard}(2019)}]{bar19}
{Baron}, D. \& {M{\'e}nard}, B. 2019, \mnras, 487, 3404

\bibitem[{{Becker} {et~al.}(1995){Becker}, {White}, \& {Helfand}}]{bec95}
{Becker}, R.~H., {White}, R.~L., \& {Helfand}, D.~J. 1995, \apj, 450, 559

\bibitem[{{Best} \& {Heckman}(2012)}]{bes12}
{Best}, P.~N. \& {Heckman}, T.~M. 2012, \mnras, 421, 1569

\bibitem[{{Best} {et~al.}(2005){Best}, {Kauffmann}, {Heckman}, \&
  {Ivezi{\'c}}}]{bes05}
{Best}, P.~N., {Kauffmann}, G., {Heckman}, T.~M., \& {Ivezi{\'c}}, {\v{Z}}.
  2005, \mnras, 362, 9

\bibitem[{{Birrer} \& {Amara}(2018)}]{bir18}
{Birrer}, S. \& {Amara}, A. 2018, Physics of the Dark Universe, 22, 189

\bibitem[{{Birrer} {et~al.}(2015){Birrer}, {Amara}, \& {Refregier}}]{bir15}
{Birrer}, S., {Amara}, A., \& {Refregier}, A. 2015, \apj, 813, 102

\bibitem[{{B{\^\i}rzan} {et~al.}(2008){B{\^\i}rzan}, {McNamara}, {Nulsen},
  {Carilli}, \& {Wise}}]{bir08}
{B{\^\i}rzan}, L., {McNamara}, B.~R., {Nulsen}, P.~E.~J., {Carilli}, C.~L., \&
  {Wise}, M.~W. 2008, \apj, 686, 859

\bibitem[{{Blandford} {et~al.}(2019){Blandford}, {Meier}, \&
  {Readhead}}]{bla19}
{Blandford}, R., {Meier}, D., \& {Readhead}, A. 2019, \araa, 57, 467

\bibitem[{{Blundell} {et~al.}(1999){Blundell}, {Rawlings}, \&
  {Willott}}]{blu99}
{Blundell}, K.~M., {Rawlings}, S., \& {Willott}, C.~J. 1999, \aj, 117, 677

\bibitem[{{Boquien} {et~al.}(2019){Boquien}, {Burgarella}, {Roehlly}, {Buat},
  {Ciesla}, {Corre}, {Inoue}, \& {Salas}}]{boq19}
{Boquien}, M., {Burgarella}, D., {Roehlly}, Y., {et~al.} 2019, \aap, 622, A103

\bibitem[{{Bosch} {et~al.}(2018){Bosch}, {Armstrong}, {Bickerton}, {Furusawa},
  {Ikeda}, {Koike}, {Lupton}, {Mineo}, {Price}, {Takata}, {Tanaka}, {Yasuda},
  {AlSayyad}, {Becker}, {Coulton}, {Coupon}, {Garmilla}, {Huang}, {Krughoff},
  {Lang}, {Leauthaud}, {Lim}, {Lust}, {MacArthur}, {Mand elbaum}, {Miyatake},
  {Miyazaki}, {Murata}, {More}, {Okura}, {Owen}, {Swinbank}, {Strauss},
  {Yamada}, \& {Yamanoi}}]{bos18}
{Bosch}, J., {Armstrong}, R., {Bickerton}, S., {et~al.} 2018, \pasj, 70, S5

\bibitem[{{Brandt} \& {Alexander}(2015)}]{bra15}
{Brandt}, W.~N. \& {Alexander}, D.~M. 2015, \aapr, 23, 1

\bibitem[{{Brinkmann} {et~al.}(2000){Brinkmann}, {Laurent-Muehleisen}, {Voges},
  {Siebert}, {Becker}, {Brotherton}, {White}, \& {Gregg}}]{bri00}
{Brinkmann}, W., {Laurent-Muehleisen}, S.~A., {Voges}, W., {et~al.} 2000, \aap,
  356, 445

\bibitem[{{Brunner} {et~al.}(2021){Brunner}, {Liu}, {Lamer}, {Georgakakis},
  {Merloni}, {Brusa}, {Bulbul}, {Dennerl}, {Friedrich}, {Liu}, {Maitra},
  {Nandra}, {Ramos-Ceja}, {Sanders}, {Stewart}, {Boller}, {Buchner}, {Clerc},
  {Comparat}, {Dwelly}, {Eckert}, {Finoguenov}, {Freyberg}, {Ghirardini},
  {Gueguen}, {Haberl}, {Kreykenbohm}, {Krumpe}, {Osterhage}, {Pacaud},
  {Predehl}, {Reiprich}, {Robrade}, {Salvato}, {Santangelo}, {Schrabback},
  {Schwope}, \& {Wilms}}]{bru21}
{Brunner}, H., {Liu}, T., {Lamer}, G., {et~al.} 2021, arXiv e-prints,
  arXiv:2106.14517

\bibitem[{{Brusa} {et~al.}(2021){Brusa}, {Urrutia}, {Toba}, {Buchner}, {Li},
  {Liu}, {Perna}, {Salvato}, {Merloni}, {Musiimenta}, {Nandra}, {Wolf},
  {Arcodia}, {Dwelly}, {Georgakakis}, {Goulding}, {Matsuoka}, {Nagao},
  {Schramm}, {Silverman}, \& {Terashima}}]{bru21a}
{Brusa}, M., {Urrutia}, T., {Toba}, Y., {et~al.} 2021, arXiv e-prints,
  arXiv:2106.14525

\bibitem[{{Bruzual} \& {Charlot}(2003)}]{bru03}
{Bruzual}, G. \& {Charlot}, S. 2003, \mnras, 344, 1000

\bibitem[{{Buchner} {et~al.}(2015){Buchner}, {Georgakakis}, {Nandra},
  {Brightman}, {Menzel}, {Liu}, {Hsu}, {Salvato}, {Rangel}, {Aird}, {Merloni},
  \& {Ross}}]{buc15}
{Buchner}, J., {Georgakakis}, A., {Nandra}, K., {et~al.} 2015, \apj, 802, 89

\bibitem[{{Calzetti} {et~al.}(2000){Calzetti}, {Armus}, {Bohlin}, {Kinney},
  {Koornneef}, \& {Storchi-Bergmann}}]{cal00}
{Calzetti}, D., {Armus}, L., {Bohlin}, R.~C., {et~al.} 2000, \apj, 533, 682

\bibitem[{{Capetti} {et~al.}(2017){Capetti}, {Massaro}, \& {Baldi}}]{cap17}
{Capetti}, A., {Massaro}, F., \& {Baldi}, R.~D. 2017, \aap, 598, A49

\bibitem[{{Cavagnolo} {et~al.}(2010){Cavagnolo}, {McNamara}, {Nulsen},
  {Carilli}, {Jones}, \& {B{\^\i}rzan}}]{cav10}
{Cavagnolo}, K.~W., {McNamara}, B.~R., {Nulsen}, P.~E.~J., {et~al.} 2010, \apj,
  720, 1066

\bibitem[{{Chabrier}(2003)}]{cha03}
{Chabrier}, G. 2003, \pasp, 115, 763

\bibitem[{{Chen} {et~al.}(2017){Chen}, {Brandt}, {Reines}, {Lansbury}, {Stern},
  {Alexander}, {Bauer}, {Del Moro}, {Gandhi}, {Harrison}, {Hickox}, {Koss},
  {Lanz}, {Luo}, {Mullaney}, {Ricci}, \& {Trump}}]{che17}
{Chen}, C. T.~J., {Brandt}, W.~N., {Reines}, A.~E., {et~al.} 2017, \apj, 837,
  48

\bibitem[{{Chen} {et~al.}(2020){Chen}, {Ichikawa}, {Noda}, {Kawamuro},
  {Kawaguchi}, {Toba}, \& {Akiyama}}]{che20}
{Chen}, X., {Ichikawa}, K., {Noda}, H., {et~al.} 2020, \apjl, 905, L2

\bibitem[{{Chiaberge} \& {Marconi}(2011)}]{chi11}
{Chiaberge}, M. \& {Marconi}, A. 2011, \mnras, 416, 917

\bibitem[{{Ciotti} {et~al.}(2009){Ciotti}, {Ostriker}, \& {Proga}}]{cio09}
{Ciotti}, L., {Ostriker}, J.~P., \& {Proga}, D. 2009, \apj, 699, 89

\bibitem[{{Condon}(1992)}]{con92}
{Condon}, J.~J. 1992, \araa, 30, 575

\bibitem[{{Corbel} {et~al.}(2003){Corbel}, {Nowak}, {Fender}, {Tzioumis}, \&
  {Markoff}}]{cor03}
{Corbel}, S., {Nowak}, M.~A., {Fender}, R.~P., {Tzioumis}, A.~K., \& {Markoff},
  S. 2003, \aap, 400, 1007

\bibitem[{{Coriat} {et~al.}(2012){Coriat}, {Fender}, \& {Dubus}}]{cor12}
{Coriat}, M., {Fender}, R.~P., \& {Dubus}, G. 2012, \mnras, 424, 1991

\bibitem[{{Croom} {et~al.}(2009){Croom}, {Richards}, {Shanks}, {Boyle},
  {Sharp}, {Bland-Hawthorn}, {Bridges}, {Brunner}, {Cannon}, {Carson}, {Chiu},
  {Colless}, {Couch}, {De Propris}, {Drinkwater}, {Edge}, {Fine}, {Loveday},
  {Miller}, {Myers}, {Nichol}, {Outram}, {Pimbblet}, {Roseboom}, {Ross},
  {Schneider}, {Smith}, {Stoughton}, {Strauss}, \& {Wake}}]{cro09}
{Croom}, S.~M., {Richards}, G.~T., {Shanks}, T., {et~al.} 2009, \mnras, 392, 19

\bibitem[{{Daly} {et~al.}(2012){Daly}, {Sprinkle}, {O'Dea}, {Kharb}, \&
  {Baum}}]{dal12}
{Daly}, R.~A., {Sprinkle}, T.~B., {O'Dea}, C.~P., {Kharb}, P., \& {Baum}, S.~A.
  2012, \mnras, 423, 2498

\bibitem[{{de Gasperin} {et~al.}(2018){de Gasperin}, {Intema}, \&
  {Frail}}]{deg18}
{de Gasperin}, F., {Intema}, H.~T., \& {Frail}, D.~A. 2018, \mnras, 474, 5008

\bibitem[{{Delvecchio} {et~al.}(2014){Delvecchio}, {Gruppioni}, {Pozzi},
  {Berta}, {Zamorani}, {Cimatti}, {Lutz}, {Scott}, {Vignali}, {Cresci},
  {Feltre}, {Cooray}, {Vaccari}, {Fritz}, {Le Floc'h}, {Magnelli}, {Popesso},
  {Oliver}, {Bock}, {Carollo}, {Contini}, {Le F{\'e}vre}, {Lilly}, {Mainieri},
  {Renzini}, \& {Scodeggio}}]{del14}
{Delvecchio}, I., {Gruppioni}, C., {Pozzi}, F., {et~al.} 2014, \mnras, 439,
  2736

\bibitem[{{Delvecchio} {et~al.}(2018){Delvecchio}, {Smol{\v{c}}i{\'c}},
  {Zamorani}, {Rosario}, {Bondi}, {Marchesi}, {Miyaji}, {Novak}, {Sargent},
  {Alexander}, \& {Delhaize}}]{del18}
{Delvecchio}, I., {Smol{\v{c}}i{\'c}}, V., {Zamorani}, G., {et~al.} 2018,
  \mnras, 481, 4971

\bibitem[{{Dey} {et~al.}(2019){Dey}, {Schlegel}, {Lang}, {Blum}, {Burleigh},
  {Fan}, {Findlay}, {Finkbeiner}, {Herrera}, {Juneau}, {Landriau}, {Levi},
  {McGreer}, {Meisner}, {Myers}, {Moustakas}, {Nugent}, {Patej}, {Schlafly},
  {Walker}, {Valdes}, {Weaver}, {Y{\`e}che}, {Zou}, {Zhou}, {Abareshi},
  {Abbott}, {Abolfathi}, {Aguilera}, {Alam}, {Allen}, {Alvarez}, {Annis},
  {Ansarinejad}, {Aubert}, {Beechert}, {Bell}, {BenZvi}, {Beutler}, {Bielby},
  {Bolton}, {Brice{\~n}o}, {Buckley-Geer}, {Butler}, {Calamida}, {Carlberg},
  {Carter}, {Casas}, {Castander}, {Choi}, {Comparat}, {Cukanovaite}, {Delubac},
  {DeVries}, {Dey}, {Dhungana}, {Dickinson}, {Ding}, {Donaldson}, {Duan},
  {Duckworth}, {Eftekharzadeh}, {Eisenstein}, {Etourneau}, {Fagrelius},
  {Farihi}, {Fitzpatrick}, {Font-Ribera}, {Fulmer}, {G{\"a}nsicke},
  {Gaztanaga}, {George}, {Gerdes}, {Gontcho}, {Gorgoni}, {Green}, {Guy},
  {Harmer}, {Hernandez}, {Honscheid}, {Huang}, {James}, {Jannuzi}, {Jiang},
  {Joyce}, {Karcher}, {Karkar}, {Kehoe}, {Kneib}, {Kueter-Young}, {Lan},
  {Lauer}, {Le Guillou}, {Le Van Suu}, {Lee}, {Lesser}, {Perreault Levasseur},
  {Li}, {Mann}, {Marshall}, {Mart{\'\i}nez-V{\'a}zquez}, {Martini}, {du Mas des
  Bourboux}, {McManus}, {Meier}, {M{\'e}nard}, {Metcalfe},
  {Mu{\~n}oz-Guti{\'e}rrez}, {Najita}, {Napier}, {Narayan}, {Newman}, {Nie},
  {Nord}, {Norman}, {Olsen}, {Paat}, {Palanque-Delabrouille}, {Peng},
  {Poppett}, {Poremba}, {Prakash}, {Rabinowitz}, {Raichoor}, {Rezaie},
  {Robertson}, {Roe}, {Ross}, {Ross}, {Rudnick}, {Safonova}, {Saha},
  {S{\'a}nchez}, {Savary}, {Schweiker}, {Scott}, {Seo}, {Shan}, {Silva},
  {Slepian}, {Soto}, {Sprayberry}, {Staten}, {Stillman}, {Stupak}, {Summers},
  {Sien Tie}, {Tirado}, {Vargas-Maga{\~n}a}, {Vivas}, {Wechsler}, {Williams},
  {Yang}, {Yang}, {Yapici}, {Zaritsky}, {Zenteno}, {Zhang}, {Zhang}, {Zhou}, \&
  {Zhou}}]{dey19}
{Dey}, A., {Schlegel}, D.~J., {Lang}, D., {et~al.} 2019, \aj, 157, 168

\bibitem[{{Donato} {et~al.}(2001){Donato}, {Ghisellini}, {Tagliaferri}, \&
  {Fossati}}]{don01}
{Donato}, D., {Ghisellini}, G., {Tagliaferri}, G., \& {Fossati}, G. 2001, \aap,
  375, 739

\bibitem[{{Dong} \& {Wu}(2015)}]{don15}
{Dong}, A.-J. \& {Wu}, Q. 2015, \mnras, 453, 3447

\bibitem[{{Drinkwater} {et~al.}(2018){Drinkwater}, {Byrne}, {Blake},
  {Glazebrook}, {Brough}, {Colless}, {Couch}, {Croton}, {Croom}, {Davis},
  {Forster}, {Gilbank}, {Hinton}, {Jelliffe}, {Jurek}, {Li}, {Martin},
  {Pimbblet}, {Poole}, {Pracy}, {Sharp}, {Smillie}, {Spolaor}, {Wisnioski},
  {Woods}, {Wyder}, \& {Yee}}]{dri18}
{Drinkwater}, M.~J., {Byrne}, Z.~J., {Blake}, C., {et~al.} 2018, \mnras, 474,
  4151

\bibitem[{{Dunlop} \& {Peacock}(1990)}]{dun90}
{Dunlop}, J.~S. \& {Peacock}, J.~A. 1990, \mnras, 247, 19

\bibitem[{{Fabian}(2012)}]{fab12}
{Fabian}, A.~C. 2012, \araa, 50, 455

\bibitem[{{Falcke} {et~al.}(2004){Falcke}, {K{\"o}rding}, \& {Markoff}}]{fal04}
{Falcke}, H., {K{\"o}rding}, E., \& {Markoff}, S. 2004, \aap, 414, 895

\bibitem[{{Finlez} {et~al.}(2022){Finlez}, {Treister}, {Bauer}, {Keel}, {Koss},
  {Nagar}, {Sartori}, {Maksym}, {Venturi}, {Tub{\'\i}n}, \& {Harvey}}]{fin22}
{Finlez}, C., {Treister}, E., {Bauer}, F., {et~al.} 2022, \apj, 936, 88

\bibitem[{{Fossati} {et~al.}(1998){Fossati}, {Maraschi}, {Celotti}, {Comastri},
  \& {Ghisellini}}]{fos98}
{Fossati}, G., {Maraschi}, L., {Celotti}, A., {Comastri}, A., \& {Ghisellini},
  G. 1998, \mnras, 299, 433

\bibitem[{{Furusawa} {et~al.}(2018){Furusawa}, {Koike}, {Takata}, {Okura},
  {Miyatake}, {Lupton}, {Bickerton}, {Price}, {Bosch}, {Yasuda}, {Mineo},
  {Yamada}, {Miyazaki}, {Nakata}, {Koshida}, {Komiyama}, {Utsumi},
  {Kawanomoto}, {Jeschke}, {Noumaru}, {Schubert}, {Iwata}, {Finet},
  {Fujiyoshi}, {Tajitsu}, {Terai}, \& {Lee}}]{fur18}
{Furusawa}, H., {Koike}, M., {Takata}, T., {et~al.} 2018, \pasj, 70, S3

\bibitem[{{Gaia Collaboration} {et~al.}(2018){Gaia Collaboration}, {Brown},
  {Vallenari}, {Prusti}, {de Bruijne}, {Babusiaux}, {Bailer-Jones}, {Biermann},
  {Evans}, {Eyer}, \& et~al.}]{gai18}
{Gaia Collaboration}, {Brown}, A.~G.~A., {Vallenari}, A., {et~al.} 2018, \aap,
  616, A1

\bibitem[{{Gallo} {et~al.}(2003){Gallo}, {Fender}, \& {Pooley}}]{gal03}
{Gallo}, E., {Fender}, R.~P., \& {Pooley}, G.~G. 2003, \mnras, 344, 60

\bibitem[{{Gandhi} {et~al.}(2009){Gandhi}, {Horst}, {Smette}, {H{\"o}nig},
  {Comastri}, {Gilli}, {Vignali}, \& {Duschl}}]{gan09}
{Gandhi}, P., {Horst}, H., {Smette}, A., {et~al.} 2009, \aap, 502, 457

\bibitem[{{Ghisellini} {et~al.}(2017){Ghisellini}, {Righi}, {Costamante}, \&
  {Tavecchio}}]{ghi17}
{Ghisellini}, G., {Righi}, C., {Costamante}, L., \& {Tavecchio}, F. 2017,
  \mnras, 469, 255

\bibitem[{{Ghisellini} {et~al.}(2014){Ghisellini}, {Tavecchio}, {Maraschi},
  {Celotti}, \& {Sbarrato}}]{ghi14}
{Ghisellini}, G., {Tavecchio}, F., {Maraschi}, L., {Celotti}, A., \&
  {Sbarrato}, T. 2014, \nat, 515, 376

\bibitem[{{G{\"u}ltekin} {et~al.}(2019){G{\"u}ltekin}, {King}, {Cackett},
  {Nyland}, {Miller}, {Di Matteo}, {Markoff}, \& {Rupen}}]{gul19}
{G{\"u}ltekin}, K., {King}, A.~L., {Cackett}, E.~M., {et~al.} 2019, \apj, 871,
  80

\bibitem[{{Haardt} \& {Maraschi}(1991)}]{haa91}
{Haardt}, F. \& {Maraschi}, L. 1991, \apjl, 380, L51

\bibitem[{{Hamann} {et~al.}(2017){Hamann}, {Zakamska}, {Ross}, {Paris},
  {Alexandroff}, {Villforth}, {Richards}, {Herbst}, {Brandt}, {Cook}, {Denney},
  {Greene}, {Schneider}, \& {Strauss}}]{ham17}
{Hamann}, F., {Zakamska}, N.~L., {Ross}, N., {et~al.} 2017, \mnras, 464, 3431

\bibitem[{{Hannikainen} {et~al.}(1998){Hannikainen}, {Hunstead},
  {Campbell-Wilson}, \& {Sood}}]{han98}
{Hannikainen}, D.~C., {Hunstead}, R.~W., {Campbell-Wilson}, D., \& {Sood},
  R.~K. 1998, \aap, 337, 460

\bibitem[{{Hardcastle}(2018)}]{har18}
{Hardcastle}, M.~J. 2018, \mnras, 475, 2768

\bibitem[{{Hardcastle} {et~al.}(2019){Hardcastle}, {Williams}, {Best},
  {Croston}, {Duncan}, {R{\"o}ttgering}, {Sabater}, {Shimwell}, {Tasse},
  {Callingham}, {Cochrane}, {de Gasperin}, {G{\"u}rkan}, {Jarvis}, {Mahatma},
  {Miley}, {Mingo}, {Mooney}, {Morabito}, {O'Sullivan}, {Prandoni},
  {Shulevski}, \& {Smith}}]{har19}
{Hardcastle}, M.~J., {Williams}, W.~L., {Best}, P.~N., {et~al.} 2019, \aap,
  622, A12

\bibitem[{{Harwood} {et~al.}(2013){Harwood}, {Hardcastle}, {Croston}, \&
  {Goodger}}]{har13}
{Harwood}, J.~J., {Hardcastle}, M.~J., {Croston}, J.~H., \& {Goodger}, J.~L.
  2013, \mnras, 435, 3353

\bibitem[{{Hasinger} {et~al.}(2005){Hasinger}, {Miyaji}, \& {Schmidt}}]{has05}
{Hasinger}, G., {Miyaji}, T., \& {Schmidt}, M. 2005, \aap, 441, 417

\bibitem[{{Hawley} \& {Balbus}(2002)}]{haw02}
{Hawley}, J.~F. \& {Balbus}, S.~A. 2002, \apj, 573, 738

\bibitem[{{Heckman} \& {Best}(2014)}]{hec14}
{Heckman}, T.~M. \& {Best}, P.~N. 2014, \araa, 52, 589

\bibitem[{{Heinz} \& {Merloni}(2004)}]{hei04}
{Heinz}, S. \& {Merloni}, A. 2004, \mnras, 355, L1

\bibitem[{{Heinz} \& {Sunyaev}(2003)}]{hei03}
{Heinz}, S. \& {Sunyaev}, R.~A. 2003, \mnras, 343, L59

\bibitem[{{Helfand} {et~al.}(2015){Helfand}, {White}, \& {Becker}}]{hel15}
{Helfand}, D.~J., {White}, R.~L., \& {Becker}, R.~H. 2015, \apj, 801, 26

\bibitem[{{HI4PI Collaboration} {et~al.}(2016){HI4PI Collaboration}, {Ben
  Bekhti}, {Fl{\"o}er}, {Keller}, {Kerp}, {Lenz}, {Winkel}, {Bailin},
  {Calabretta}, {Dedes}, {Ford}, {Gibson}, {Haud}, {Janowiecki}, {Kalberla},
  {Lockman}, {McClure-Griffiths}, {Murphy}, {Nakanishi}, {Pisano}, \&
  {Staveley-Smith}}]{ben16}
{HI4PI Collaboration}, {Ben Bekhti}, N., {Fl{\"o}er}, L., {et~al.} 2016, \aap,
  594, A116

\bibitem[{{Hickox} {et~al.}(2009){Hickox}, {Jones}, {Forman}, {Murray},
  {Kochanek}, {Eisenstein}, {Jannuzi}, {Dey}, {Brown}, {Stern}, {Eisenhardt},
  {Gorjian}, {Brodwin}, {Narayan}, {Cool}, {Kenter}, {Caldwell}, \&
  {Anderson}}]{hic09}
{Hickox}, R.~C., {Jones}, C., {Forman}, W.~R., {et~al.} 2009, \apj, 696, 891

\bibitem[{{Ho}(2008)}]{ho08}
{Ho}, L.~C. 2008, \araa, 46, 475

\bibitem[{Holt {et~al.}(2008)Holt, Tadhunter, \& Morganti}]{hol08}
Holt, J., Tadhunter, C.~N., \& Morganti, R. 2008, Mon. Not. R. Astron. Soc.,
  387, 639

\bibitem[{{Huang} {et~al.}(2018){Huang}, {Leauthaud}, {Murata}, {Bosch},
  {Price}, {Lupton}, {Mand elbaum}, {Lackner}, {Bickerton}, {Miyazaki},
  {Coupon}, \& {Tanaka}}]{hua18}
{Huang}, S., {Leauthaud}, A., {Murata}, R., {et~al.} 2018, \pasj, 70, S6

\bibitem[{{Hwang} {et~al.}(2018){Hwang}, {Zakamska}, {Alexandroff}, {Hamann},
  {Greene}, {Perrotta}, \& {Richards}}]{hwa18}
{Hwang}, H.-C., {Zakamska}, N.~L., {Alexandroff}, R.~M., {et~al.} 2018, \mnras,
  477, 830

\bibitem[{{Ichikawa} {et~al.}(2019{\natexlab{a}}){Ichikawa}, {Kawamuro},
  {Shidatsu}, {Ricci}, {Bae}, {Matsuoka}, {Shin}, {Toba}, {Ueda}, \&
  {Ueda}}]{ich19c}
{Ichikawa}, K., {Kawamuro}, T., {Shidatsu}, M., {et~al.} 2019{\natexlab{a}},
  \apjl, 883, L13

\bibitem[{{Ichikawa} {et~al.}(2015){Ichikawa}, {Packham}, {Ramos Almeida},
  {Asensio Ramos}, {Alonso-Herrero}, {Gonz{\'a}lez-Mart{\'{\i}}n},
  {Lopez-Rodriguez}, {Ueda}, {D{\'{\i}}az-Santos}, {Elitzur}, {H{\"o}nig},
  {Imanishi}, {Levenson}, {Mason}, {Perlman}, \& {Alsip}}]{ich15}
{Ichikawa}, K., {Packham}, C., {Ramos Almeida}, C., {et~al.} 2015, \apj, 803,
  57

\bibitem[{{Ichikawa} {et~al.}(2019{\natexlab{b}}){Ichikawa}, {Ricci}, {Ueda},
  {Bauer}, {Kawamuro}, {Koss}, {Oh}, {Rosario}, {Shimizu}, {Stalevski},
  {Fuller}, {Packham}, \& {Trakhtenbrot}}]{ich19a}
{Ichikawa}, K., {Ricci}, C., {Ueda}, Y., {et~al.} 2019{\natexlab{b}}, \apj,
  870, 31

\bibitem[{{Ichikawa} {et~al.}(2017){Ichikawa}, {Ricci}, {Ueda}, {Matsuoka},
  {Toba}, {Kawamuro}, {Trakhtenbrot}, \& {Koss}}]{ich17a}
{Ichikawa}, K., {Ricci}, C., {Ueda}, Y., {et~al.} 2017, \apj, 835, 74

\bibitem[{{Ichikawa} {et~al.}(2016){Ichikawa}, {Ueda}, {Shidatsu}, {Kawamuro},
  \& {Matsuoka}}]{ich16}
{Ichikawa}, K., {Ueda}, J., {Shidatsu}, M., {Kawamuro}, T., \& {Matsuoka}, K.
  2016, \pasj, 68, 9

\bibitem[{{Ichikawa} {et~al.}(2012){Ichikawa}, {Ueda}, {Terashima}, {Oyabu},
  {Gandhi}, {Matsuta}, \& {Nakagawa}}]{ich12}
{Ichikawa}, K., {Ueda}, Y., {Terashima}, Y., {et~al.} 2012, \apj, 754, 45

\bibitem[{{Ichikawa} {et~al.}(2021){Ichikawa}, {Yamashita}, {Toba}, {Nagao},
  {Inayoshi}, {Charisi}, {He}, {Wagner}, {Akiyama}, {Vijarnwannaluk}, {Chen},
  {Kajisawa}, {Kawamuro}, {Lee}, {Matsuoka}, {Schramm}, {Suh}, {Tanaka},
  {Uchiyama}, {Ueda}, {Pflugradt}, \& {Fukuchi}}]{ich21}
{Ichikawa}, K., {Yamashita}, T., {Toba}, Y., {et~al.} 2021, arXiv e-prints,
  arXiv:2108.02781

\bibitem[{{Ichimaru}(1977)}]{ich77}
{Ichimaru}, S. 1977, \apj, 214, 840

\bibitem[{{Igumenshchev} \& {Abramowicz}(1999)}]{igu99}
{Igumenshchev}, I.~V. \& {Abramowicz}, M.~A. 1999, \mnras, 303, 309

\bibitem[{{Ilbert} {et~al.}(2006){Ilbert}, {Arnouts}, {McCracken},
  {Bolzonella}, {Bertin}, {Le F{\`e}vre}, {Mellier}, {Zamorani}, {Pell{\`o}},
  {Iovino}, {Tresse}, {Le Brun}, {Bottini}, {Garilli}, {Maccagni}, {Picat},
  {Scaramella}, {Scodeggio}, {Vettolani}, {Zanichelli}, {Adami}, {Bardelli},
  {Cappi}, {Charlot}, {Ciliegi}, {Contini}, {Cucciati}, {Foucaud}, {Franzetti},
  {Gavignaud}, {Guzzo}, {Marano}, {Marinoni}, {Mazure}, {Meneux}, {Merighi},
  {Paltani}, {Pollo}, {Pozzetti}, {Radovich}, {Zucca}, {Bondi}, {Bongiorno},
  {Busarello}, {de La Torre}, {Gregorini}, {Lamareille}, {Mathez}, {Merluzzi},
  {Ripepi}, {Rizzo}, \& {Vergani}}]{ilb06}
{Ilbert}, O., {Arnouts}, S., {McCracken}, H.~J., {et~al.} 2006, \aap, 457, 841

\bibitem[{{Inayoshi} {et~al.}(2016){Inayoshi}, {Haiman}, \&
  {Ostriker}}]{ina16b}
{Inayoshi}, K., {Haiman}, Z., \& {Ostriker}, J.~P. 2016, \mnras, 459, 3738

\bibitem[{{Inayoshi} {et~al.}(2018){Inayoshi}, {Ichikawa}, \& {Haiman}}]{ina18}
{Inayoshi}, K., {Ichikawa}, K., \& {Haiman}, Z. 2018, \apjl, 863, L36

\bibitem[{{Inayoshi} {et~al.}(2019){Inayoshi}, {Ichikawa}, {Ostriker}, \&
  {Kuiper}}]{ina19}
{Inayoshi}, K., {Ichikawa}, K., {Ostriker}, J.~P., \& {Kuiper}, R. 2019,
  \mnras, 486, 5377

\bibitem[{{Ineson} {et~al.}(2017){Ineson}, {Croston}, {Hardcastle}, \&
  {Mingo}}]{ine17}
{Ineson}, J., {Croston}, J.~H., {Hardcastle}, M.~J., \& {Mingo}, B. 2017,
  \mnras, 467, 1586

\bibitem[{{Inoue} {et~al.}(2017){Inoue}, {Doi}, {Tanaka}, {Sikora}, \&
  {Madejski}}]{ino17}
{Inoue}, Y., {Doi}, A., {Tanaka}, Y.~T., {Sikora}, M., \& {Madejski}, G.~M.
  2017, \apj, 840, 46

\bibitem[{{Inoue} \& {Totani}(2009)}]{ino09}
{Inoue}, Y. \& {Totani}, T. 2009, \apj, 702, 523

\bibitem[{{Ishino} {et~al.}(2020){Ishino}, {Matsuoka}, {Koyama}, {Saeda},
  {Strauss}, {Goulding}, {Imanishi}, {Kawaguchi}, {Minezaki}, {Nagao},
  {Noboriguchi}, {Schramm}, {Silverman}, {Taniguchi}, \& {Toba}}]{ish20}
{Ishino}, T., {Matsuoka}, Y., {Koyama}, S., {et~al.} 2020, \pasj, 72, 83

\bibitem[{{Isobe} {et~al.}(1990){Isobe}, {Feigelson}, {Akritas}, \&
  {Babu}}]{iso90}
{Isobe}, T., {Feigelson}, E.~D., {Akritas}, M.~G., \& {Babu}, G.~J. 1990, \apj,
  364, 104

\bibitem[{{Ivezi{\'c}} {et~al.}(2002){Ivezi{\'c}}, {Menou}, {Knapp}, {Strauss},
  {Lupton}, {Vand en Berk}, {Richards}, {Tremonti}, {Weinstein}, {Anderson},
  {Bahcall}, {Becker}, {Bernardi}, {Blanton}, {Eisenstein}, {Fan},
  {Finkbeiner}, {Finlator}, {Frieman}, {Gunn}, {Hall}, {Kim}, {Kinkhabwala},
  {Narayanan}, {Rockosi}, {Schlegel}, {Schneider}, {Strateva}, {SubbaRao},
  {Thakar}, {Voges}, {White}, {Yanny}, {Brinkmann}, {Doi}, {Fukugita},
  {Hennessy}, {Munn}, {Nichol}, \& {York}}]{ive02}
{Ivezi{\'c}}, {\v{Z}}., {Menou}, K., {Knapp}, G.~R., {et~al.} 2002, \aj, 124,
  2364

\bibitem[{{Jimenez-Gallardo} {et~al.}(2019){Jimenez-Gallardo}, {Massaro},
  {Capetti}, {Prieto}, {Paggi}, {Baldi}, {Grossova}, {Ostorero},
  {Siemiginowska}, \& {Viada}}]{jim19}
{Jimenez-Gallardo}, A., {Massaro}, F., {Capetti}, A., {et~al.} 2019, \aap, 627,
  A108

\bibitem[{{Jones} {et~al.}(2009){Jones}, {Read}, {Saunders}, {Colless},
  {Jarrett}, {Parker}, {Fairall}, {Mauch}, {Sadler}, {Watson}, {Burton},
  {Campbell}, {Cass}, {Croom}, {Dawe}, {Fiegert}, {Frankcombe}, {Hartley},
  {Huchra}, {James}, {Kirby}, {Lahav}, {Lucey}, {Mamon}, {Moore}, {Peterson},
  {Prior}, {Proust}, {Russell}, {Safouris}, {Wakamatsu}, {Westra}, \&
  {Williams}}]{jon09}
{Jones}, D.~H., {Read}, M.~A., {Saunders}, W., {et~al.} 2009, \mnras, 399, 683

\bibitem[{{Jun} {et~al.}(2015){Jun}, {Im}, {Lee}, {Ohyama}, {Woo}, {Fan},
  {Goto}, {Kim}, {Kim}, {Kim}, {Lee}, {Nakagawa}, {Pearson}, \&
  {Serjeant}}]{jun15}
{Jun}, H.~D., {Im}, M., {Lee}, H.~M., {et~al.} 2015, \apj, 806, 109

\bibitem[{{Jurlin} {et~al.}(2021){Jurlin}, {Brienza}, {Morganti}, {Wadadekar},
  {Ishwara-Chandra}, {Maddox}, \& {Mahatma}}]{jur21}
{Jurlin}, N., {Brienza}, M., {Morganti}, R., {et~al.} 2021, \aap, 653, A110

\bibitem[{{Kaiser} {et~al.}(1997){Kaiser}, {Dennett-Thorpe}, \&
  {Alexander}}]{kai97}
{Kaiser}, C.~R., {Dennett-Thorpe}, J., \& {Alexander}, P. 1997, \mnras, 292,
  723

\bibitem[{{Kawanomoto} {et~al.}(2018){Kawanomoto}, {Uraguchi}, {Komiyama},
  {Miyazaki}, {Furusawa}, {Finet}, {Hattori}, {Wang}, {Yasuda}, \&
  {Suzuki}}]{kaw18}
{Kawanomoto}, S., {Uraguchi}, F., {Komiyama}, Y., {et~al.} 2018, \pasj, 70, 66

\bibitem[{{Keel} {et~al.}(2017){Keel}, {Lintott}, {Maksym}, {Bennert},
  {Chojnowski}, {Moiseev}, {Smirnova}, {Schawinski}, {Sartori}, {Urry},
  {Pancoast}, {Schirmer}, {Scott}, {Showley}, \& {Flatland}}]{kee17}
{Keel}, W.~C., {Lintott}, C.~J., {Maksym}, W.~P., {et~al.} 2017, \apj, 835, 256

\bibitem[{{Keel} {et~al.}(2015){Keel}, {Maksym}, {Bennert}, {Lintott},
  {Chojnowski}, {Moiseev}, {Smirnova}, {Schawinski}, {Urry}, {Evans},
  {Pancoast}, {Scott}, {Showley}, \& {Flatland}}]{kee15}
{Keel}, W.~C., {Maksym}, W.~P., {Bennert}, V.~N., {et~al.} 2015, \aj, 149, 155

\bibitem[{{Kellermann} {et~al.}(1989){Kellermann}, {Sramek}, {Schmidt},
  {Shaffer}, \& {Green}}]{kel89}
{Kellermann}, K.~I., {Sramek}, R., {Schmidt}, M., {Shaffer}, D.~B., \& {Green},
  R. 1989, \aj, 98, 1195

\bibitem[{{Kimball} {et~al.}(2011){Kimball}, {Kellermann}, {Condon},
  {Ivezi{\'c}}, \& {Perley}}]{kim11}
{Kimball}, A.~E., {Kellermann}, K.~I., {Condon}, J.~J., {Ivezi{\'c}}, {\v{Z}}.,
  \& {Perley}, R.~A. 2011, \apjl, 739, L29

\bibitem[{{Komissarov} \& {Gubanov}(1994)}]{kom94}
{Komissarov}, S.~S. \& {Gubanov}, A.~G. 1994, \aap, 285, 27

\bibitem[{{Komiyama} {et~al.}(2018){Komiyama}, {Obuchi}, {Nakaya}, {Kamata},
  {Kawanomoto}, {Utsumi}, {Miyazaki}, {Uraguchi}, {Furusawa}, {Morokuma},
  {Uchida}, {Miyatake}, {Mineo}, {Fujimori}, {Aihara}, {Karoji}, {Gunn}, \&
  {Wang}}]{kom18}
{Komiyama}, Y., {Obuchi}, Y., {Nakaya}, H., {et~al.} 2018, \pasj, 70, S2

\bibitem[{{Kormendy} \& {Ho}(2013)}]{kor13}
{Kormendy}, J. \& {Ho}, L.~C. 2013, \araa, 51, 511

\bibitem[{{Li} {et~al.}(2021){Li}, {Silverman}, {Ding}, {Strauss}, {Goulding},
  {Birrer}, {Yesuf}, {Xue}, {Kawinwanichakij}, {Matsuoka}, {Toba}, {Nagao},
  {Schramm}, \& {Inayoshi}}]{li21}
{Li}, J., {Silverman}, J.~D., {Ding}, X., {et~al.} 2021, arXiv e-prints,
  arXiv:2105.06568

\bibitem[{{Li} {et~al.}(2008){Li}, {Wu}, \& {Wang}}]{li08}
{Li}, Z.-Y., {Wu}, X.-B., \& {Wang}, R. 2008, \apj, 688, 826

\bibitem[{{Liska} {et~al.}(2019){Liska}, {Tchekhovskoy}, {Ingram}, \& {van der
  Klis}}]{lis19}
{Liska}, M., {Tchekhovskoy}, A., {Ingram}, A., \& {van der Klis}, M. 2019,
  \mnras, 487, 550

\bibitem[{{Liu} {et~al.}(2021){Liu}, {Buchner}, {Nandra}, {Merloni}, {Dwelly},
  {Sanders}, {Salvato}, {Arcodia}, {Brusa}, {Wolf}, {Georgakakis}, {Boller},
  {Krumpe}, {Lamer}, {Waddell}, {Urrutia}, {Schwope}, {Robrade}, {Wilms},
  {Dauser}, {Comparat}, {Toba}, {Ichikawa}, {Iwasawa}, {Shen}, \& {Ibarra
  Medel}}]{liu21}
{Liu}, T., {Buchner}, J., {Nandra}, K., {et~al.} 2021, arXiv e-prints,
  arXiv:2106.14522

\bibitem[{{Lopez-Rodriguez} {et~al.}(2018){Lopez-Rodriguez}, {Alonso-Herrero},
  {Diaz-Santos}, {Gonzalez-Martin}, {Ichikawa}, {Levenson}, {Martinez-Paredes},
  {Nikutta}, {Packham}, {Perlman}, {Ramos Almeida}, {Rodriguez-Espinosa}, \&
  {Telesco}}]{lop18}
{Lopez-Rodriguez}, E., {Alonso-Herrero}, A., {Diaz-Santos}, T., {et~al.} 2018,
  \mnras, 478, 2350

\bibitem[{{Lopez-Rodriguez} {et~al.}(2014){Lopez-Rodriguez}, {Packham},
  {Tadhunter}, {Mason}, {Perlman}, {Alonso-Herrero}, {Ramos Almeida},
  {Ichikawa}, {Levenson}, {Rodr{\'\i}guez-Espinosa}, {{\'A}lvarez},
  {Ram{\'\i}rez}, \& {Telesco}}]{lop14}
{Lopez-Rodriguez}, E., {Packham}, C., {Tadhunter}, C., {et~al.} 2014, \apj,
  793, 81

\bibitem[{{Lubow} {et~al.}(1994){Lubow}, {Papaloizou}, \& {Pringle}}]{lub94}
{Lubow}, S.~H., {Papaloizou}, J.~C.~B., \& {Pringle}, J.~E. 1994, \mnras, 267,
  235

\bibitem[{{Luo} {et~al.}(2015){Luo}, {Zhao}, {Zhao}, {Deng}, {Liu}, {Jing},
  {Wang}, {Zhang}, {Shi}, {Cui}, {Chu}, {Li}, {Bai}, {Wu}, {Cai}, {Cao}, {Cao},
  {Carlin}, {Chen}, {Chen}, {Chen}, {Chen}, {Chen}, {Chen}, {Chen},
  {Christlieb}, {Chu}, {Cui}, {Dong}, {Du}, {Fan}, {Feng}, {Fu}, {Gao}, {Gong},
  {Gu}, {Guo}, {Han}, {He}, {Hou}, {Hou}, {Hou}, {Hu}, {Hu}, {Hu}, {Huo},
  {Jia}, {Jiang}, {Jiang}, {Jiang}, {Jin}, {Kong}, {Kong}, {Lei}, {Li}, {Li},
  {Li}, {Li}, {Li}, {Li}, {Li}, {Li}, {Li}, {Li}, {Li}, {Li}, {Liang}, {Lin},
  {Liu}, {Liu}, {Liu}, {Liu}, {Lu}, {Luo}, {Mao}, {Newberg}, {Ni}, {Qi}, {Qi},
  {Shen}, {Shi}, {Song}, {Song}, {Su}, {Su}, {Tang}, {Tao}, {Tian}, {Wang},
  {Wang}, {Wang}, {Wang}, {Wang}, {Wang}, {Wang}, {Wang}, {Wang}, {Wang},
  {Wang}, {Wang}, {Wang}, {Wang}, {Wang}, {Wang}, {Wang}, {Wang}, {Wang},
  {Wang}, {Wei}, {Wei}, {Wu}, {Wu}, {Wu}, {Wu}, {Xing}, {Xu}, {Xu}, {Xu},
  {Yan}, {Yang}, {Yang}, {Yang}, {Yang}, {Yao}, {Yu}, {Yuan}, {Yuan}, {Yuan},
  {Yuan}, {Zhai}, {Zhang}, {Zhang}, {Zhang}, {Zhang}, {Zhang}, {Zhang},
  {Zhang}, {Zhang}, {Zhao}, {Zhou}, {Zhou}, {Zhu}, {Zhu}, {Zou}, \&
  {Zuo}}]{luo15}
{Luo}, A.~L., {Zhao}, Y.-H., {Zhao}, G., {et~al.} 2015, Research in Astronomy
  and Astrophysics, 15, 1095

\bibitem[{{Maraschi} \& {Rovetti}(1994)}]{mar94}
{Maraschi}, L. \& {Rovetti}, F. 1994, \apj, 436, 79

\bibitem[{{Marconi} {et~al.}(2004){Marconi}, {Risaliti}, {Gilli}, {Hunt},
  {Maiolino}, \& {Salvati}}]{mar04}
{Marconi}, A., {Risaliti}, G., {Gilli}, R., {et~al.} 2004, \mnras, 351, 169

\bibitem[{{Mason} {et~al.}(2012){Mason}, {Lopez-Rodriguez}, {Packham},
  {Alonso-Herrero}, {Levenson}, {Radomski}, {Ramos Almeida}, {Colina},
  {Elitzur}, {Aretxaga}, {Roche}, \& {Oi}}]{mas12}
{Mason}, R.~E., {Lopez-Rodriguez}, E., {Packham}, C., {et~al.} 2012, \aj, 144,
  11

\bibitem[{{Massaro} {et~al.}(2015){Massaro}, {Maselli}, {Leto}, {Marchegiani},
  {Perri}, {Giommi}, \& {Piranomonte}}]{mas15}
{Massaro}, E., {Maselli}, A., {Leto}, C., {et~al.} 2015, \apss, 357, 75

\bibitem[{{Mateos} {et~al.}(2015){Mateos}, {Carrera}, {Alonso-Herrero},
  {Rovilos}, {Hern{\'a}n-Caballero}, {Barcons}, {Blain}, {Caccianiga}, {Della
  Ceca}, \& {Severgnini}}]{mat15}
{Mateos}, S., {Carrera}, F.~J., {Alonso-Herrero}, A., {et~al.} 2015, \mnras,
  449, 1422

\bibitem[{{Matsuta} {et~al.}(2012){Matsuta}, {Gandhi}, {Dotani}, {Nakagawa},
  {Isobe}, {Ueda}, {Ichikawa}, {Terashima}, {Oyabu}, {Yamamura}, \&
  {Stawarz}}]{mat12a}
{Matsuta}, K., {Gandhi}, P., {Dotani}, T., {et~al.} 2012, \apj, 753, 104

\bibitem[{{Mauch} \& {Sadler}(2007)}]{mau07}
{Mauch}, T. \& {Sadler}, E.~M. 2007, \mnras, 375, 931

\bibitem[{{McConnell} {et~al.}(2011){McConnell}, {Ma}, {Gebhardt}, {Wright},
  {Murphy}, {Lauer}, {Graham}, \& {Richstone}}]{mcc11}
{McConnell}, N.~J., {Ma}, C.-P., {Gebhardt}, K., {et~al.} 2011, \nat, 480, 215

\bibitem[{{McKinney} {et~al.}(2014){McKinney}, {Tchekhovskoy}, {Sadowski}, \&
  {Narayan}}]{mck14}
{McKinney}, J.~C., {Tchekhovskoy}, A., {Sadowski}, A., \& {Narayan}, R. 2014,
  \mnras, 441, 3177

\bibitem[{{McNamara} \& {Nulsen}(2007)}]{mcn07}
{McNamara}, B.~R. \& {Nulsen}, P.~E.~J. 2007, \araa, 45, 117

\bibitem[{{Merloni} {et~al.}(2014){Merloni}, {Bongiorno}, {Brusa}, {Iwasawa},
  {Mainieri}, {Magnelli}, {Salvato}, {Berta}, {Cappelluti}, {Comastri},
  {Fiore}, {Gilli}, {Koekemoer}, {Le Floc'h}, {Lusso}, {Lutz}, {Miyaji},
  {Pozzi}, {Riguccini}, {Rosario}, {Silverman}, {Symeonidis}, {Treister},
  {Vignali}, \& {Zamorani}}]{mer14}
{Merloni}, A., {Bongiorno}, A., {Brusa}, M., {et~al.} 2014, \mnras, 437, 3550

\bibitem[{{Merloni} {et~al.}(2003){Merloni}, {Heinz}, \& {di Matteo}}]{mer03}
{Merloni}, A., {Heinz}, S., \& {di Matteo}, T. 2003, \mnras, 345, 1057

\bibitem[{{Merloni} {et~al.}(2012){Merloni}, {Predehl}, {Becker},
  {B{\"o}hringer}, {Boller}, {Brunner}, {Brusa}, {Dennerl}, {Freyberg},
  {Friedrich}, {Georgakakis}, {Haberl}, {Hasinger}, {Meidinger}, {Mohr},
  {Nandra}, {Rau}, {Reiprich}, {Robrade}, {Salvato}, {Santangelo}, {Sasaki},
  {Schwope}, {Wilms}, \& {German eROSITA Consortium}}]{mer12}
{Merloni}, A., {Predehl}, P., {Becker}, W., {et~al.} 2012, arXiv e-prints,
  arXiv:1209.3114

\bibitem[{{Miyazaki} {et~al.}(2018){Miyazaki}, {Komiyama}, {Kawanomoto}, {Doi},
  {Furusawa}, {Hamana}, {Hayashi}, {Ikeda}, {Kamata}, {Karoji}, {Koike},
  {Kurakami}, {Miyama}, {Morokuma}, {Nakata}, {Namikawa}, {Nakaya}, {Nariai},
  {Obuchi}, {Oishi}, {Okada}, {Okura}, {Tait}, {Takata}, {Tanaka}, {Tanaka},
  {Terai}, {Tomono}, {Uraguchi}, {Usuda}, {Utsumi}, {Yamada}, {Yamanoi},
  {Aihara}, {Fujimori}, {Mineo}, {Miyatake}, {Oguri}, {Uchida}, {Tanaka},
  {Yasuda}, {Takada}, {Murayama}, {Nishizawa}, {Sugiyama}, {Chiba}, {Futamase},
  {Wang}, {Chen}, {Ho}, {Liaw}, {Chiu}, {Ho}, {Lai}, {Lee}, {Jeng}, {Iwamura},
  {Armstrong}, {Bickerton}, {Bosch}, {Gunn}, {Lupton}, {Loomis}, {Price},
  {Smith}, {Strauss}, {Turner}, {Suzuki}, {Miyazaki}, {Muramatsu}, {Yamamoto},
  {Endo}, {Ezaki}, {Ito}, {Kawaguchi}, {Sofuku}, {Taniike}, {Akutsu}, {Dojo},
  {Kasumi}, {Matsuda}, {Imoto}, {Miwa}, {Suzuki}, {Takeshi}, \&
  {Yokota}}]{miy18}
{Miyazaki}, S., {Komiyama}, Y., {Kawanomoto}, S., {et~al.} 2018, \pasj, 70, S1

\bibitem[{{Morgan} {et~al.}(2010){Morgan}, {Kochanek}, {Morgan}, \&
  {Falco}}]{mor10}
{Morgan}, C.~W., {Kochanek}, C.~S., {Morgan}, N.~D., \& {Falco}, E.~E. 2010,
  \apj, 712, 1129

\bibitem[{{Morganti} {et~al.}(2021){Morganti}, {Jurlin}, {Oosterloo},
  {Brienza}, {Orru'}, {Kutkin}, {Prandoni}, {Adams}, {Denes}, {Hess},
  {Shulevski}, {van der Hulst}, \& {Ziemke}}]{mor21}
{Morganti}, R., {Jurlin}, N., {Oosterloo}, T., {et~al.} 2021, arXiv e-prints,
  arXiv:2111.04776

\bibitem[{Morganti {et~al.}(2005)Morganti, Tadhunter, \& Oosterloo}]{mor05}
Morganti, R., Tadhunter, C.~N., \& Oosterloo, T.~a. 2005, Astron. Astrophys.,
  444, L9

\bibitem[{{Mortlock} {et~al.}(2011){Mortlock}, {Warren}, {Venemans}, {Patel},
  {Hewett}, {McMahon}, {Simpson}, {Theuns}, {Gonz{\'a}les-Solares}, {Adamson},
  {Dye}, {Hambly}, {Hirst}, {Irwin}, {Kuiper}, {Lawrence}, \&
  {R{\"o}ttgering}}]{mor11}
{Mortlock}, D.~J., {Warren}, S.~J., {Venemans}, B.~P., {et~al.} 2011, \nat,
  474, 616

\bibitem[{{Mullaney} {et~al.}(2011){Mullaney}, {Alexander}, {Goulding}, \&
  {Hickox}}]{mul11}
{Mullaney}, J.~R., {Alexander}, D.~M., {Goulding}, A.~D., \& {Hickox}, R.~C.
  2011, \mnras, 414, 1082

\bibitem[{{Mullaney} {et~al.}(2012){Mullaney}, {Daddi}, {B{\'e}thermin},
  {Elbaz}, {Juneau}, {Pannella}, {Sargent}, {Alexander}, \& {Hickox}}]{mul12}
{Mullaney}, J.~R., {Daddi}, E., {B{\'e}thermin}, M., {et~al.} 2012, \apjl, 753,
  L30

\bibitem[{{Narayan} {et~al.}(2000){Narayan}, {Igumenshchev}, \&
  {Abramowicz}}]{nar00}
{Narayan}, R., {Igumenshchev}, I.~V., \& {Abramowicz}, M.~A. 2000, \apj, 539,
  798

\bibitem[{{Narayan} \& {Yi}(1994)}]{nar94}
{Narayan}, R. \& {Yi}, I. 1994, \apjl, 428, L13

\bibitem[{{Narayan} \& {Yi}(1995)}]{nar95}
{Narayan}, R. \& {Yi}, I. 1995, \apj, 444, 231

\bibitem[{{Nemmen} \& {Tchekhovskoy}(2015)}]{nem15}
{Nemmen}, R.~S. \& {Tchekhovskoy}, A. 2015, \mnras, 449, 316

\bibitem[{Nesvadba {et~al.}(2017)Nesvadba, De~Breuck, Lehnert, Best, \&
  Collet}]{nes17}
Nesvadba, N. P.~H., De~Breuck, C., Lehnert, M.~D., Best, P.~N., \& Collet, C.
  2017, Astron. Astrophys. Suppl. Ser., 599, A123

\bibitem[{{Nishizawa} {et~al.}(2020){Nishizawa}, {Hsieh}, {Tanaka}, \&
  {Takata}}]{nis20}
{Nishizawa}, A.~J., {Hsieh}, B.-C., {Tanaka}, M., \& {Takata}, T. 2020, arXiv
  e-prints, arXiv:2003.01511

\bibitem[{{Ohsuga} {et~al.}(2009){Ohsuga}, {Mineshige}, {Mori}, \&
  {Kato}}]{ohs09}
{Ohsuga}, K., {Mineshige}, S., {Mori}, M., \& {Kato}, Y. 2009, \pasj, 61, L7

\bibitem[{{Ohsuga} {et~al.}(2005){Ohsuga}, {Mori}, {Nakamoto}, \&
  {Mineshige}}]{ohs05}
{Ohsuga}, K., {Mori}, M., {Nakamoto}, T., \& {Mineshige}, S. 2005, \apj, 628,
  368

\bibitem[{{Onoue} {et~al.}(2019){Onoue}, {Kashikawa}, {Matsuoka}, {Kato},
  {Izumi}, {Nagao}, {Strauss}, {Harikane}, {Imanishi}, {Ito}, {Iwasawa},
  {Kawaguchi}, {Lee}, {Noboriguchi}, {Suh}, {Tanaka}, \& {Toba}}]{ono19}
{Onoue}, M., {Kashikawa}, N., {Matsuoka}, Y., {et~al.} 2019, \apj, 880, 77

\bibitem[{{Orr} \& {Browne}(1982)}]{orr82}
{Orr}, M.~J.~L. \& {Browne}, I.~W.~A. 1982, \mnras, 200, 1067

\bibitem[{{Padovani}(2016)}]{pad16}
{Padovani}, P. 2016, \aapr, 24, 13

\bibitem[{{Pflugradt} {et~al.}(2022){Pflugradt}, {Ichikawa}, {Akiyama},
  {Kokubo}, {Vijarnwannaluk}, {Noda}, \& {Chen}}]{pfl22}
{Pflugradt}, J., {Ichikawa}, K., {Akiyama}, M., {et~al.} 2022, \apj, 938, 75

\bibitem[{{Predehl} {et~al.}(2021){Predehl}, {Andritschke}, {Arefiev},
  {Babyshkin}, {Batanov}, {Becker}, {B{\"o}hringer}, {Bogomolov}, {Boller},
  {Borm}, {Bornemann}, {Br{\"a}uninger}, {Br{\"u}ggen}, {Brunner}, {Brusa},
  {Bulbul}, {Buntov}, {Burwitz}, {Burkert}, {Clerc}, {Churazov}, {Coutinho},
  {Dauser}, {Dennerl}, {Doroshenko}, {Eder}, {Emberger}, {Eraerds},
  {Finoguenov}, {Freyberg}, {Friedrich}, {Friedrich}, {F{\"u}rmetz},
  {Georgakakis}, {Gilfanov}, {Granato}, {Grossberger}, {Gueguen}, {Gureev},
  {Haberl}, {H{\"a}lker}, {Hartner}, {Hasinger}, {Huber}, {Ji}, {Kienlin},
  {Kink}, {Korotkov}, {Kreykenbohm}, {Lamer}, {Lomakin}, {Lapshov}, {Liu},
  {Maitra}, {Meidinger}, {Menz}, {Merloni}, {Mernik}, {Mican}, {Mohr},
  {M{\"u}ller}, {Nandra}, {Nazarov}, {Pacaud}, {Pavlinsky}, {Perinati},
  {Pfeffermann}, {Pietschner}, {Ramos-Ceja}, {Rau}, {Reiffers}, {Reiprich},
  {Robrade}, {Salvato}, {Sanders}, {Santangelo}, {Sasaki}, {Scheuerle},
  {Schmid}, {Schmitt}, {Schwope}, {Shirshakov}, {Steinmetz}, {Stewart},
  {Str{\"u}der}, {Sunyaev}, {Tenzer}, {Tiedemann}, {Tr{\"u}mper}, {Voron},
  {Weber}, {Wilms}, \& {Yaroshenko}}]{pre21}
{Predehl}, P., {Andritschke}, R., {Arefiev}, V., {et~al.} 2021, \aap, 647, A1

\bibitem[{{Privon} {et~al.}(2012){Privon}, {Baum}, {O'Dea}, {Gallimore},
  {Noel-Storr}, {Axon}, \& {Robinson}}]{pri12}
{Privon}, G.~C., {Baum}, S.~A., {O'Dea}, C.~P., {et~al.} 2012, \apj, 747, 46

\bibitem[{{Rafferty} {et~al.}(2006){Rafferty}, {McNamara}, {Nulsen}, \&
  {Wise}}]{raf06}
{Rafferty}, D.~A., {McNamara}, B.~R., {Nulsen}, P.~E.~J., \& {Wise}, M.~W.
  2006, \apj, 652, 216

\bibitem[{{Rakshit} {et~al.}(2020){Rakshit}, {Stalin}, \& {Kotilainen}}]{rak20}
{Rakshit}, S., {Stalin}, C.~S., \& {Kotilainen}, J. 2020, \apjs, 249, 17

\bibitem[{{Ramos Almeida} {et~al.}(2009){Ramos Almeida}, {Levenson},
  {Rodr{\'\i}guez Espinosa}, {Alonso-Herrero}, {Asensio Ramos}, {Radomski},
  {Packham}, {Fisher}, \& {Telesco}}]{ram09}
{Ramos Almeida}, C., {Levenson}, N.~A., {Rodr{\'\i}guez Espinosa}, J.~M.,
  {et~al.} 2009, \apj, 702, 1127

\bibitem[{{Ricci} {et~al.}(2017){Ricci}, {Trakhtenbrot}, {Koss}, {Ueda}, {Del
  Vecchio}, {Treister}, {Schawinski}, {Paltani}, {Oh}, {Lamperti}, {Berney},
  {Gandhi}, {Ichikawa}, {Bauer}, {Ho}, {Asmus}, {Beckmann}, {Soldi},
  {Balokovi{\'c}}, {Gehrels}, \& {Markwardt}}]{ric17}
{Ricci}, C., {Trakhtenbrot}, B., {Koss}, M.~J., {et~al.} 2017, \apjs, 233, 17

\bibitem[{{Ricci} {et~al.}(2015){Ricci}, {Ueda}, {Koss}, {Trakhtenbrot},
  {Bauer}, \& {Gandhi}}]{ric15}
{Ricci}, C., {Ueda}, Y., {Koss}, M.~J., {et~al.} 2015, \apjl, 815, L13

\bibitem[{{Ross} {et~al.}(2018){Ross}, {Ford}, {Graham}, {McKernan}, {Stern},
  {Meisner}, {Assef}, {Dey}, {Drake}, {Jun}, \& {Lang}}]{ros18}
{Ross}, N.~P., {Ford}, K.~E.~S., {Graham}, M., {et~al.} 2018, \mnras, 480, 4468

\bibitem[{{Ross} {et~al.}(2015){Ross}, {Hamann}, {Zakamska}, {Richards},
  {Villforth}, {Strauss}, {Greene}, {Alexandroff}, {Brandt}, {Liu}, {Myers},
  {P{\^a}ris}, \& {Schneider}}]{ros15}
{Ross}, N.~P., {Hamann}, F., {Zakamska}, N.~L., {et~al.} 2015, \mnras, 453,
  3932

\bibitem[{{Ruiz} {et~al.}(2018){Ruiz}, {Corral}, {Mountrichas}, \&
  {Georgantopoulos}}]{rui18}
{Ruiz}, A., {Corral}, A., {Mountrichas}, G., \& {Georgantopoulos}, I. 2018,
  \aap, 618, A52

\bibitem[{{Ryan} {et~al.}(2017){Ryan}, {Ressler}, {Dolence}, {Tchekhovskoy},
  {Gammie}, \& {Quataert}}]{rya17}
{Ryan}, B.~R., {Ressler}, S.~M., {Dolence}, J.~C., {et~al.} 2017, \apjl, 844,
  L24

\bibitem[{{Saade} {et~al.}(2022){Saade}, {Brightman}, {Stern}, {Malkan}, \&
  {Garc{\'\i}a}}]{saa22}
{Saade}, M.~L., {Brightman}, M., {Stern}, D., {Malkan}, M.~A., \&
  {Garc{\'\i}a}, J.~A. 2022, \apj, 936, 162

\bibitem[{{Sadler} {et~al.}(2002){Sadler}, {Jackson}, {Cannon}, {McIntyre},
  {Murphy}, {Bland-Hawthorn}, {Bridges}, {Cole}, {Colless}, {Collins}, {Couch},
  {Dalton}, {De Propris}, {Driver}, {Efstathiou}, {Ellis}, {Frenk},
  {Glazebrook}, {Lahav}, {Lewis}, {Lumsden}, {Maddox}, {Madgwick}, {Norberg},
  {Peacock}, {Peterson}, {Sutherland}, \& {Taylor}}]{sad02}
{Sadler}, E.~M., {Jackson}, C.~A., {Cannon}, R.~D., {et~al.} 2002, \mnras, 329,
  227

\bibitem[{{Salvato} {et~al.}(2018){Salvato}, {Buchner}, {Budav{\'a}ri},
  {Dwelly}, {Merloni}, {Brusa}, {Rau}, {Fotopoulou}, \& {Nandra}}]{sal18}
{Salvato}, M., {Buchner}, J., {Budav{\'a}ri}, T., {et~al.} 2018, \mnras, 473,
  4937

\bibitem[{{Salvato} {et~al.}(2009){Salvato}, {Hasinger}, {Ilbert}, {Zamorani},
  {Brusa}, {Scoville}, {Rau}, {Capak}, {Arnouts}, {Aussel}, {Bolzonella},
  {Buongiorno}, {Cappelluti}, {Caputi}, {Civano}, {Cook}, {Elvis}, {Gilli},
  {Jahnke}, {Kartaltepe}, {Impey}, {Lamareille}, {Le Floc'h}, {Lilly},
  {Mainieri}, {McCarthy}, {McCracken}, {Mignoli}, {Mobasher}, {Murayama},
  {Sasaki}, {Sanders}, {Schiminovich}, {Shioya}, {Shopbell}, {Silverman},
  {Smol{\v{c}}i{\'c}}, {Surace}, {Taniguchi}, {Thompson}, {Trump}, {Urry}, \&
  {Zamojski}}]{sal09}
{Salvato}, M., {Hasinger}, G., {Ilbert}, O., {et~al.} 2009, \apj, 690, 1250

\bibitem[{{Salvato} {et~al.}(2011){Salvato}, {Ilbert}, {Hasinger}, {Rau},
  {Civano}, {Zamorani}, {Brusa}, {Elvis}, {Vignali}, {Aussel}, {Comastri},
  {Fiore}, {Le Floc'h}, {Mainieri}, {Bardelli}, {Bolzonella}, {Bongiorno},
  {Capak}, {Caputi}, {Cappelluti}, {Carollo}, {Contini}, {Garilli}, {Iovino},
  {Fotopoulou}, {Fruscione}, {Gilli}, {Halliday}, {Kneib}, {Kakazu},
  {Kartaltepe}, {Koekemoer}, {Kovac}, {Ideue}, {Ikeda}, {Impey}, {Le Fevre},
  {Lamareille}, {Lanzuisi}, {Le Borgne}, {Le Brun}, {Lilly}, {Maier},
  {Manohar}, {Masters}, {McCracken}, {Messias}, {Mignoli}, {Mobasher}, {Nagao},
  {Pello}, {Puccetti}, {Perez-Montero}, {Renzini}, {Sargent}, {Sanders},
  {Scodeggio}, {Scoville}, {Shopbell}, {Silvermann}, {Taniguchi}, {Tasca},
  {Tresse}, {Trump}, \& {Zucca}}]{sal11}
{Salvato}, M., {Ilbert}, O., {Hasinger}, G., {et~al.} 2011, \apj, 742, 61

\bibitem[{{Salvato} {et~al.}(2021){Salvato}, {Wolf}, {Dwelly}, {Georgakakis},
  {Brusa}, {Merloni}, {Liu}, {Toba}, {Nandra}, {Lamer}, {Buchner}, {Schneider},
  {Freund}, {Rau}, {Schwope}, {Nishizawa}, {Klein}, {Arcodia}, {Comparat},
  {Musiimenta}, {Nagao}, {Brunner}, {Malyali}, {Finoguenov}, {Anderson},
  {Shen}, {Ibarra-Mendel}, {Trump}, {Brandt}, {Urry}, {Rivera}, {Krumpe},
  {Urrutia}, {Miyaji}, {Ichikawa}, {Schneider}, {Fresco}, {Wilms}, {Boller},
  {Haase}, {Brownstein}, {Lane}, {Bizyaev}, \& {Nitschelm}}]{sal21}
{Salvato}, M., {Wolf}, J., {Dwelly}, T., {et~al.} 2021, arXiv e-prints,
  arXiv:2106.14520

\bibitem[{{Sartori} {et~al.}(2018){Sartori}, {Schawinski}, {Koss}, {Ricci},
  {Treister}, {Stern}, {Lansbury}, {Maksym}, {Balokovi{\'c}}, {Gandhi}, {Keel},
  \& {Ballantyne}}]{sar18a}
{Sartori}, L.~F., {Schawinski}, K., {Koss}, M.~J., {et~al.} 2018, \mnras, 474,
  2444

\bibitem[{{Schawinski} {et~al.}(2010){Schawinski}, {Evans}, {Virani}, {Urry},
  {Keel}, {Natarajan}, {Lintott}, {Manning}, {Coppi}, {Kaviraj}, {Bamford},
  {J{\'o}zsa}, {Garrett}, {van Arkel}, {Gay}, \& {Fortson}}]{sch10}
{Schawinski}, K., {Evans}, D.~A., {Virani}, S., {et~al.} 2010, \apjl, 724, L30

\bibitem[{{Schinnerer} {et~al.}(2007){Schinnerer}, {Smol{\v{c}}i{\'c}},
  {Carilli}, {Bondi}, {Ciliegi}, {Jahnke}, {Scoville}, {Aussel}, {Bertoldi},
  {Blain}, {Impey}, {Koekemoer}, {Le Fevre}, \& {Urry}}]{sch07}
{Schinnerer}, E., {Smol{\v{c}}i{\'c}}, V., {Carilli}, C.~L., {et~al.} 2007,
  \apjs, 172, 46

\bibitem[{{Schirmer} {et~al.}(2016){Schirmer}, {Malhotra}, {Levenson}, {Fu},
  {Davies}, {Keel}, {Torrey}, {Bennert}, {Pancoast}, \& {Turner}}]{sch16}
{Schirmer}, M., {Malhotra}, S., {Levenson}, N.~A., {et~al.} 2016, \mnras, 463,
  1554

\bibitem[{{Schlafly} {et~al.}(2019){Schlafly}, {Meisner}, \& {Green}}]{sch19}
{Schlafly}, E.~F., {Meisner}, A.~M., \& {Green}, G.~M. 2019, \apjs, 240, 30

\bibitem[{{Schmitt}(1985)}]{sch85}
{Schmitt}, J.~H.~M.~M. 1985, \apj, 293, 178

\bibitem[{{Shabala} \& {Godfrey}(2013)}]{sha13}
{Shabala}, S.~S. \& {Godfrey}, L.~E.~H. 2013, \apj, 769, 129

\bibitem[{{Shabala} {et~al.}(2020){Shabala}, {Jurlin}, {Morganti}, {Brienza},
  {Hardcastle}, {Godfrey}, {Krause}, \& {Turner}}]{sha20}
{Shabala}, S.~S., {Jurlin}, N., {Morganti}, R., {et~al.} 2020, \mnras, 496,
  1706

\bibitem[{{Shen} {et~al.}(2011){Shen}, {Richards}, {Strauss}, {Hall},
  {Schneider}, {Snedden}, {Bizyaev}, {Brewington}, {Malanushenko},
  {Malanushenko}, {Oravetz}, {Pan}, \& {Simmons}}]{she11}
{Shen}, Y., {Richards}, G.~T., {Strauss}, M.~A., {et~al.} 2011, \apjs, 194, 45

\bibitem[{{Sikora} {et~al.}(2007){Sikora}, {Stawarz}, \& {Lasota}}]{sik07}
{Sikora}, M., {Stawarz}, {\L}., \& {Lasota}, J.-P. 2007, \apj, 658, 815

\bibitem[{{Smol{\v{c}}i{\'c}} {et~al.}(2017{\natexlab{a}}){Smol{\v{c}}i{\'c}},
  {Delvecchio}, {Zamorani}, {Baran}, {Novak}, {Delhaize}, {Schinnerer},
  {Berta}, {Bondi}, {Ciliegi}, {Capak}, {Civano}, {Karim}, {Le Fevre},
  {Ilbert}, {Laigle}, {Marchesi}, {McCracken}, {Tasca}, {Salvato}, \&
  {Vardoulaki}}]{smo17b}
{Smol{\v{c}}i{\'c}}, V., {Delvecchio}, I., {Zamorani}, G., {et~al.}
  2017{\natexlab{a}}, \aap, 602, A2

\bibitem[{{Smol{\v{c}}i{\'c}} {et~al.}(2017{\natexlab{b}}){Smol{\v{c}}i{\'c}},
  {Novak}, {Bondi}, {Ciliegi}, {Mooley}, {Schinnerer}, {Zamorani}, {Navarrete},
  {Bourke}, {Karim}, {Vardoulaki}, {Leslie}, {Delhaize}, {Carilli}, {Myers},
  {Baran}, {Delvecchio}, {Miettinen}, {Banfield}, {Balokovi{\'c}}, {Bertoldi},
  {Capak}, {Frail}, {Hallinan}, {Hao}, {Herrera Ruiz}, {Horesh}, {Ilbert},
  {Intema}, {Jeli{\'c}}, {Kl{\"o}ckner}, {Krpan}, {Kulkarni}, {McCracken},
  {Laigle}, {Middleberg}, {Murphy}, {Sargent}, {Scoville}, \& {Sheth}}]{smo17a}
{Smol{\v{c}}i{\'c}}, V., {Novak}, M., {Bondi}, M., {et~al.} 2017{\natexlab{b}},
  \aap, 602, A1

\bibitem[{{Soltan}(1982)}]{sol82}
{Soltan}, A. 1982, \mnras, 200, 115

\bibitem[{{Stern}(2015)}]{ste15}
{Stern}, D. 2015, \apj, 807, 129

\bibitem[{{Stone} \& {Pringle}(2001)}]{sto01}
{Stone}, J.~M. \& {Pringle}, J.~E. 2001, \mnras, 322, 461

\bibitem[{{Stone} {et~al.}(1999){Stone}, {Pringle}, \& {Begelman}}]{sto99}
{Stone}, J.~M., {Pringle}, J.~E., \& {Begelman}, M.~C. 1999, \mnras, 310, 1002

\bibitem[{{Sunyaev} {et~al.}(2021){Sunyaev}, {Arefiev}, {Babyshkin},
  {Bogomolov}, {Borisov}, {Buntov}, {Brunner}, {Burenin}, {Churazov},
  {Coutinho}, {Eder}, {Eismont}, {Freyberg}, {Gilfanov}, {Gureyev}, {Hasinger},
  {Khabibullin}, {Kolmykov}, {Komovkin}, {Krivonos}, {Lapshov}, {Levin},
  {Lomakin}, {Lutovinov}, {Medvedev}, {Merloni}, {Mernik}, {Mikhailov},
  {Molodzov}, {Mzhelsky}, {Mueller}, {Nandra}, {Nazarov}, {Pavlinsky},
  {Poghodin}, {Predehl}, {Robrade}, {Sazonov}, {Scheuerle}, {Shirshakov},
  {Tkachenko}, \& {Voron}}]{sun21}
{Sunyaev}, R., {Arefiev}, V., {Babyshkin}, V., {et~al.} 2021, arXiv e-prints,
  arXiv:2104.13267

\bibitem[{{Tadhunter}(2016)}]{tad16}
{Tadhunter}, C. 2016, \aapr, 24, 10

\bibitem[{{Takeo} {et~al.}(2020){Takeo}, {Inayoshi}, \& {Mineshige}}]{tak20}
{Takeo}, E., {Inayoshi}, K., \& {Mineshige}, S. 2020, arXiv e-prints,
  arXiv:2002.07187

\bibitem[{{Tanaka} {et~al.}(2018){Tanaka}, {Coupon}, {Hsieh}, {Mineo},
  {Nishizawa}, {Speagle}, {Furusawa}, {Miyazaki}, \& {Murayama}}]{tan18}
{Tanaka}, M., {Coupon}, J., {Hsieh}, B.-C., {et~al.} 2018, \pasj, 70, S9

\bibitem[{{Tchekhovskoy} {et~al.}(2011){Tchekhovskoy}, {Narayan}, \&
  {McKinney}}]{tch11}
{Tchekhovskoy}, A., {Narayan}, R., \& {McKinney}, J.~C. 2011, \mnras, 418, L79

\bibitem[{{Toba} {et~al.}(2021){Toba}, {Liu}, {Urrutia}, {Salvato}, {Li},
  {Ueda}, {Brusa}, {Yutani}, {Wada}, {Nishizawa}, {Buchner}, {Nagao},
  {Merloni}, {Akiyama}, {Arcodia}, {Hsieh}, {Ichikawa}, {Imanishi}, {Inoue},
  {Kawaguchi}, {Lamer}, {Nandra}, {Silverman}, \& {Terashima}}]{tob21}
{Toba}, Y., {Liu}, T., {Urrutia}, T., {et~al.} 2021, arXiv e-prints,
  arXiv:2106.14527

\bibitem[{{Toba} {et~al.}(2019){Toba}, {Yamashita}, {Nagao}, {Wang}, {Ueda},
  {Ichikawa}, {Kawaguchi}, {Akiyama}, {Hsieh}, {Kajisawa}, {Lee}, {Matsuoka},
  {Noboriguchi}, {Onoue}, {Schramm}, {Tanaka}, \& {Komiyama}}]{tob19a}
{Toba}, Y., {Yamashita}, T., {Nagao}, T., {et~al.} 2019, \apjs, 243, 15

\bibitem[{{Trakhtenbrot}(2014)}]{tra14}
{Trakhtenbrot}, B. 2014, \apjl, 789, L9

\bibitem[{{Truemper}(1993)}]{tru93}
{Truemper}, J. 1993, Science, 260, 1769

\bibitem[{{Uchiyama} {et~al.}(2021){Uchiyama}, {Yamashita}, {Toshikawa},
  {Kashikawa}, {Ichikawa}, {Kubo}, {Ito}, {Kawakatu}, {Nagao}, {Toba}, {Ono},
  {Harikane}, {Imanishi}, {Kajisawa}, {Lee}, \& {Liang}}]{uch21}
{Uchiyama}, H., {Yamashita}, T., {Toshikawa}, J., {et~al.} 2021, arXiv
  e-prints, arXiv:2112.01684

\bibitem[{{Ueda} {et~al.}(2014){Ueda}, {Iono}, {Yun}, {Crocker}, {Narayanan},
  {Komugi}, {Espada}, {Hatsukade}, {Kaneko}, {Matsuda}, {Tamura}, {Wilner},
  {Kawabe}, \& {Pan}}]{ued14}
{Ueda}, J., {Iono}, D., {Yun}, M.~S., {et~al.} 2014, \apjs, 214, 1

\bibitem[{{Urry} \& {Padovani}(1995)}]{urr95}
{Urry}, C.~M. \& {Padovani}, P. 1995, \pasp, 107, 803

\bibitem[{{van Velzen} \& {Falcke}(2013)}]{van13b}
{van Velzen}, S. \& {Falcke}, H. 2013, \aap, 557, L7

\bibitem[{{Villar-Mart{\'{\i}}n} {et~al.}(2018){Villar-Mart{\'{\i}}n},
  {Cabrera-Lavers}, {Humphrey}, {Silva}, {Ramos Almeida}, {Piqueras-L{\'o}pez},
  \& {Emonts}}]{vil18}
{Villar-Mart{\'{\i}}n}, M., {Cabrera-Lavers}, A., {Humphrey}, A., {et~al.}
  2018, \mnras, 474, 2302

\bibitem[{{Wang} {et~al.}(2006){Wang}, {Wu}, \& {Kong}}]{wan06}
{Wang}, R., {Wu}, X.-B., \& {Kong}, M.-Z. 2006, \apj, 645, 890

\bibitem[{{White} {et~al.}(1997){White}, {Becker}, {Helfand}, \&
  {Gregg}}]{whi97}
{White}, R.~L., {Becker}, R.~H., {Helfand}, D.~J., \& {Gregg}, M.~D. 1997,
  \apj, 475, 479

\bibitem[{{Willott} {et~al.}(1999){Willott}, {Rawlings}, {Blundell}, \&
  {Lacy}}]{wil99}
{Willott}, C.~J., {Rawlings}, S., {Blundell}, K.~M., \& {Lacy}, M. 1999,
  \mnras, 309, 1017

\bibitem[{{Willott} {et~al.}(2001){Willott}, {Rawlings}, {Blundell}, {Lacy}, \&
  {Eales}}]{wil01}
{Willott}, C.~J., {Rawlings}, S., {Blundell}, K.~M., {Lacy}, M., \& {Eales},
  S.~A. 2001, \mnras, 322, 536

\bibitem[{{Wright} {et~al.}(2010){Wright}, {Eisenhardt}, {Mainzer}, {Ressler},
  {Cutri}, {Jarrett}, {Kirkpatrick}, {Padgett}, {McMillan}, {Skrutskie},
  {Stanford}, {Cohen}, {Walker}, {Mather}, {Leisawitz}, {Gautier}, {McLean},
  {Benford}, {Lonsdale}, {Blain}, {Mendez}, {Irace}, {Duval}, {Liu}, {Royer},
  {Heinrichsen}, {Howard}, {Shannon}, {Kendall}, {Walsh}, {Larsen}, {Cardon},
  {Schick}, {Schwalm}, {Abid}, {Fabinsky}, {Naes}, \& {Tsai}}]{wri10}
{Wright}, E.~L., {Eisenhardt}, P.~R.~M., {Mainzer}, A.~K., {et~al.} 2010, \aj,
  140, 1868

\bibitem[{{Wu} {et~al.}(2015){Wu}, {Wang}, {Fan}, {Yi}, {Zuo}, {Bian}, {Jiang},
  {McGreer}, {Wang}, {Yang}, {Yang}, {Thompson}, \& {Beletsky}}]{wu15}
{Wu}, X.-B., {Wang}, F., {Fan}, X., {et~al.} 2015, \nat, 518, 512

\bibitem[{{Wylezalek} {et~al.}(2018){Wylezalek}, {Zakamska}, {Greene},
  {Riffel}, {Drory}, {Andrews}, {Merloni}, \& {Thomas}}]{wyl18}
{Wylezalek}, D., {Zakamska}, N.~L., {Greene}, J.~E., {et~al.} 2018, \mnras,
  474, 1499

\bibitem[{{Xie} \& {Yuan}(2012)}]{xie12}
{Xie}, F.-G. \& {Yuan}, F. 2012, \mnras, 427, 1580

\bibitem[{{Yamashita} {et~al.}(2018){Yamashita}, {Nagao}, {Akiyama}, {He},
  {Ikeda}, {Tanaka}, {Niida}, {Kajisawa}, {Matsuoka}, {Nobuhara}, {Lee},
  {Morokuma}, {Toba}, {Kawaguchi}, \& {Noboriguchi}}]{yam18}
{Yamashita}, T., {Nagao}, T., {Akiyama}, M., {et~al.} 2018, \apj, 866, 140

\bibitem[{{Yamashita} {et~al.}(2020){Yamashita}, {Nagao}, {Ikeda}, {Toba},
  {Kajisawa}, {Ono}, {Tanaka}, {Akiyama}, {Harikane}, {Ichikawa}, {Kawaguchi},
  {Kawamuro}, {Kohno}, {Lee}, {Lee}, {Matsuoka}, {Niida}, {Ogura}, {Onoue}, \&
  {Uchiyama}}]{yam20}
{Yamashita}, T., {Nagao}, T., {Ikeda}, H., {et~al.} 2020, \aj, 160, 60

\bibitem[{{Yang} {et~al.}(2020){Yang}, {Wang}, {Fan}, {Hennawi}, {Davies},
  {Yue}, {Banados}, {Wu}, {Venemans}, {Barth}, {Bian}, {Boutsia}, {Decarli},
  {Farina}, {Green}, {Jiang}, {Li}, {Mazzucchelli}, \& {Walter}}]{yan20}
{Yang}, J., {Wang}, F., {Fan}, X., {et~al.} 2020, \apjl, 897, L14

\bibitem[{{Zakamska} \& {Greene}(2014)}]{zak14}
{Zakamska}, N.~L. \& {Greene}, J.~E. 2014, \mnras, 442, 784

\bibitem[{{Zakamska} {et~al.}(2016){Zakamska}, {Hamann}, {P{\^a}ris}, {Brandt},
  {Greene}, {Strauss}, {Villforth}, {Wylezalek}, {Alexandroff}, \&
  {Ross}}]{zak16}
{Zakamska}, N.~L., {Hamann}, F., {P{\^a}ris}, I., {et~al.} 2016, \mnras, 459,
  3144

\end{thebibliography}
\end{document}